\documentclass[usenatbib]{mnras}

\usepackage{natbib}
\usepackage{delarray}
\usepackage{xcolor}
\usepackage{amsmath}
\usepackage{amssymb}
\usepackage{here}
\usepackage{graphicx}
\usepackage[utf8]{inputenc}
\usepackage{float}
\usepackage{epsfig}
\usepackage{tensor}
\usepackage{xspace}
\usepackage{multirow}
\usepackage[]{units}
\usepackage{newtxtext,newtxmath}

\usepackage[T1]{fontenc}
\usepackage{ae,aecompl}

\newcommand{\appropto}{\mathrel{\vcenter{
  \offinterlineskip\halign{\hfil$##$\cr
    \propto\cr\noalign{\kern2pt}\sim\cr\noalign{\kern-2pt}}}}}

\newcommand{\mach}{\mathcal{M}}

\newcommand{\be}{\begin{equation}} \newcommand{\ee}{\end{equation}}

\newcommand{\solarmass}{\,{\rm M}_\odot}
\newcommand{\msun}{\solarmass}

\newcommand{\dderiv}{\mathrm{d}}

\newcommand{\Bvector}{\mathbf{B}}

\newcommand{\pc}{\,{\rm pc }}

\newcommand{\K}{\,{\rm K }}
\newcommand{\kelvin}{\K}
\newcommand{\percc}{\mathrm{cm}^{-3}}

\newcommand{\acknowledgments}{\begin{small}\section*{Acknowledgments}\end{small}}
\newcommand\altaffilmark[1]{$^{#1}$}
\newcommand\altaffiltext[1]{$^{#1}$}

\newcommand{\myquote}[1]{``#1''}

\usepackage[normalem]{ulem}

\title[Cloud evolution across cosmic time]{Evolution of giant molecular clouds across cosmic time}

\author[Guszejnov, Grudi{\'c}, Offner, Boylan-Kolchin,  Faucher-Gig{\`e}re, Wetzel, Benincasa \&\ Loebman]{
\parbox[t]{\textwidth}{ D\'avid Guszejnov\altaffilmark{1}\thanks{E-mail:guszejnov@utexas.edu}, Michael Y. Grudi{\'c}\altaffilmark{2}, Stella S. R. Offner\altaffilmark{1}, Michael Boylan-Kolchin\altaffilmark{1}, Claude-Andr{\'e} Faucher-Gigu{\`e}re\altaffilmark{3}, Andrew  Wetzel\altaffilmark{4}, Samantha M. Benincasa\altaffilmark{4}, and Sarah Loebman\altaffilmark{4}}
\vspace*{6pt} \\
\altaffiltext{1}{Department of Astronomy, The University of Texas at Austin, Austin, TX 78712, USA} \\
\altaffiltext{2}{TAPIR, MC 350-17, California Institute of Technology, Pasadena, CA 91125, USA} \\
\altaffiltext{3}{CIERA and Department of Physics and Astronomy, Northwestern University, 2145 Sheridan Road, Evanston, IL 60208, USA}\\
\altaffiltext{4}{Department of Physics, University of California, Davis, CA, USA 95616}
}

\date{To be submitted to MNRAS, \today \vspace{-0.6cm}}
\begin{document}
\maketitle
\label{firstpage}

\begin{abstract}
Giant molecular clouds (GMCs) are well-studied in the local Universe, however, exactly how their properties vary during galaxy evolution is poorly understood due to challenging resolution requirements, both observational and computational. We present the first time-dependent analysis of giant molecular clouds in a Milky Way-like galaxy and an LMC-like dwarf galaxy of the FIRE-2 (Feedback In Realistic Environments) simulation suite, which have sufficient resolution to predict the bulk properties of GMCs in cosmological galaxy formation self-consistently. We show explicitly that the majority of star formation outside the galactic center occurs within self-gravitating gas structures that have properties consistent with observed bound GMCs. We find that the typical cloud bulk properties such as mass and surface density do not vary more than a factor of 2 in any systematic way after the first Gyr of cosmic evolution within a given galaxy from its progenitor.
While the median properties are constant, the tails of the distributions can briefly undergo drastic changes, which can produce very massive and dense self-gravitating gas clouds. Once the galaxy forms, we identify only two systematic trends in bulk properties over cosmic time: a steady increase in metallicity produced by previous stellar populations and a weak decrease in bulk cloud temperatures. With the exception of metallicity we find no significant differences in cloud properties between the Milky Way-like and dwarf galaxies. These results have important implications for cosmological star and star cluster formation and put especially strong constraints on theories relating the stellar initial mass function to cloud properties.
\end{abstract}

\begin{keywords}
ISM: clouds -- stars: formation -- galaxies: ISM -- galaxies: star formation -- turbulence -- cosmology: theory
\vspace{-1.0cm}
\end{keywords}

\section{Introduction}\label{sec:intro}

In our Galaxy, the majority of the molecular gas in the interstellar medium (ISM) is found in giant molecular clouds (GMCs). GMCs are the dominant sites of star formation in the local Universe \citep[see reviews of][]{mckee_star_formation, Dobbs_2014_GMC_review}, and  their properties therefore provide the initial condition for star formation. 

Over the years, many theories have been proposed that link the properties of star formation (e.g., rate, initial mass function) to the initial conditions of the star formation process \citep[e.g.][]{batebonell2005, hc08, padoan_nordlund_2011_imf, krumholz_stellar_mass_origin}. For example, a systematic change in the initial mass function would lead to different supernova rates and 
metallicities, with major effects on the evolution of the galaxy. Understanding the evolution of GMC properties over cosmic time could thus provide invaluable insight into the star formation histories of galaxies.

A crucial yet fraught element in the study of GMCs is their identification and characterization. The gas in GMCs is cold ($<\!100\,\kelvin$) and dense ($>\!100\,\percc$), making it unfeasible to observe them directly from $\mathrm{H_{2}}$ emission lines. Instead, observations rely on emission from tracer molecules (mostly CO) to identify clouds (see the review of \citealt{heyer_dame_2015} for an observational overview). Identifying clouds from these emission maps is a non-trivial exercise as these maps only contain position-position-velocity information, giving an incomplete picture of the inherently 6 dimensional data. Many observers rely on dendrogram methods that identify nested structures around a local intensity maximum in either 2D or 3D; essentially, each pixel is assigned to the lowest density structure it resides in \citep[e.g.,][]{rosolowsky2008_dendogram}. GMCs are selected based on a choice of dendrogram parameters (e.g., maximum number of substructures) that are set in a way to recover previously identified, \myquote{well-known} GMCs \citep[i.e., in ][]{rice2016_mw_gmc_catalogue}. More advanced schemes (e.g., SCIMES, see \citealt{Colombo_2015_GMC_identification_dendo}) accomplish the same task by using grouping algorithms like spectral clustering to identify individual GMCs. Alternatively clouds can be identified by an appropriately chosen iso-temperature surface in the emission map \citep{Rosolowsky_2006_GMC_CPROPS}.

Connecting observations of GMCs to a theoretical framework for their understanding is also both pressing and difficult. Several studies have attempted to simulate GMCs in galaxies and compare their properties with observations through synthetic observations \citep[e.g.,][]{Pan2015_synth_obs_GMC,Ward_2016_simulated_clouds,Duarte_Cabral_2016_synth_obs_GMC,Riching_2016_GMC_chemical_evol_in_sims, Grisdale_2018_simulated_clouds, Lakhlani_2019_GMC_FIRE_present}. Recently, advances in numerical methods led to cosmological scale simulations that can resolve GMC scale objects ($\sim 10^{5}\,\msun$), allowing a more faithful comparison, and potentially allowing us to follow their evolution through cosmic time and account for the effects of events such as galaxy mergers. So far, only a few such studies have been done, most of which concentrate on comparing the properties of clouds identified in the simulations to the present day observable GMCs (e.g., \citealt{Pettitt_2018_GMC_galaxy_sim, Dobbs_2019_M33_GMC_FoF}, see \citealt{Oklopvic_2017_FIRE_highz_clumps} for a high-redshift comparison).

With current observations, it is extremely challenging to observe GMCs at higher redshifts, mainly due to the relatively small size of GMCs compared to the resolution of observations \citep[e.g.,][]{Dessauges_Zavadsky_2015_highz_clumps}.
Preliminary results from surveys that exploit gravitational lensing to enhance their resolution \citep[e.g.,][]{Cava_2018_giant_clump_highz, Sharma_2018_hires_lensed_highz_galaxy} are beginning to inform our understanding of properties of molecular gas and star-forming regions at early cosmic times. The ISM of these galaxies shows an increase in velocity dispersion \citep{Tacconi_2013_massive_galaxy_gas, Wisnioski_2015_KMOS3D} and a decrease in molecular gas fraction \citep{Dessauges_Zavadsky_2017_molecular_gas_highz} and star formation efficiency \citep{Pavesi_2018_low_sfe_high_z_galaxies} with increasing redshift. It should be noted that the galaxies observed in these measurements are not \myquote{median} MW progenitors at their prospective redshifts -- they had already reached the mass of the MW at redshifts 2-4, making them relatively rare galaxies (more likely to evolve into the present day ellipticals). The median progenitor of present-day MW-mass systems (what we seek to study here) was likely closer to the present-day LMC in mass scale, making it extremely challenging to observe with current instruments.




In this paper, we study the cosmic evolution of GMCs in a simulated MW-like spiral and a dwarf galaxy within the FIRE-2 simulations \cite{hopkins2014_fire, hopkins_fire2}. Since resolving all but the most massive GMCs in galaxies at high redshifts is beyond the capabilities of current telescopes, comparing with direct observations is not possible. Thus our aim in this paper is to investigate how the  \emph{initial conditions of star formation evolve over time}. That is why in Section we define GMCs as the largest self-gravitating clouds of the ISM, a definition motivated by the physics of star formation \citep[e.g.,][]{rosolowsky2008_dendogram, hopkins_2012_excursion_set}. We will show that this definition reveals a population of objects in the simulations whose statistical properties are broadly similar to the populations of GMCs observed at low redshift. For a more direct comparison with observation see \cite{Lakhlani_2019_GMC_FIRE_present}.

The layout of the paper is as follows: in Section \ref{sec:methods} we define the bulk cloud properties our study focuses on and give a brief summary of the FIRE simulations we utilize, while Section \ref{sec:cloudphinder} discusses the motivation behind our adopted GMC definition. In Section \ref{sec:results} we show that the bulk properties of GMCs in a MW-like simulated galaxy show essentially no trend over cosmic time, with the exception of metallicity and a related factor of 2 change in bulk temperature. Section \ref{sec:discussion} discusses the implications of these results, while Appendix \ref{sec:m11q} contains the results for a simulated dwarf galaxy.

\section{Methods}\label{sec:methods}
\subsection{The case for studying bound clouds}
\label{sec:methods:boundclouds}
We wish to study the properties of the gas structures that can be understood as the direct progenitors of stellar associations and clusters and determine how these change throughout the cosmological evolution of galaxies. In essence, we seek to organize the ISM into 
self-contained units that can be mapped onto stars to a reasonable degree of approximation. It is important to note here that such a picture is not likely to be entirely correct or rigorous: the formation, evolution and dispersion of GMCs is thought to be a highly dynamic process in which ongoing accretion and cloud-cloud mergers can defy the notion of isolated units of star-forming gas \citep{dobbs_2013, ibanez_2017}. Nevertheless, it is reasonable to presume the existence of  sub-regions of the ISM within which the internal evolution occurs over shorter timescales than external processes and the system behaves in an approximately self-contained manner\footnote{Note that while the FIRE simulations we are using do resolve all GMCs down to about an order of magnitude above our resolution limit ($7100\,\msun$), they are not treated as isolated units. We only treat them as separate entities during post-processing when we apply our cloud identification algorithm.}.

When cataloguing such systems, we wish to avoid definitions that impose a characteristic scale upon the system, either in length, density, surface density, mass, or any other dimensional quantity. The motivation for this is the observation that the (cold) ISM is supersonically turbulent, which has approximately scale-free behaviour \citep[e.g.,][]{elmegreen_1996_fractal,mckee_star_formation,guszejnov_scaling_laws}, and hence should produce a population of clouds that cannot be assumed a priori to have some scale apart those set by the initial conditions, which may change over cosmic time. This criterion is violated by previously-used cloud-finding algorithms such as ``Friends-of-Friends" or ``watershed" methods that identify islands above a certain 2D or 3D density cut. These methods require the value of the density cut, surface density cut, or linking length as an input parameter. They have been used in previous studies of galaxy simulations, and it has generally been found that for an {\it appropriate choice} of these parameters, one recovers cloud properties that are in good agreement with observations in the local Universe \citep{dobbs_2011, Hopkins_2012_galaxy_structure, dobbs_2013, ibanez_2017, hopkins_fire2, Dobbs_2019_M33_GMC_FoF, fujimoto_2019}. However, a cloud definition that is valid for the relatively narrow range of ISM conditions found in nearby galaxies where most GMCs are catalogued \citep{bolatto_2008} may not generalize well to high-redshift conditions. Therefore, to study the properties of star-forming clouds across cosmic time, we must adopt a more general, scale-free and physically-motivated definition.

One scale-free definition for clouds is the set of gravitationally-bound ISM structures. This definition has some motivation as a proxy for GMCs, both observationally and theoretically. In observations \citep[see e.g.,][]{kauffmann_pillai_2013,heyer_dame_2015}, the importance of self-gravity is quantified in GMCs by measuring the cloud-scale virial parameter \citep{bertoldi_mckee_1992}:
\begin{equation}
    \alpha_{\rm BM92} = \frac{5 R \sigma_{\rm 1D}^2}{G M},
\end{equation}
where $\sigma_{1D}$ is the velocity dispersion of the cloud measuring along the line of sight, $M$ is the mass of the cloud, and $R$ is its radius. For a uniform sphere with no internal size-linewidth relation, this reduces to the ratio $2E_{\rm kin}/|E_{\rm grav}|$, such that $\alpha_{\rm BM92} = 2$ is the threshold of gravitational boundedness. More generally, the threshold of gravitational boundedness is merely of this order depending on assumptions about the internal structure and kinematics of the cloud. Whenever $\alpha_{\rm BM92}$ is measured in a population of GMCs, a wide range (0.1-10) of virial parameters tends to be found, but in all GMC catalogues that we are aware of, the distribution is peaked at a value of the same order as $2$ \citep[see references in][]{Dobbs_2014_GMC_review}. It is unlikely that {\it all} of the clouds observed are gravitationally-bound, but the observation of a characteristic virial parameter hints strongly that the properties of GMCs are deeply connected to their self-gravity.

On the theoretical front, \citet{hopkins_2012_excursion_set} used the excursion-set formalism to calculate the properties of the largest self-gravitating gas structures within the turbulent ISM. Here, ``largest" refers to so-called ``first-crossing" objects, which are bound gas structures that are not contained within some larger bound gas structure. \citet{hopkins_2012_excursion_set} showed that the properties of GMCs in nearby galaxies are largely consistent with the hypothesis that they are merely tracers of this underlying population of first-crossing objects. The predictions of this model have been validated quantitatively in numerical simulations of isolated galaxies \citep{Hopkins_2012_galaxy_structure}. 

Thus, the criterion of self-gravity provides a scale-free cloud definition that we expect to recover the properties of objects commonly referred to as GMCs in the local Universe. Henceforth in this paper, the term ``GMC", or more generally ``cloud", will be used interchangeably with the definition proposed here: the family of self-gravitating gas clouds that are not part of any larger self-gravitating structure, i.e., the first-crossing objects described in \citet{hopkins_2012_excursion_set}. Note that clouds that have recently undergone star formation and are disrupted by feedback are not covered by this definition.

\subsection{Simulations}\label{sec:sim}

We utilize several simulated galaxies from the Feedback in Realistic Environments (FIRE) project (\citealt{hopkins2014_fire})\footnote{\url{http://fire.northwestern.edu}}. These galaxies have been presented in detail in \cite{hopkins_fire2,Hopkins_FIRE2MHD_2019}. For full numerical details, the reader is referred to \citet{hopkins_fire2}. These are cosmological \myquote{zoom-in} simulations: the simulation starts from a large cosmological box that is later rerun with increased resolution in areas of matter concentration (\myquote{zooms-in} on galaxies). The simulations proceed from $z>100$ to present day. They are run using the GIZMO code \citep{Hopkins2015_GIZMO}\footnote{\url{http://www.tapir.caltech.edu/~phopkins/Site/GIZMO.html}}, with the mesh-free Godunov ``MFM'' method for the hydrodynamics \citep{Hopkins2015_GIZMO}. Self-gravity is included with fully-adaptive force and hydrodynamic resolution. The simulations include detailed metallicity-dependent cooling and heating processes from $T=10-10^{10}\,$K, including photo-ionization/recombination, thermal bremsstrahlung, Compton, photoelectric, metal line (following \citealt{Wiersma2009_cooling}), molecular, fine structure (following \citealt{CLOUDY}), dust collisional and cosmic ray processes, including both a meta-galactic UV background and a local source term from each star particle.
Note that unlike the \myquote{basic} FIRE2 simulations \citep{hopkins_fire2} these include magnetic fields through an expanded version of the idealized magneto-hydrodynamic (MHD) equations that include anisotropic Spitzer-Braginskii conduction and viscosity \citep[see][]{hopkins_gizmo_mhd}. The mass resolution for individual simulations is fixed at $M_{\rm min}=7100\,\msun$ (see Table \ref{tab:galaxies_input}); however, we have partially rerun them with increased resolution to check for convergence (see Figure \ref{fig:GMC_MF}) .

The resolution of these cosmological simulations is not high enough to resolve the formation of individual stars ($M_{\rm min}\gg 0.01\,\msun$). Instead, gas cells are converted to star particles representing simple stellar populations, according to a star formation prescription. In general, gas cells are converted to star particles stochastically \citep{katz96}, such that the cell has an average star formation rate
\begin{equation}
\dot{M}_\star = \epsilon_{\rm ff,res} f_{\rm mol} m_{\rm gas} / t_{\rm ff},
\label{eq:sfr}
\end{equation}
where $\epsilon_{\rm ff,res}$ is the per-freefall star formation efficiency within a single resolution element, $f_{\rm mol}$ is the fraction of the gas that is molecular according to the \citet{krumholz_2011_self_shield} prescription, $m_{\rm gas}$ is the mass of the gas cell, and $t_{\rm ff} = \sqrt{\frac{3 \mathrm{\pi}}{32 G \rho}}$ is the local free-fall time. Note $\epsilon_{\rm ff,res}$ is set to zero for all gas that does not exceed the density threshold $n_{\rm crit}$ (see Table \ref{tab:galaxies_input}). We set $\epsilon_{\rm ff,res}=1$ for gas that is self-gravitating at the resolution scale according to a virial criterion \citep{hopkins_2013_sf_criteria}\footnote{Note that clouds at all scales are observed to have star formation efficiencies significantly lower than unity \protect\citep{Krumholz_2012_SF_law}, in the simulations  $\epsilon_{\rm ff,res}=1$ is used to ensure that star formation is only regulated by the galactic scale feedback, not an arbitrarily chosen parameter. \cite{orr_fire_ks} showed that this prescription does lead to galaxy-wide star formation efficiencies consistent with observations}. For gas that is above the density threshold but not self-gravitating, FIRE simulations adopt $\epsilon_{\rm ff,res}=0.0015$ to prevent rare cases where very dense gas leads to extremely small timesteps, greatly slowing down the simulation until it becomes self-gravitating and turns into stars. A consequence of this choice is that a fraction of star formation (star particle spawning) takes place \myquote{prematurely} in not-yet bound structures, such as in clouds located in the galactic center.
 
Once formed, each of these star particles represents a stellar population with the same formation properties (age, metallicity etc.) and are assumed to have a well-sampled, universal \citet{kroupa_imf} IMF. They inject feedback into the surrounding gas via OB \&\ AGB mass-loss, SNe Ia \&\ II, and multi-wavelength photo-heating and radiation pressure; with inputs taken directly from stellar evolution models \citep{1Leitherer_1999_Starburst99}.

\subsubsection{Simulated galaxies}

In this paper we utilize simulated galaxies with two different sets of initial conditions (see Table \ref{tab:galaxies_input} for details) that lead to different types of galaxies\footnote{Note that these are the same galaxies used by \cite{guszejnov_extragal_imf_var}.}:
\begin{itemize}
    \item \textbf{m12i}: A simulated spiral galaxy with similar properties to the Milky Way (see \textbf{m12i} with \textit{MHD+} physics in \citealt{Hopkins_FIRE2MHD_2019}) with a mass resolution of $7100\,\msun$. This is the primary focus of this paper. 
    \item \textbf{m11q}: An isolated dwarf galaxy that is similar in mass to the Large Magellanic Cloud (see \textbf{m11q} with \textit{MHD+} physics in \citealt{Hopkins_FIRE2MHD_2019}). To account for effects stemming from numerical resolution we conducted a resolution study with this galaxy by rerunning the last 1 Gyr of cosmic evolution at different mass resolutions. The reason this study was done with \textbf{m11q} instead of \textbf{m12i} is due to the enormous computational cost of a higher resolution rerun of the MW-like \textbf{m12i} galaxy.
\end{itemize}
Note that prior work has shown that FIRE galaxies provide realistic analogues to observed galaxies as they follow the observed stellar-to-halo mass relation \citep{hopkins_fire2}, have similar disk morphologies and metallicity gradients \citep{Ma_2017_fire_morphology,Garrison_Kimmel_2018_galaxy_morphology}, have similar atomic/molecular gas kinematics at present day \citep{El_Badry_2018_FIRE_gas,El_Badry_2018_FIRE_gas_HI} and similar evolution over time \citep{Hung_2019_gas_evolution}, and they reproduce the  Kennicutt-Schmidt relation \citep{orr_fire_ks}. The studies listed above were carried out on versions of the simulations that do not include MHD effects; however the effects of magnetic fields on galaxy-wide properties have been shown to be weak \citep{kungyi_weak_mhd_2016,Hopkins_FIRE2MHD_2019}.

\begin{table*}
	\centering
		\setlength\tabcolsep{3.0pt} 
		\begin{tabular}{|c|c|c|c|c|c|c|c|c|c|c|c|c|}
		\hline
		\bf Key & \bf MHD? & \bf Final Redshift & \bf $M_{\rm DM}/\msun$  & \bf $M_{\rm min}/\msun$ & \bf $n_{\rm crit}/\mathrm{cm}^{-3}$ & \bf $M_{\rm *}/\msun$ & $R_{\rm 1/2}/\mathrm{kpc}$ & \bf References \\
        \hline
        \bf m12i & Yes & $0$ & $10^{12}$ & $7100$ & $10^3$ & $6\times10^{10}$ & $3.5$ & \cite{Hopkins_FIRE2MHD_2019} \\
        \hline
        \hline
        \bf m11q & Yes & $0$  & $10^{11}$ &  $7100$ & $10^3$ & $1.5\times 10^9$ & 3.4 & \cite{Hopkins_FIRE2MHD_2019}\\
        \hline
		\end{tabular}
 \caption{Parameters of simulated galaxies from the FIRE project, including target dark matter halo virial mass $M_{\rm DM}$ (at $z=0$), gas element mass resolution $M_{\rm min}$, the critical density for star particle creation $n_{\rm crit}$, as well as the stellar mass $M_{\rm *}$ and half-mass radius $R_{\rm 1/2}$ of the galaxy in the final snapshot.} 
 \label{tab:galaxies_input}\vspace{-0.4cm}
\end{table*}

\subsection{Cloud identification}\label{sec:cloudphinder}

Clouds are identified using CloudPhinder\footnote{\url{https://github.com/omgspace/CloudPhinder}}, a new method based on the SUBFIND algorithm \citep{Springel_2001_SUBFIND}. Unlike some other popular approaches to identifying clouds, this method identifies GMCs based on the \emph{physical} definition argued for in Sec.~\ref{sec:methods:boundclouds}, by picking out the largest self-gravitating structures of gas that are present, taking gravitational, thermal, kinetic and magnetic energy into account. \footnote{Note that of the thermal, magnetic and kinetic energies, the kinetic energy is nearly always dominant, as shown in previous MHD galaxy simulations \citep{kungyi_weak_mhd_2016, Hopkins_FIRE2MHD_2019}. Essentially none of the results of this study depend on whether the thermal or magnetic energies are accounted for in the virial parameter, except perhaps the mass-to-flux ratios of the very smallest clouds resolved, which have $M/M_\Phi \sim 1$ (Figure \ref{fig:scaling}).} In principle the algorithm requires no density cut; however, for computational expedience we have limited our analysis to gas particles with $n_{\rm H} > n_{\rm min}=\unit[1]{cm^{-3}}$, so that only a small subset of the total gas present need be considered. Although the cold, molecular phase of the ISM consists largely of gas denser than $\unit[100]{cm^{-3}}$, we find that it is necessary to set this lower threshold to capture the largest bound gas structures, which can have a significant bound component in the more diffuse/neutral/warm ISM. Note that the clouds identified by this method contain a significant fraction of low-density ($n_{\rm H} <\unit[100]{cm^{-3}}$) gas, making their properties somewhat different from the clouds identified by observations (see Table \ref{tab:galaxies_result}). 

For this study using the bound cloud definition for GMCs instead of the ones commonly used by observers has two advantages: 1) we impose no characteristic scale on clouds, thus avoiding potential biases in high redshift cloud populations that might form in a very different ISM, and 2) observations can only indirectly infer gas densities through the emission maps of tracer molecules, which have critical densities above $\unit[100]{cm^{-3}}$. One disadvantage of this definition is that it can not be applied to observed data, making direct comparisons difficult. To check for potential biases in our cloud identification method, we have also applied the Friends-of-Friends and dendrogram methods commonly used in the literature to our simulated MW-like galaxy (\textbf{m12i}). We found present day cloud properties to be qualitatively similar, with many clouds having analogues in the different definitions. We will discuss in more details in a follow-up paper.

The algorithm requires exactly one input parameter: the threshold virial parameter of the clouds to be identified, $\alpha_{\rm crit}$. We chose the threshold of gravitational boundedness, identifying clouds satisfying
\begin{equation}
    \alpha_{\rm vir} \equiv \frac{2 \left(E_{\rm kin} + E_{\rm thermal} + E_{\rm mag}\right)}{|E_{\rm grav}|} \leq \alpha_{\rm crit} = 2,
    \label{eq:alphavir}
\end{equation}
where $E_{\rm kin}$, $E_{\rm thermal}$, $E_{\rm mag}$ and $E_{\rm grav}$ are the kinetic (turbulence and rotation), thermal, magnetic and gravitational bindig energies of the gas in the cloud.

Given the threshold virial parameter and the threshold density (to which the results are insensitive, provided it is low enough, $\sim \unit[1]{cm^{-3}}$), the algorithm identifies iso-density contours that satisfy Equation \ref{eq:alphavir}, walking outward from density peaks until the threshold $\alpha_{\rm crit}$ is crossed. We describe the exact algorithm for doing this in Appendix \ref{appendix:cloudphinder}.

\begin{figure*}
\begin {center}
\includegraphics[width=0.33\linewidth]{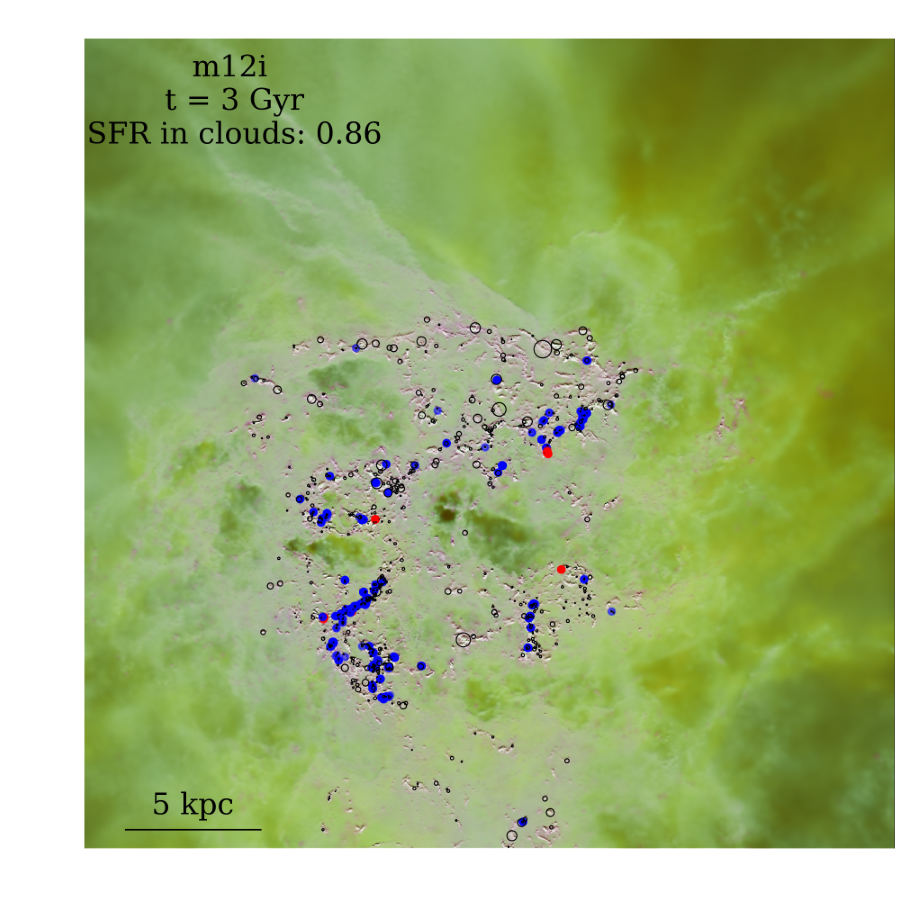}
\includegraphics[width=0.33\linewidth]{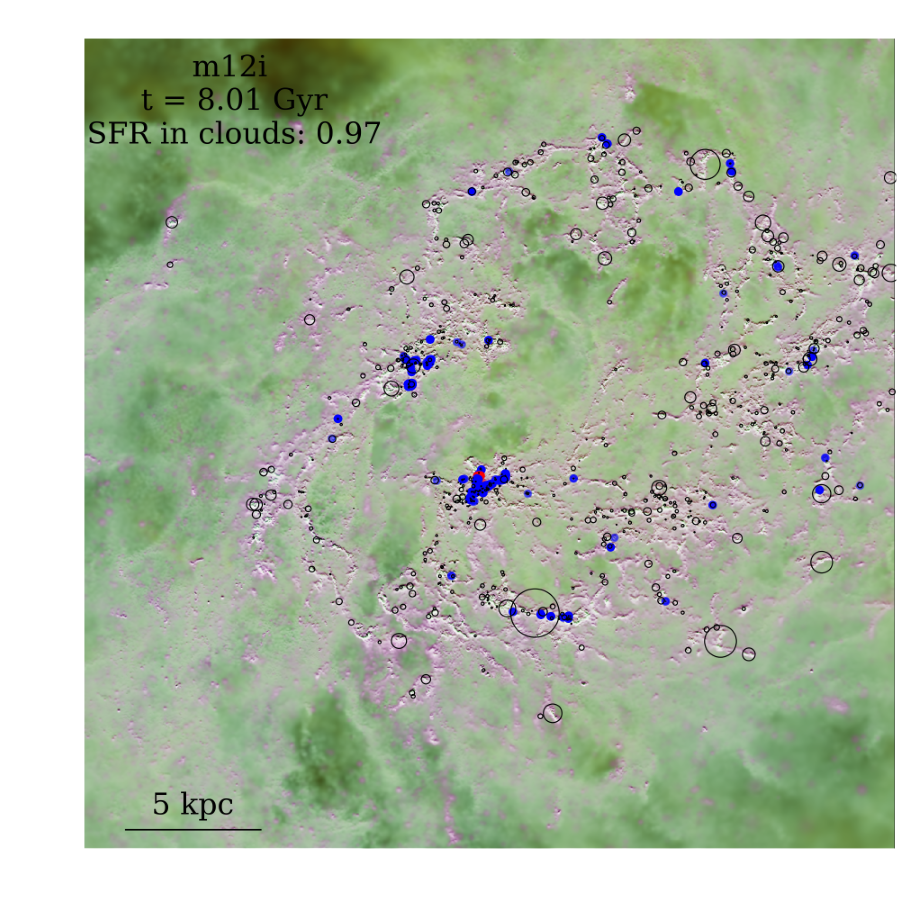}
\includegraphics[width=0.33\linewidth]{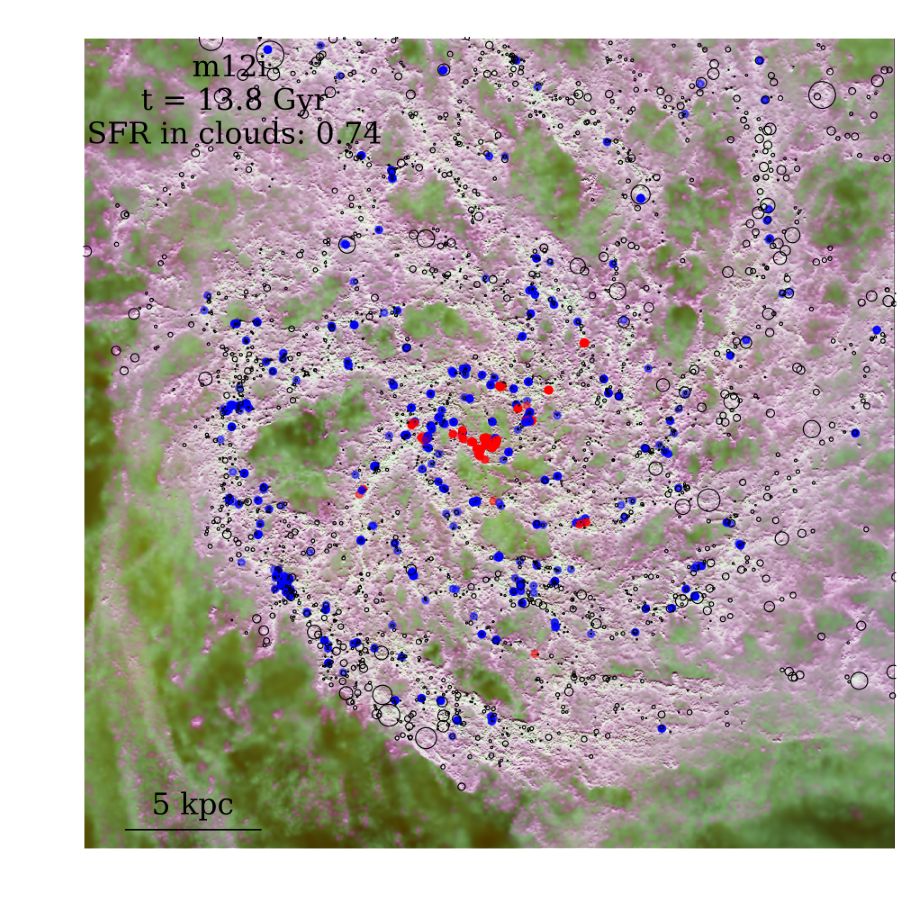}\\
\includegraphics[width=0.33\linewidth]{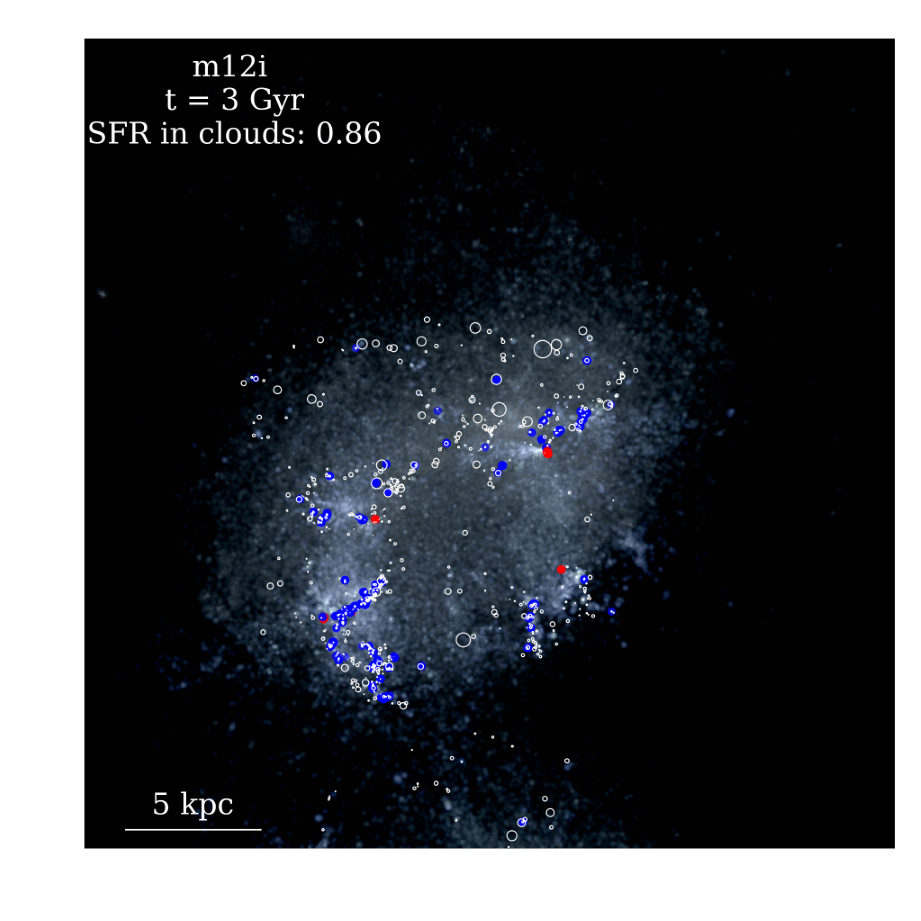}
\includegraphics[width=0.33\linewidth]{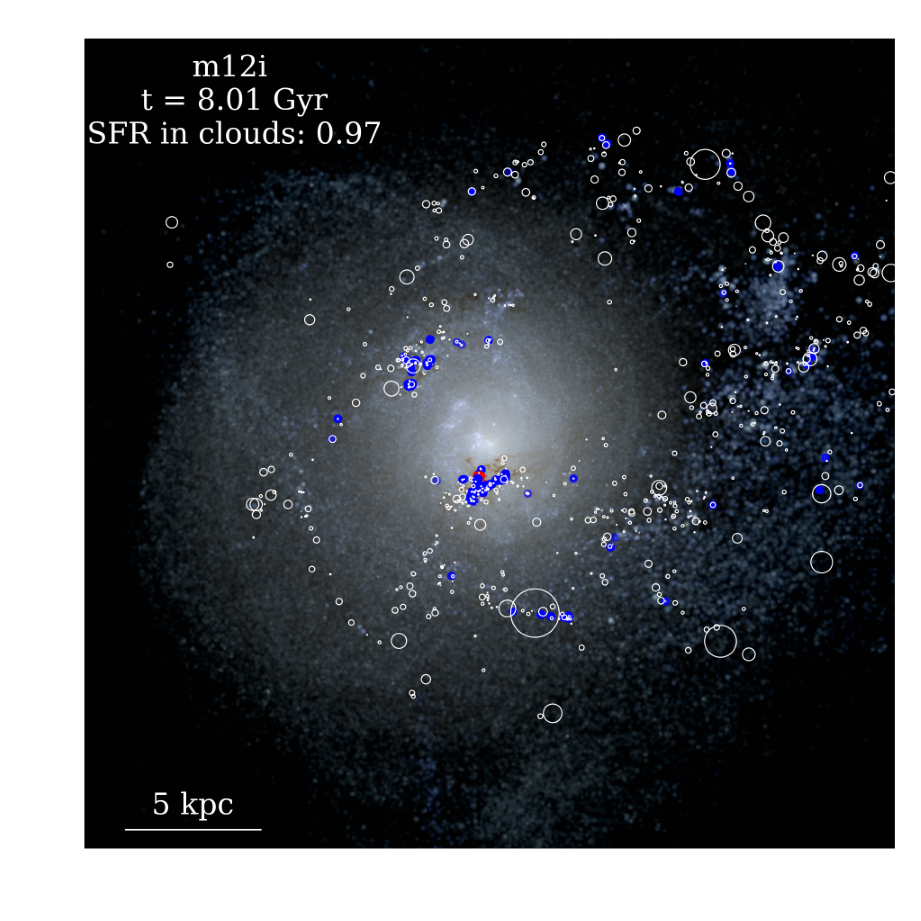}
\includegraphics[width=0.33\linewidth]{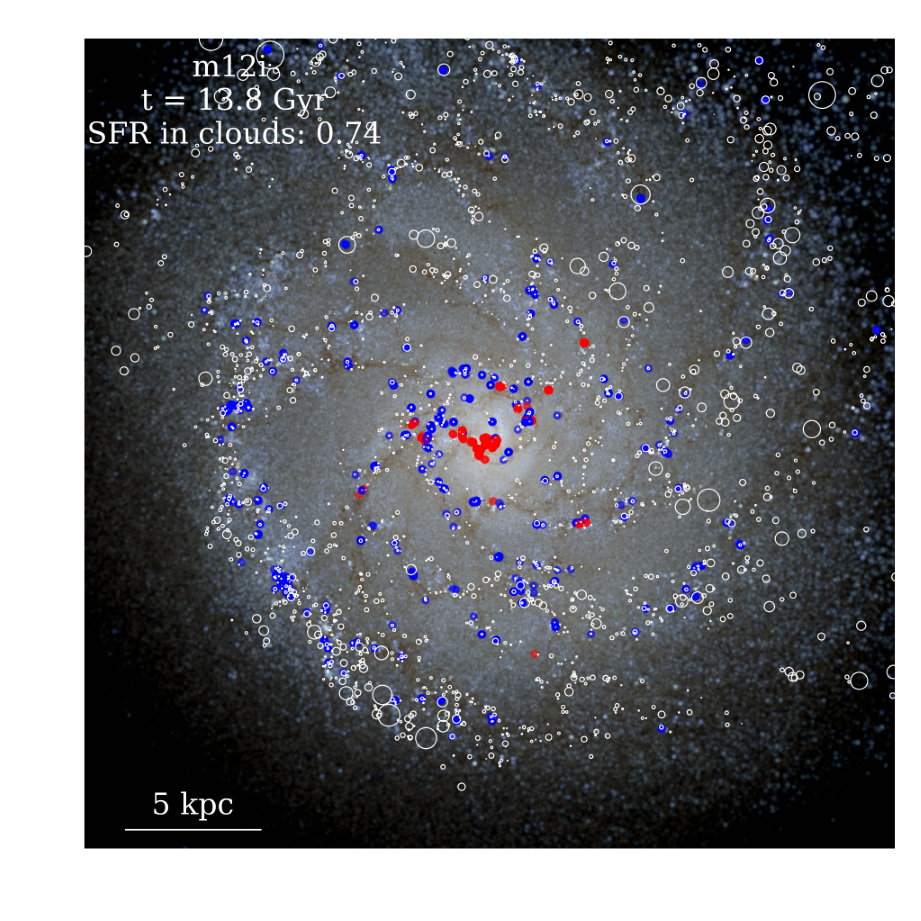}\\
\vspace{-0.5cm}
\caption{Gas temperature (Top) and attenuated starlight maps (Bottom) for the \textbf{m12i} MW-like simulated galaxy at three different times (3 Gyr, 8 Gyr and present day). Black/white circles show the location and rough extent of identified GMCs, while currently star-forming gas is marked with blue dots if within an identified GMC and red if not. The latter case happens in high-density environments where the simulation can form star particles before the gas becomes self-gravitating (see Sec. \ref{sec:methods}). This issue is mostly confined to the galactic center and does not affect our results in the outer regions, Overall, the GMCs identified by our algorithm contain all star forming gas outside this region.}
\label{fig:sfr_loc}
\vspace{-0.5cm}
\end {center}
\end{figure*}

Despite this selective criterion for grouping gas into clouds in terms of mass fraction of the total ISM, we find that the majority of star formation is in these self-gravitating clouds. An important caveat is that, owing to the specific star formation prescription and simulation resolution used (see above in Sec. \ref{sec:methods}), we do find particles that are not in any bound cloud that are nevertheless eligible to be converted to stars in the simulation (see Figure \ref{fig:sfr_loc}). This effect is most pronounced in galaxies' centres, where gas is predominantly dense ($n_{\rm H} \unit[>10^3]{cm^{-3}}$), and thus can form stars rapidly even when the virial criterion is not satisfied. As such, our catalogue is incomplete within 4 kpc of the  centres of galaxies and misses virtually all star formation within 1 kpc. We emphasize that this is not a {\it physical} effect: if stars form via gravitational collapse, it is physically necessary that they belong to {\it some} bound structure. We have confirmed that essentially all star-forming gas -- even in the centres of galaxies -- belongs to a bound structure in test runs in which we have imposed a stricter star formation criterion than the standard FIRE-2 runs \citep[see][]{grudic_2016}

\subsection{Definitions of bulk properties for GMCs}
In this study our aim is to analyze the statistics of the \emph{bulk} properties of GMCs over cosmic time in different galactic environments. First, we define the effective radius $R_{\rm eff}$ of a cloud as the radius of a sphere that would have the same moment of inertia:
\begin{equation}
    R_{\rm eff} = \sqrt{\frac{5}{3} \langle r^2 \rangle_M},
\end{equation}
where $\langle \cdot \rangle_M$ denotes a mass-weighted average over the cloud's consitituent gas particles, and $r$ is the distance of the particle from the centre of mass of the cloud. The mean cloud 2D densities are then defined accordingly:
\begin{equation}
    \Sigma_{\rm eff} = \frac{M}{\mathrm{\pi} R_{\rm eff}^2},
\end{equation}
while the 3D density $\rho$ is just volume-averaged over the cloud. We define the bulk 1D velocity dispersion $\sigma$ of a GMC as
\be
\sigma^2 = \frac{\sigma^2_{\rm x}+\sigma^2_{\rm y}+\sigma^2_{\rm z}}{3},
\label{eq:sigma1D}
\ee
where $\sigma_i^2=\langle v_i^2 \rangle_M-\langle v_i \rangle_M^2$ is the velocity dispersion in direction $i$.

To represent the thermal properties of the clouds, we take the mass-weighted average temperature of the gas assigned to it:
\be
T = \langle T_{\rm gas} \rangle_M.
\ee
Note that we have experimented with different definitions for the cloud's bulk temperature (e.g., mass-median gas temperature) and found qualitatively similar results.

To quantify the strength of turbulence in the clouds we define the ratio of turbulent to thermal energy:
 \be
 \frac{E_{\rm turb}}{E_{\rm thermal}}=\frac{M \sigma^2}{M \langle c_{\rm s, gas}^2 \rangle_M},
 \ee
where $c_{\rm s, gas}$ is the local sound speed of the gas. We choose this quantity because the clouds are not homogeneous in temperature, making it hard to define a meaningful Mach number. Nevertheless, $\frac{E_{\rm turb}}{E_{\rm thermal}}\gg 1$ corresponds to turbulence-supported supersonic clouds, while for clouds with homogeneous temperature, $\frac{E_{\rm turb}}{E_{\rm thermal}}\sim \mach^2$, where $\mach$ is the Mach number of the turbulence.


To represent the strength of the magnetic field in the cloud we introduce the effective magnetic field
\be 
B = \sqrt{\left\langle \Bvector\cdot\Bvector \right\rangle_V},
\label{eq:B_field}
\ee 
where $\langle \cdot \rangle_V$ denotes a volume-weighted average over the gas particles in the cloud. Note that we have experimented with different definitions (e.g., $B=\langle ||\Bvector|| \rangle_V$) and found qualitatively similar results.

We quantify the relative strength of magnetic support by introducing the dimensionless mass-to-flux ratio $M/M_{\Phi}$, defined as
\be
M/M_{\Phi}=\sqrt{\frac{-E_{\rm grav}}{E_{\rm mag}}},
\ee
where $E_{\rm mag}=V\frac{B^2}{2\mu_0}$ is the magnetic energy in the cloud (with $V$ as its volume), while $E_{\rm grav}$ is the gravitational binding energy (computed from the gravitational potential).

We can define the mass-weighted metallicity of the cloud in a straightforward manner as
\be 
Z = \langle z \rangle,
\ee
where $z$ is the metallicity of the individual gas particles within the cloud. Since the formation of individual stars is not resolved by the simulations, it is assumed that gravitationally bound gas on the smallest scales turns into stars (represented by star particles, see Sec. \ref{sec:sim}) on a freefall time.

\section{Results}\label{sec:results}

\begin{figure*}
\begin {center}
\includegraphics[width=0.33\linewidth]{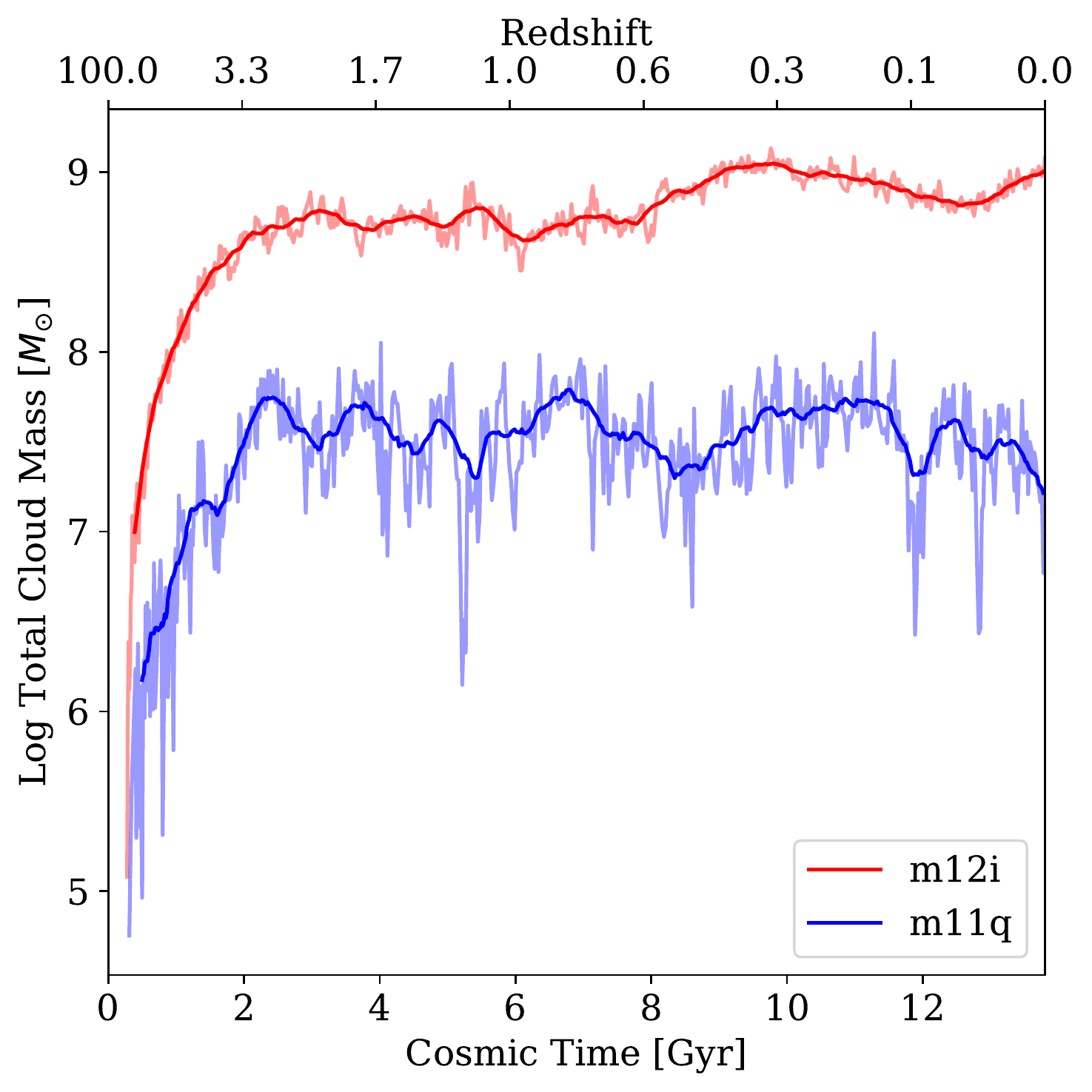}
\includegraphics[width=0.33\linewidth]{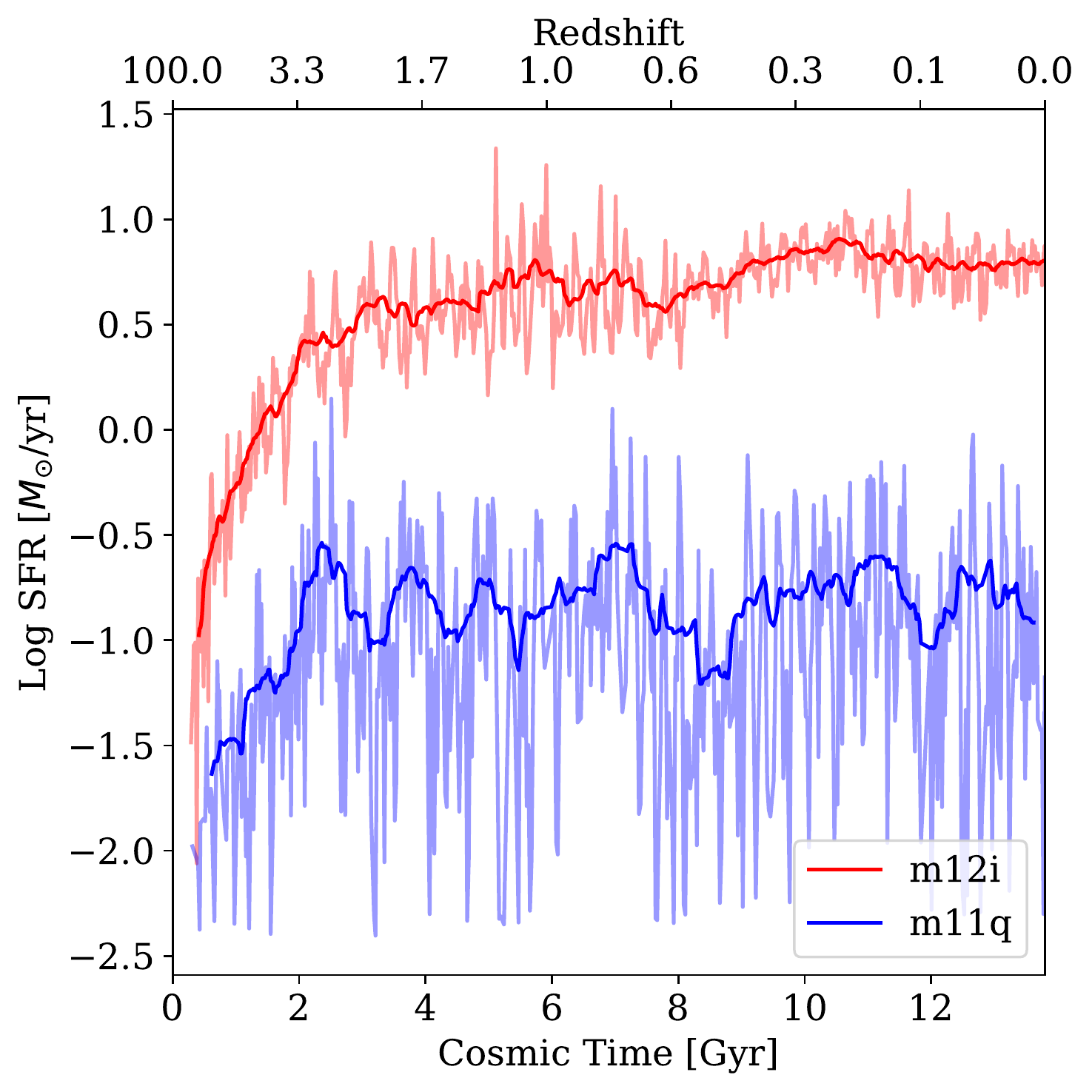}
\includegraphics[width=0.33\linewidth]{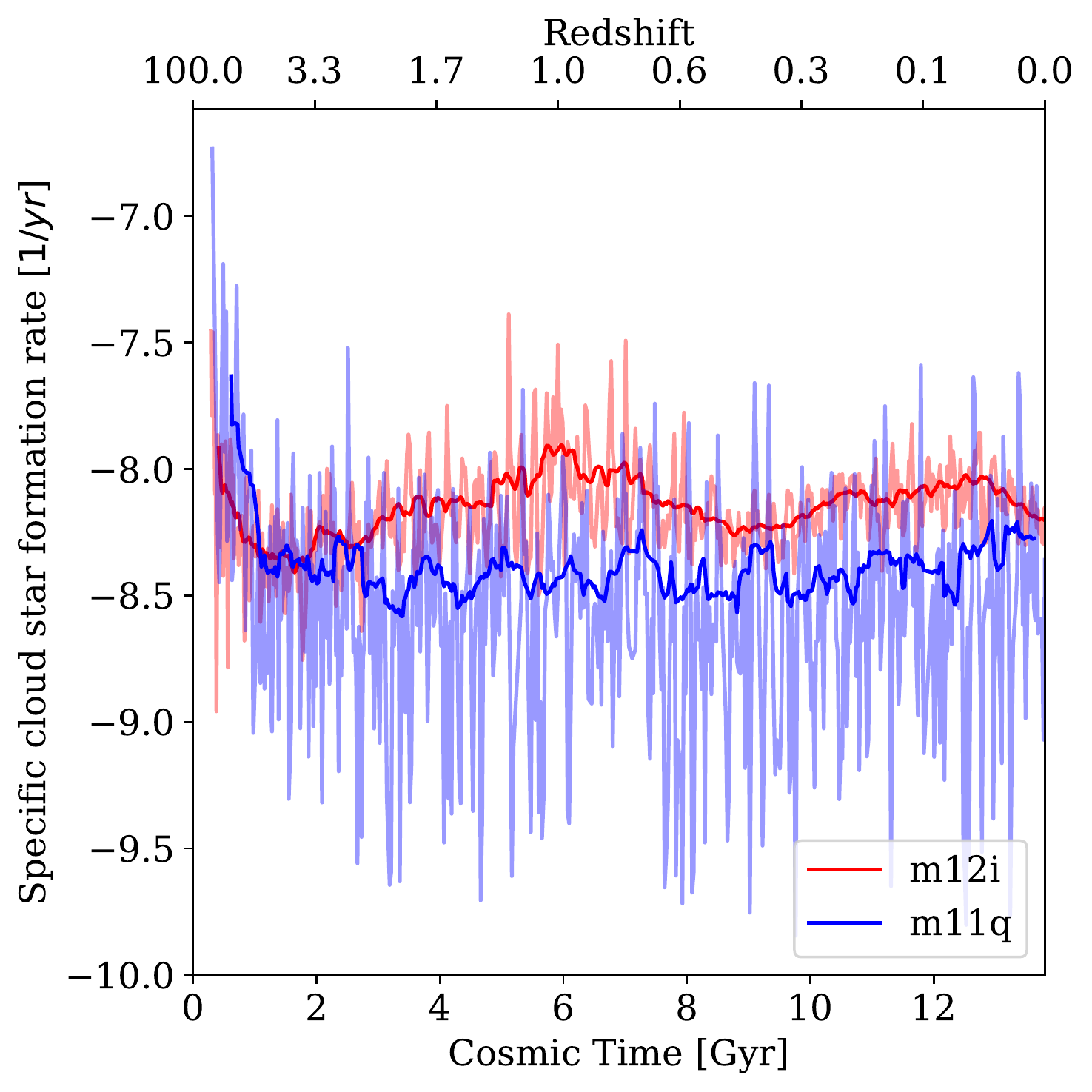}\\
\vspace{-0.4cm}
\caption{\emph{Left:} Total mass in GMCs in the simulated galaxies as a function of time. After the formation of the galaxy the time evolution of total GMC mass is close to flat with large short-lived spikes. The actual data for each snapshot is shown in transparent colors, while a smoothed version (moving average over snapshots) of it is shown with opaque lines. \emph{Middle:} Instantaneous galactic star formation rate over cosmic time in the same simulated galaxies. Since star formation primarily happens in GMCs its rate is correlated with the total GMC mass. \emph{Right:} Instantaneous specific star formation rate (defined as SFR/cloud mass) over cosmic time in the same simulated galaxies. Despite the evolving metallicity of the galaxies (see Figures \ref{fig:m12i_evolplots} and \ref{fig:m11q_evolplots}) the specific star formation rate shows no systematic time evolution but has significant fluctuations on smaller time scales.} 
\label{fig:sfr_tot}
\vspace{-0.5cm}
\end {center}
\end{figure*}

Figure \ref{fig:sfr_tot} shows the evolution of the total cloud mass in a MW-like spiral (\textbf{m12i}) and an LMC-like dwarf (\textbf{m11q}) galaxy as well as their star formation histories. As expected, both galaxies exhibit bursty star formation (similar to other FIRE galaxies, see \citealt{Sparre_2017_fire_starbursts}), but the median total cloud mass and galactic star formation rate are nearly time-invariant since $z\sim 3$, with average total cloud mass of $10^9\,\msun$ and $10^{7.5}\,\msun$ and a star formation rate (SFR) of $3\,\msun/\mathrm{yr}$ and $0.1\,\msun/\mathrm{yr}$ for \textbf{m12i} and \textbf{m11q} respectively. In the case of \textbf{m12i} there is a transition from bursty to a  more \myquote{quiescent} star formation regime around 7 Gyr, when the timescales for galactic dynamics and supernova feedback become comparable \citep[see][]{FG_2018_bursty_SFR}.

\begin{figure*}
\begin {center}
\includegraphics[width=0.33\linewidth]{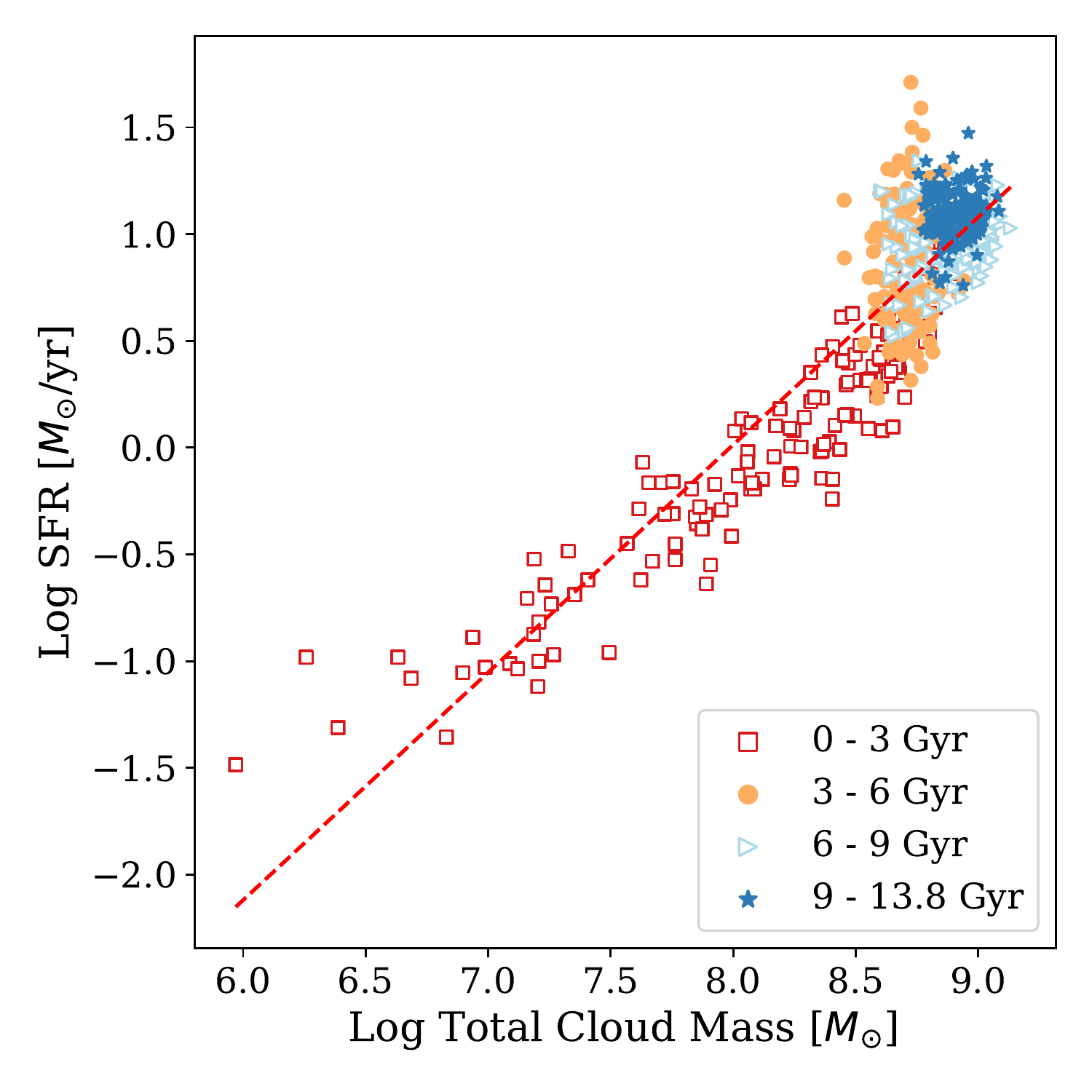}
\includegraphics[width=0.33\linewidth]{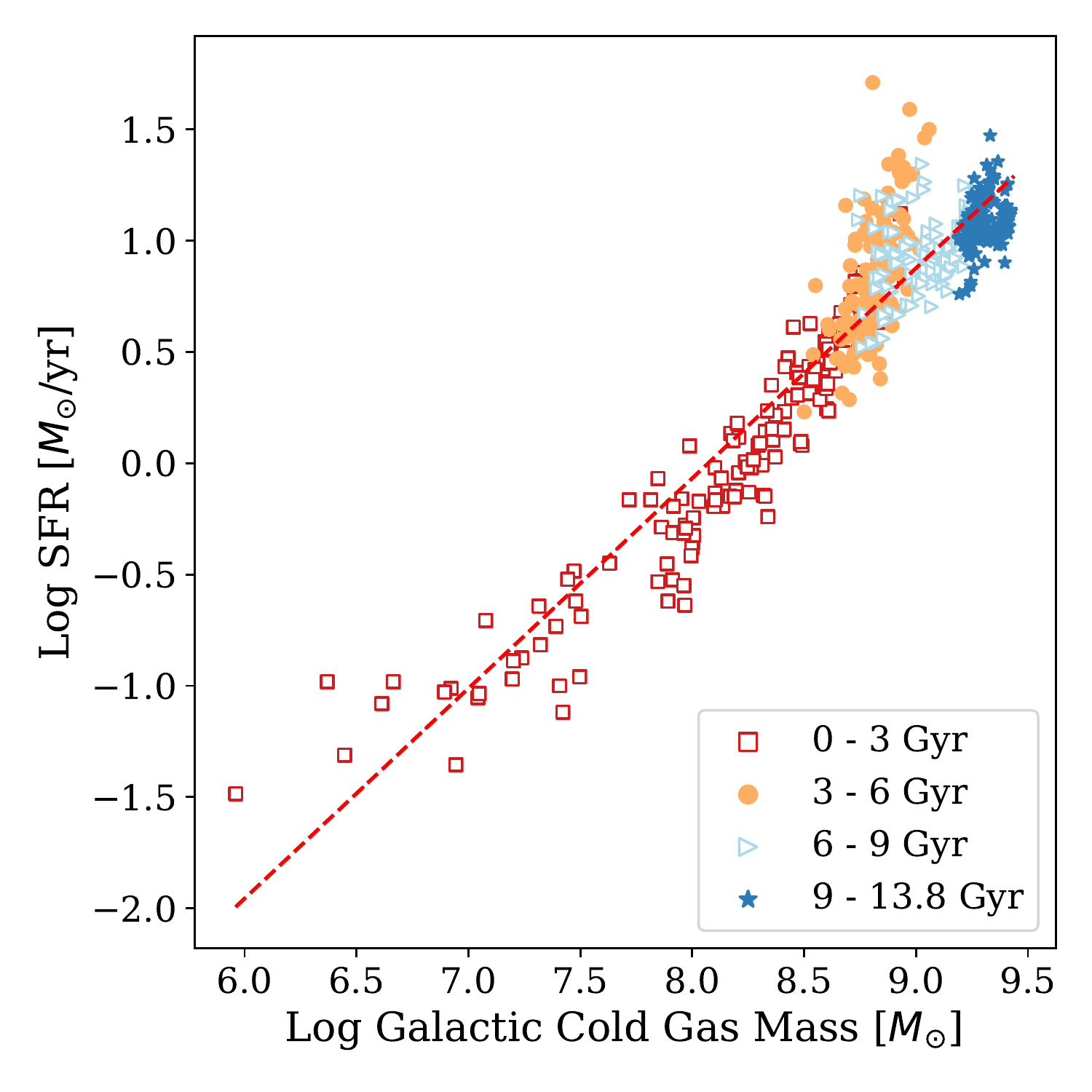}
\includegraphics[width=0.33\linewidth]{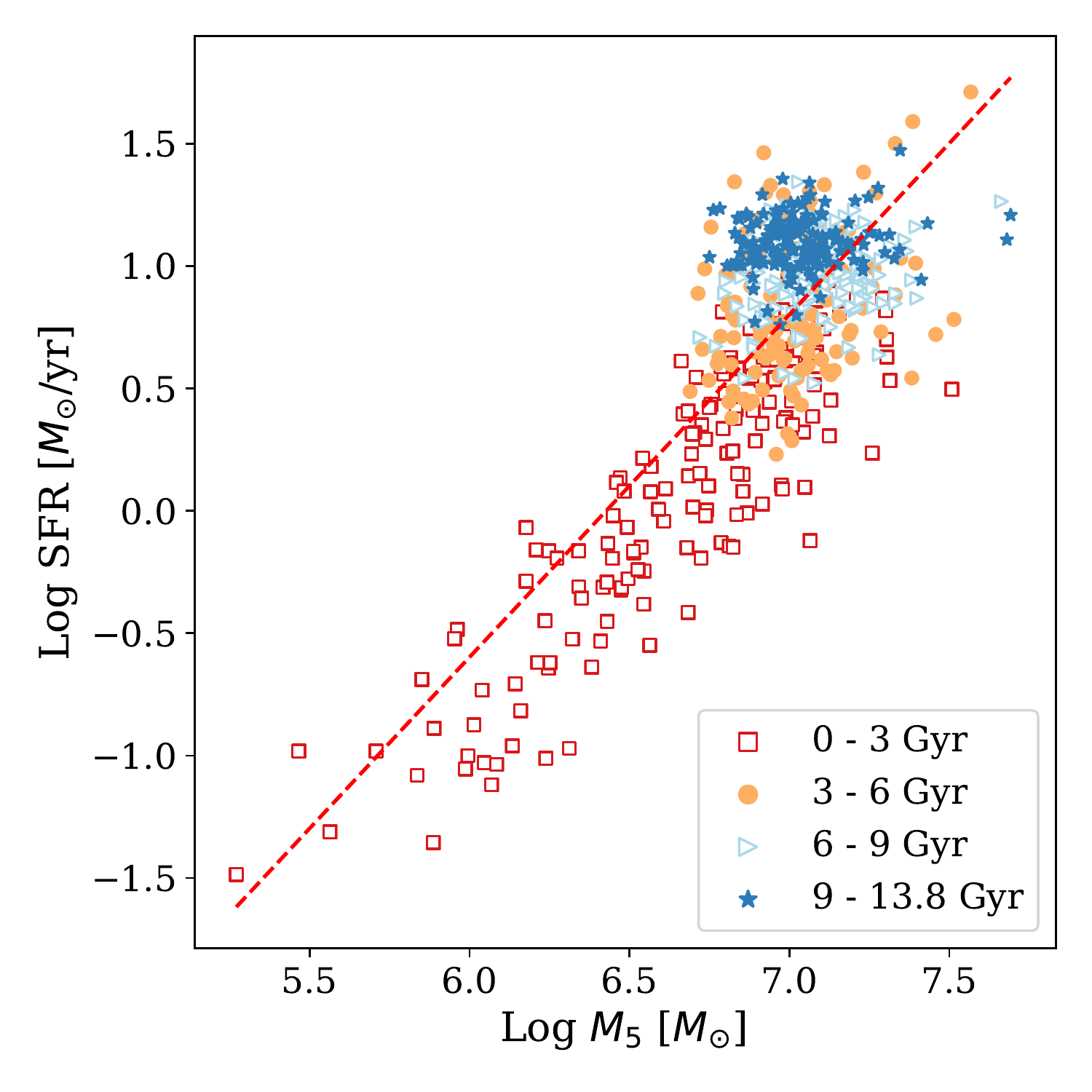}\\
\vspace{-0.5cm}
\caption{Galactic star formation rate (SFR), defined by Eq. \ref{eq:sfr} correlated versus the total cloud mass (left), the total cold gas mass (middle) and $M_5$ the mean mass of the 5 most massive clouds for \textbf{m12i}, color coded according to cosmic time. We define \myquote{cold gas} as all gas that is colder than 100 K. Each symbol represents a snapshot of the simulation (color coded according to cosmic time), while the dashed red lines are linear fits to the data.  We find that the galactic SFR correlates well with all three quantities, showing close to linear relationships. For the cold gas and cloud masses this is expected as stars form in cold, bound gas \citep{mckee_star_formation}. The correlation between the masses of the most massive clouds and SFR shows that massive clouds are tied to starburst activity. }
\label{fig:sfr_correlation}
\vspace{-0.5cm}
\end {center}
\end{figure*}

Figure \ref{fig:sfr_correlation} shows that the instantaneous galactic SFR correlates well with not only the total cold gas mass and cloud mass in the galaxy, but also the mass of the largest GMCs. This indicates that massive GMCs are the source of starburst activity. 

\begin{figure*}
\begin {center}
\includegraphics[width=0.33\linewidth]{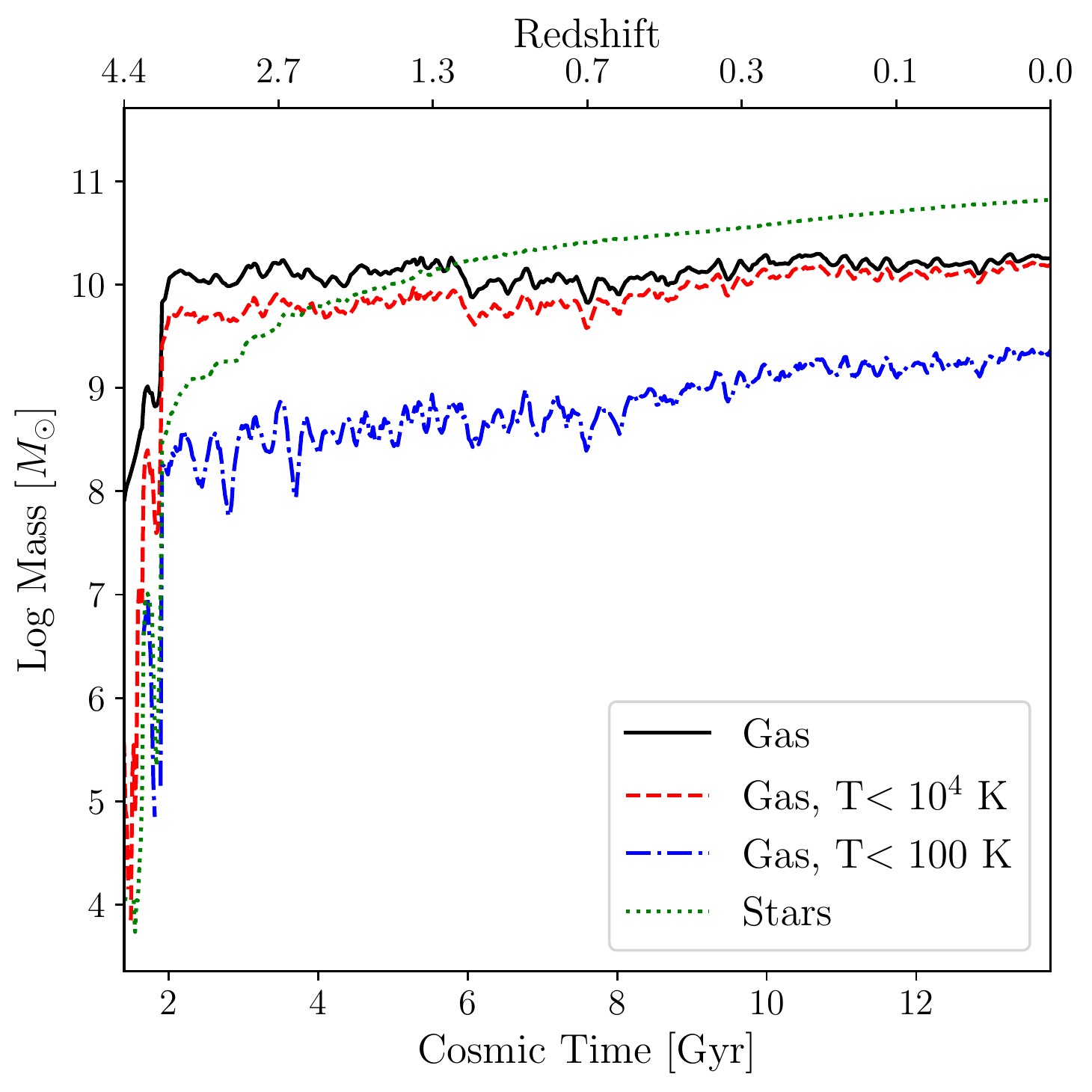}
\includegraphics[width=0.33\linewidth]{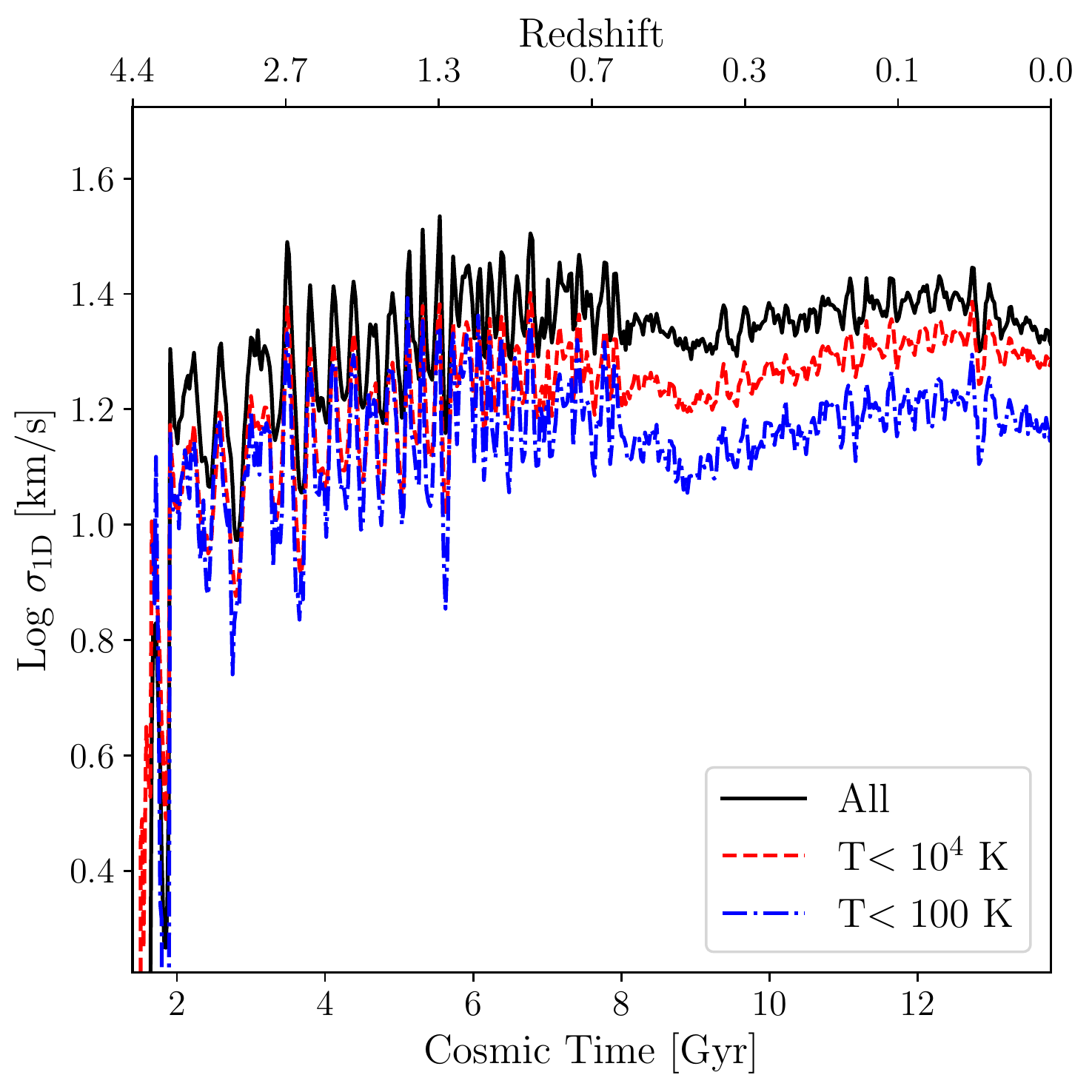}
\includegraphics[width=0.33\linewidth]{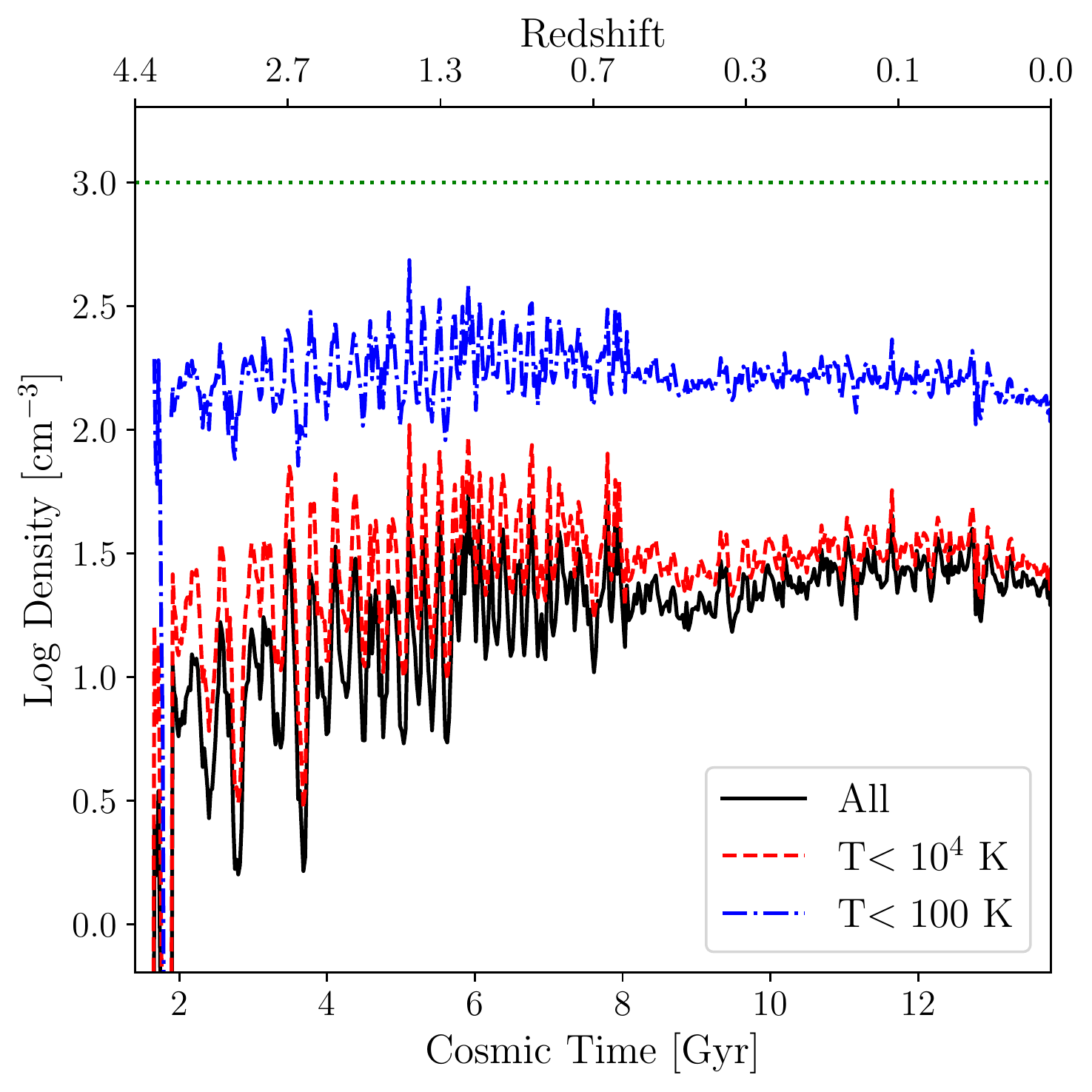}\\
\vspace{-0.4cm}
\caption{Evolution of average galactic properties in \textbf{m12i}, including galactic gas and stellar mass (left), average 1D velocity dispersion of gas on 500 pc scale (middle) and mean gas density over cosmic time (right). In the figure on the right, $n_{\rm crit}$ for star formation is marked with a horizontal line. Except for the stellar and cold gas mass these galactic properties appear to have no trend on large timescales, but both the gas density and velocity dispersion exhibit factor of 2 level variations on shorter timescales during the \myquote{bursty} star formation phase of the galaxy.} 
\label{fig:bulk_properties_evol}
\vspace{-0.5cm}
\end {center}
\end{figure*}

Figure \ref{fig:bulk_properties_evol} shows the evolution of the mass content of \textbf{m12i} as well as the average velocity dispersion and density of its gas (see \ref{fig:bulk_properties_evol_m11q} for \textbf{m11q}). Note that we define the velocity dispersion similar to Eq. \ref{eq:sigma1D} but instead of calculating it within clouds we do so in 500 pc sized cubes and take a mass-weighted average of these cube values. While the stellar and cold gas mass show an increasing trend (due to star formation and more efficient cooling due to the resulting increase in metallicity), the mass, average density and velocity dispersion of the gas appear to have no trend over cosmic time. Also, both the gas velocity dispersion and density exhibit significant variation over shorter times during the \myquote{bursty} phase of galactic star formation.

Table \ref{tab:galaxies_result} shows the mass-weighted median properties of the GMCs in the simulated galaxies over the last 200 Myr of the simulations. We find that many of the median properties of the identified GMCs in both galaxies have surprisingly similar values, this includes mass, size, turbulent velocity dispersion and magnetic properties. Still, the clouds in the \textbf{m11q} dwarf galaxy have lower metallicity, which in turn leads to less effective cooling and thus higher average temperature. Note that the mass-weighted median temperature of the \textbf{m12i} clouds is a factor of 2 higher than those with similar densities observed in the MW  \citep{heyer_dame_2015}, due to the fact that observations are sensitive to molecular line transitions with critical densities of a few $100\,\mathrm{cm}^{-3}$, while our most massive clouds have a significant portion of their mass in low density gas. To illustrate this effect we also included the properties of the clouds we get if we restrict CloudPhinder to dense gas ($n_{\rm min}=100\,\mathrm{cm}^{-3}$).

\begin{table*}
	\centering
		\setlength\tabcolsep{3.0pt} 
		\begin{tabular}{|c|c|c|c|c|c|c|c|c|c|c|c|c|}
		\hline
		\bf Key & \bf Type  & $n_{\rm min}/\mathrm{cm^{-3}}$ & $M_{\rm tot}/\msun$ & $M_{\rm cloud}/\msun$ & $R_{\rm eff}/\mathrm{pc}$ & $\Sigma/(\msun/\mathrm{pc}^2)$ & $T/\mathrm{K}$ & $B/\mathrm{\mu G}$ & $\sigma/{\rm (km/s)}$ & $E_{\rm turb}/E_{\rm thermal}$  & $M/M_{\Phi}$ & $\log(Z/Z_{\rm \sun})$ \\
        \hline
        \multirow{2}{*}{\bf m12i} & \multirow{2}{*}{Spiral, MW-like} & $1$ & $8.6\times10^8$ & $8.4\times10^5$ & 86 & 37  & 98 & 11 & 4.1 & 41 & 3.1 & 0.27\\  \cline{3-13}
         &   & $100$  & $1.3\times10^8$ & $6.1\times10^5$ & 38 & 134  & 41 & 38 & 4.5 & 103 & 2.9 & 0.20\\  \hline
        
        \multirow{2}{*}{\bf m11q} & \multirow{2}{*}{Dwarf, LMC-like}  & $1$ & $1.1\times 10^7$ & $8.4\times 10^5$ & 84 & 34 & 158  & 9 & 3.7  & 21 & 3.0 & -0.25  \\ \cline{3-13}
        &   & $100$ & $8.5\times 10^5$ & $3.9\times 10^5$ & 22 & 217 & 51  & 40 & 3.9  & 51 & 2.9 & -0.27  \\ \hline
		\end{tabular}
        \vspace{-0.1cm}
 \caption{Mass-weighted median properties of GMCs identified in the simulated galaxies time averaged over the last 200 Myr cosmic evolution. For each galaxy we show two sets of values, one with our a cut-off density $n_{\rm min}$ of $1\,\rm cm^{-3}$ (fiducial value) and with $100\,\rm cm^{-3}$ (dense gas only). These properties include the total mass of gas in clouds $M_{\rm tot}$, cloud mass $M_{\rm cloud}$, effective radius $R_{\rm eff}$, surface density $\Sigma$, temperature $T$, magnetic field strength $B$, turbulent (1D) velocity dispersion $\sigma$, turbulent to thermal energy ratio $E_{\rm turb}/E_{\rm thermal}$, mass-to-flux ratio $M/M_{\Phi}$, and metallicity $Z$.} 
 \label{tab:galaxies_result}\vspace{-0.5cm}
\end{table*}

In the following sections we focus on the GMC properties in the \textbf{m12i} MW-like spiral galaxy. The same figures for \textbf{m11q} can be found in Appendix \ref{sec:m11q}.

\subsection{Mass distribution of GMCs}

\begin{figure}
\begin {center}
\includegraphics[width=0.9\linewidth]{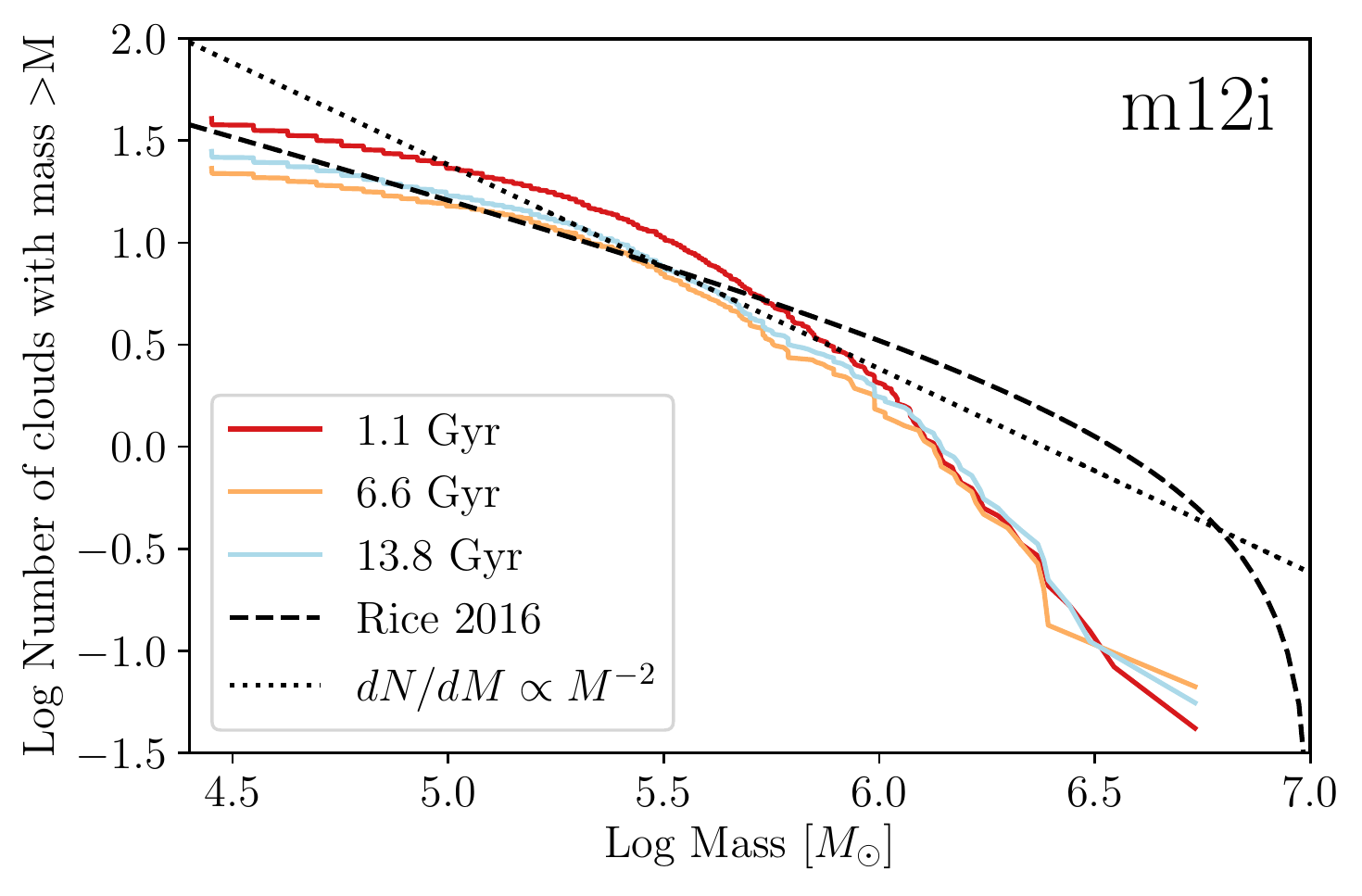}\\
\includegraphics[width=0.9\linewidth]{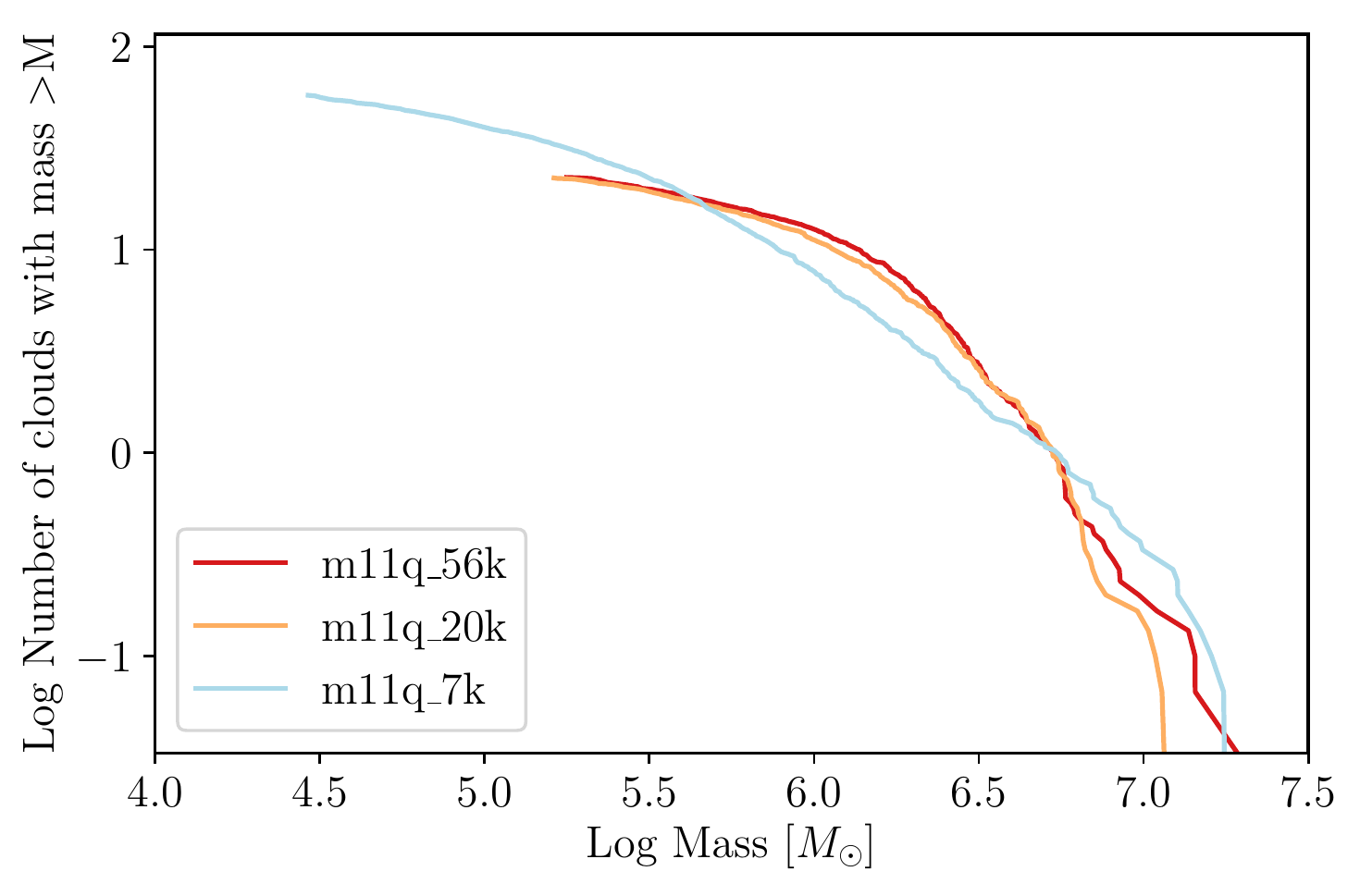}
\vspace{-0.5cm}
\caption{\emph{Top:} Evolution of the GMC mass distribution over cosmic time in our MW-like simulated galaxy \textbf{(m12i}). It shows the average number of clouds (per snapshot) with mass larger than $M$ averaged over a time window of 400 Myr. For comparison we also include the observational fit of \citealt{rice2016_mw_gmc_catalogue} and a $\dderiv N/\dderiv M\propto M^{-2}$ power law that naturally arises in scale-free structure formation processes \protect\citep{guszejnov_scaling_laws} that deposit equal mass at all scales. We find that the mass distribution of GMCs is essentially fixed throughout cosmic time, with temporary deviations at the high mass end (see spikes in the 90th percentile of GM mass in top left plot of Fig. \protect\ref{fig:m12i_evolplots}). \emph{Bottom:} The average number of GMCs above mass $M$ over the last 0.5 Gyr of cosmic time for the \textbf{m11q} simulated galaxy that has been rerun for the last 1 Gyr at different mass resolutions. As expected, we find that increasing the resolution allows us to resolve smaller GMCs. Still, the total mass in clouds >$10^5\,\msun$ in these runs is essentially constant as the mass function flattens at lower masses, so the end of the distribution (that contains the majority of the mass) is converged at the resolution of our fiducial runs ($\Delta m=7100\,\solarmass$).}
\label{fig:GMC_MF}
\vspace{-0.5cm}
\end {center}
\end{figure}

We find that the mass distribution of GMCs to be essentially invariant over cosmic time in both galaxies, with the exception of massive clouds that raise the high-mass end of the PDF (see  Figures \ref{fig:GMC_MF} and \ref{fig:m12i_evolplots} for \textbf{m12i}). During its cosmic evolution, the typical cloud in \textbf{m12i} has a mass of $\sim 10^6\,\solarmass$ and a surface density of $40\,\solarmass/\pc^2$, similar to those observed in the MW \citep[e.g.,][]{rice2016_mw_gmc_catalogue}. Figure \ref{fig:GMC_MF} shows that the overall shape of the mass function of identified GMCs is also similar to that observed for MW GMCs. While \textbf{m12i} lacks some of the most massive clouds the MW has at present day, it forms such clouds at different times during its evolution (see bottom of Figure \ref{fig:GMC_MF} for lifetime average). 

To verify the convergence of our results, we reran the \textbf{m11q} dwarf galaxy for the last 1 Gyr of its evolution at different mass resolutions ($7100$, $20000$, $56000\,\solarmass$). Figure \ref{fig:GMC_MF} shows that at our fiducial resolution of $7100\,\solarmass$, the high-mass end of the GMC mass distribution is converged. 

\subsection{Evolution of GMC properties with cosmic time}

\begin{figure*}
\begin {center}
\includegraphics[width=0.33\linewidth]{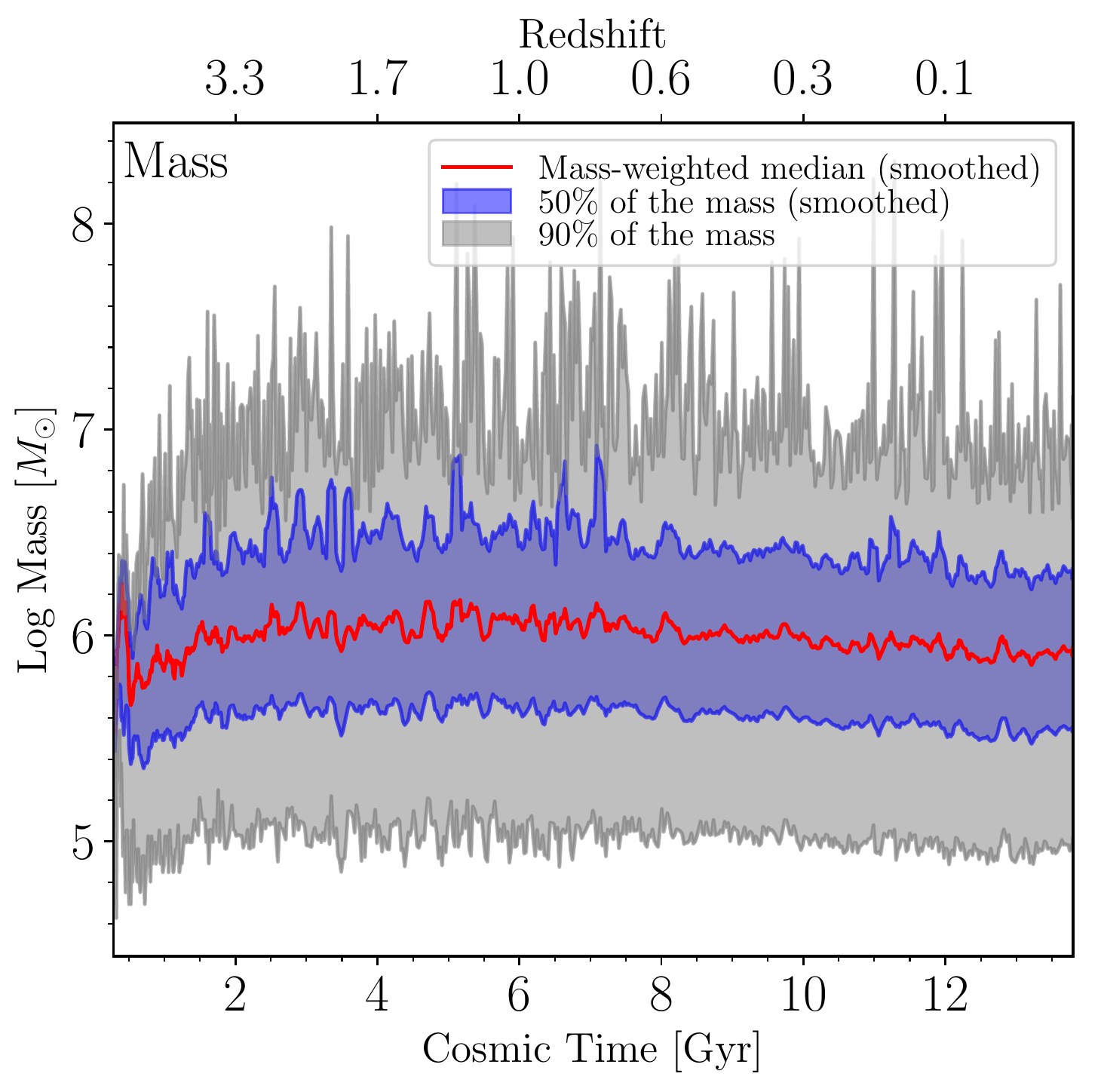}
\includegraphics[width=0.33\linewidth]{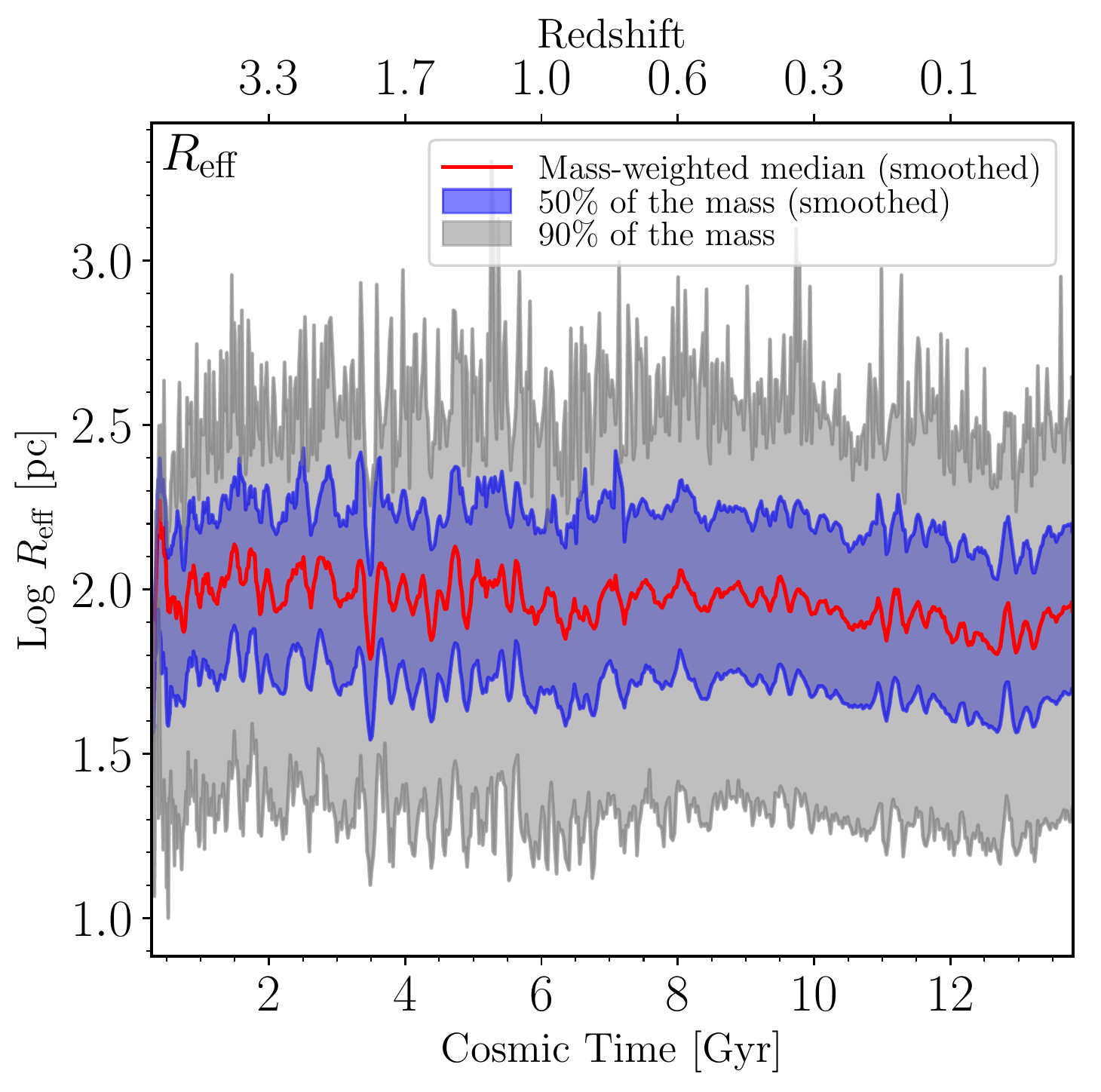}
\includegraphics[width=0.33\linewidth]{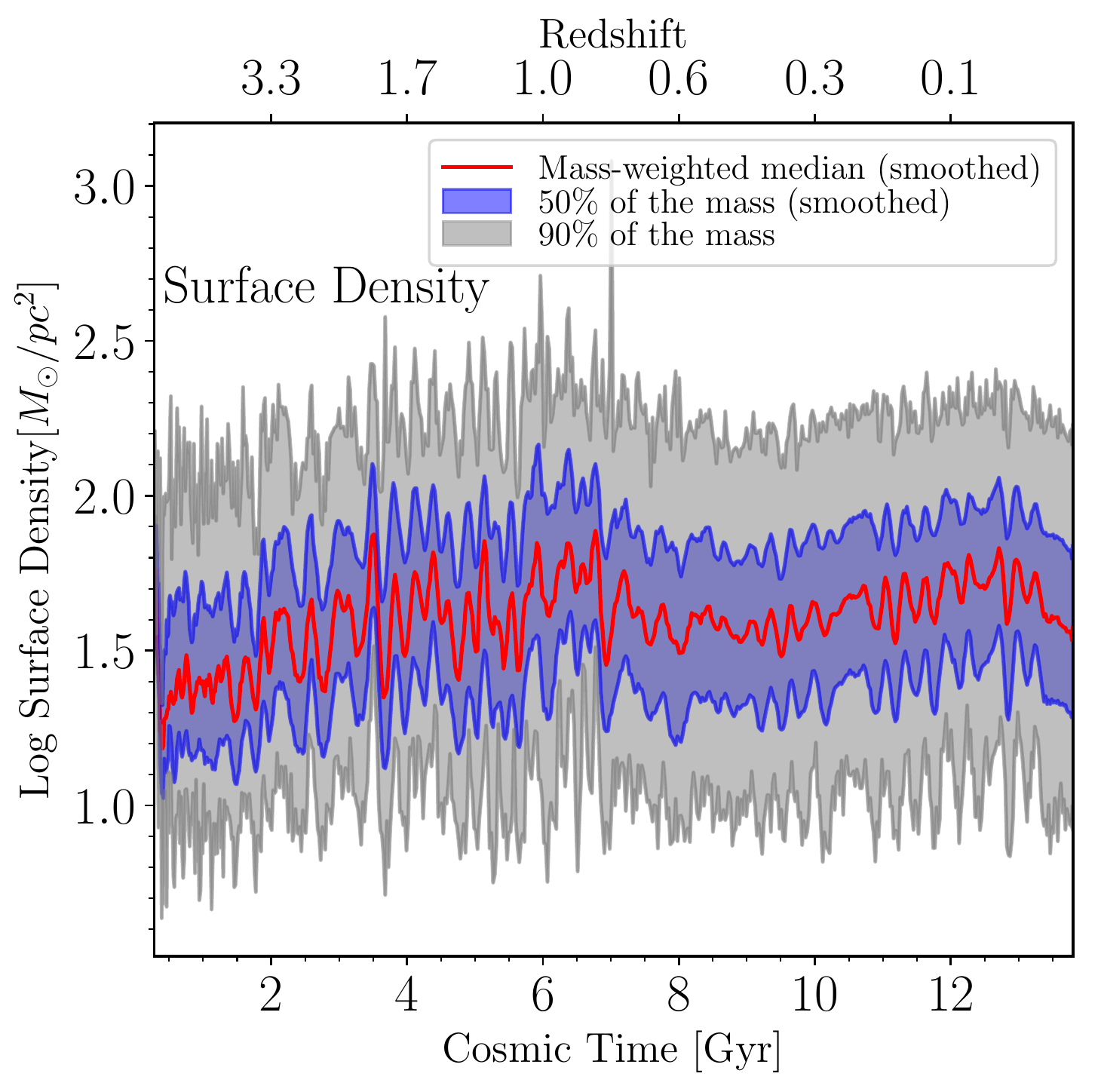} \\
\includegraphics[width=0.33\linewidth]{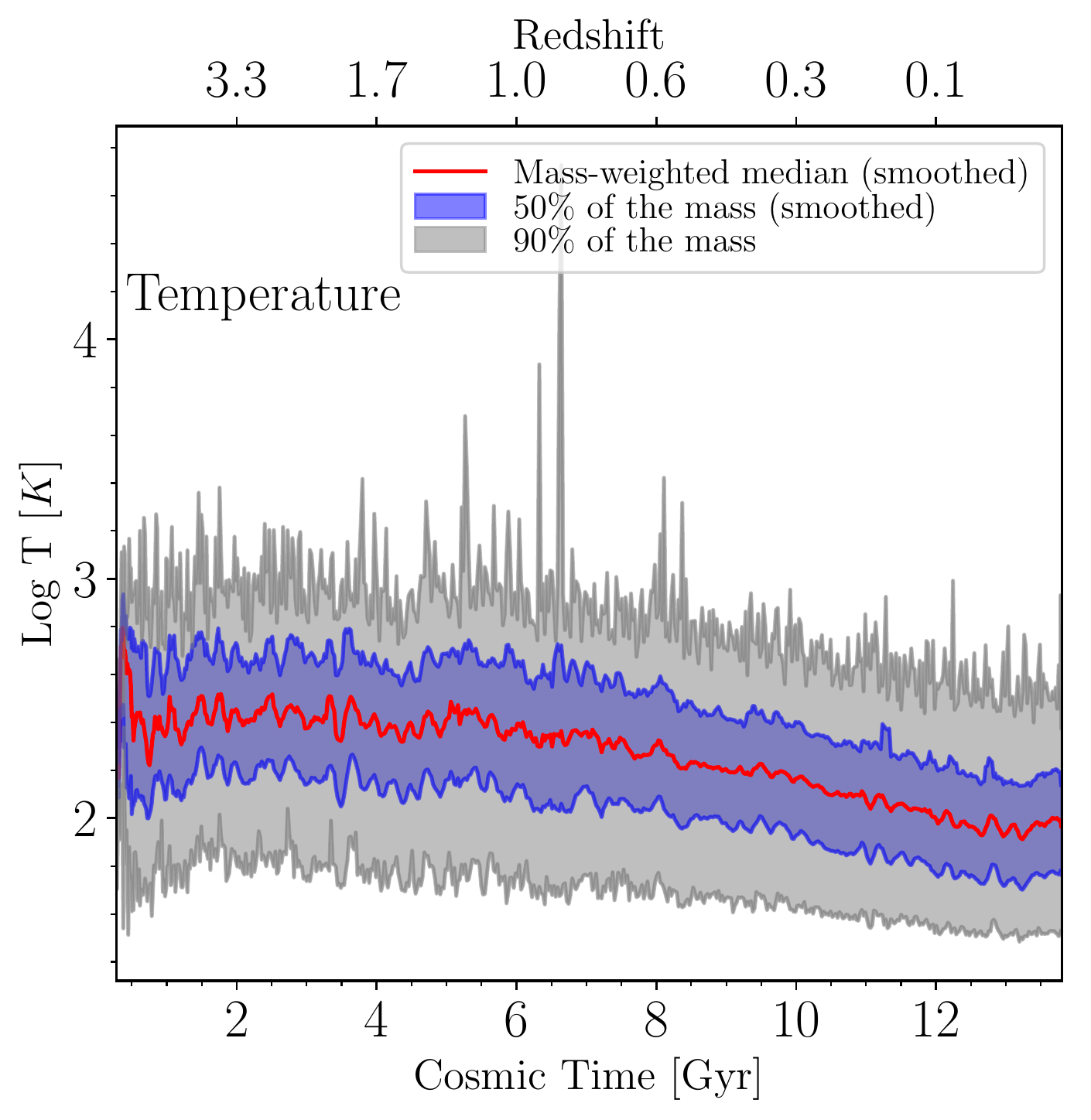}
\includegraphics[width=0.33\linewidth]{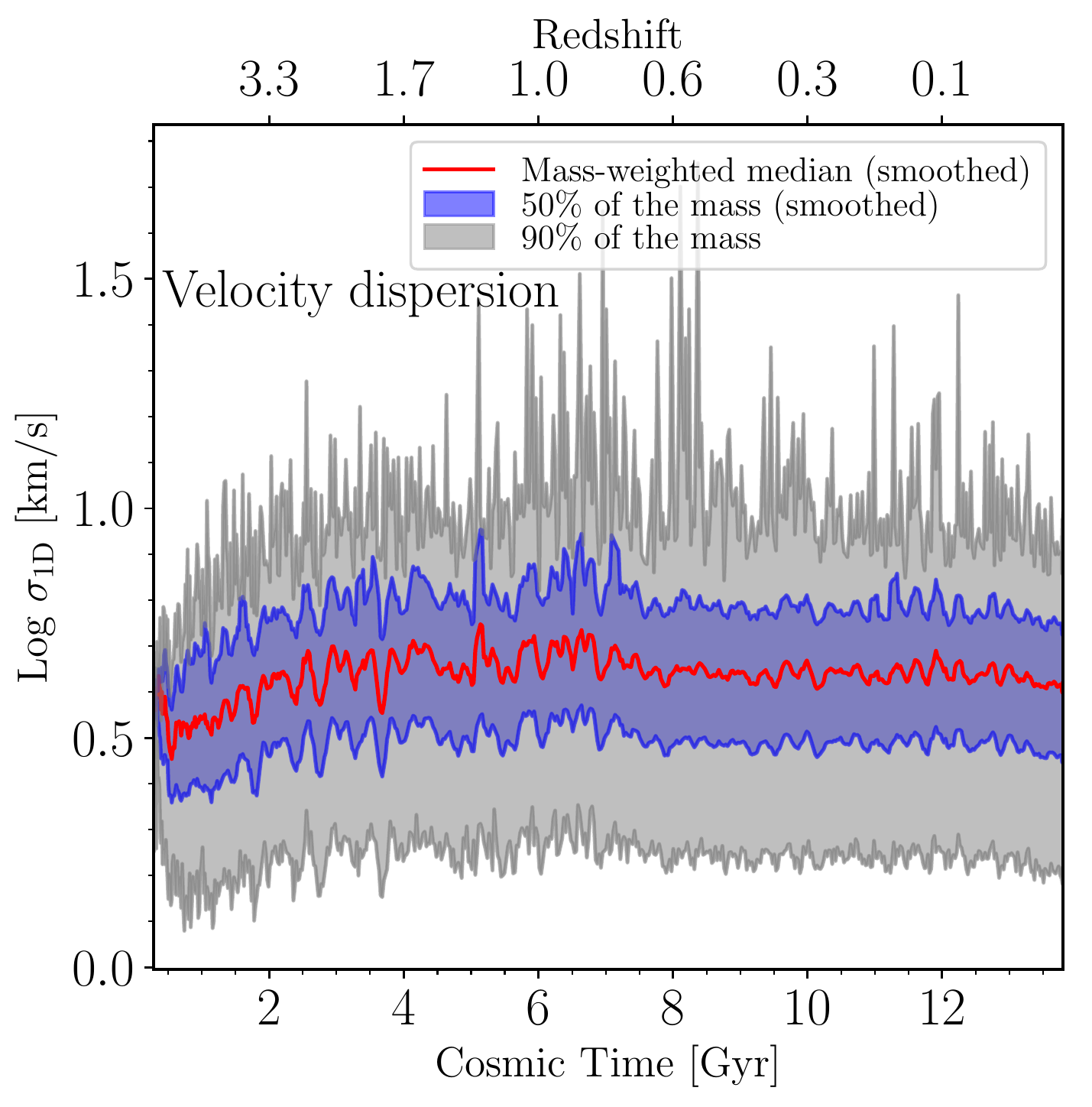}
\includegraphics[width=0.33\linewidth]{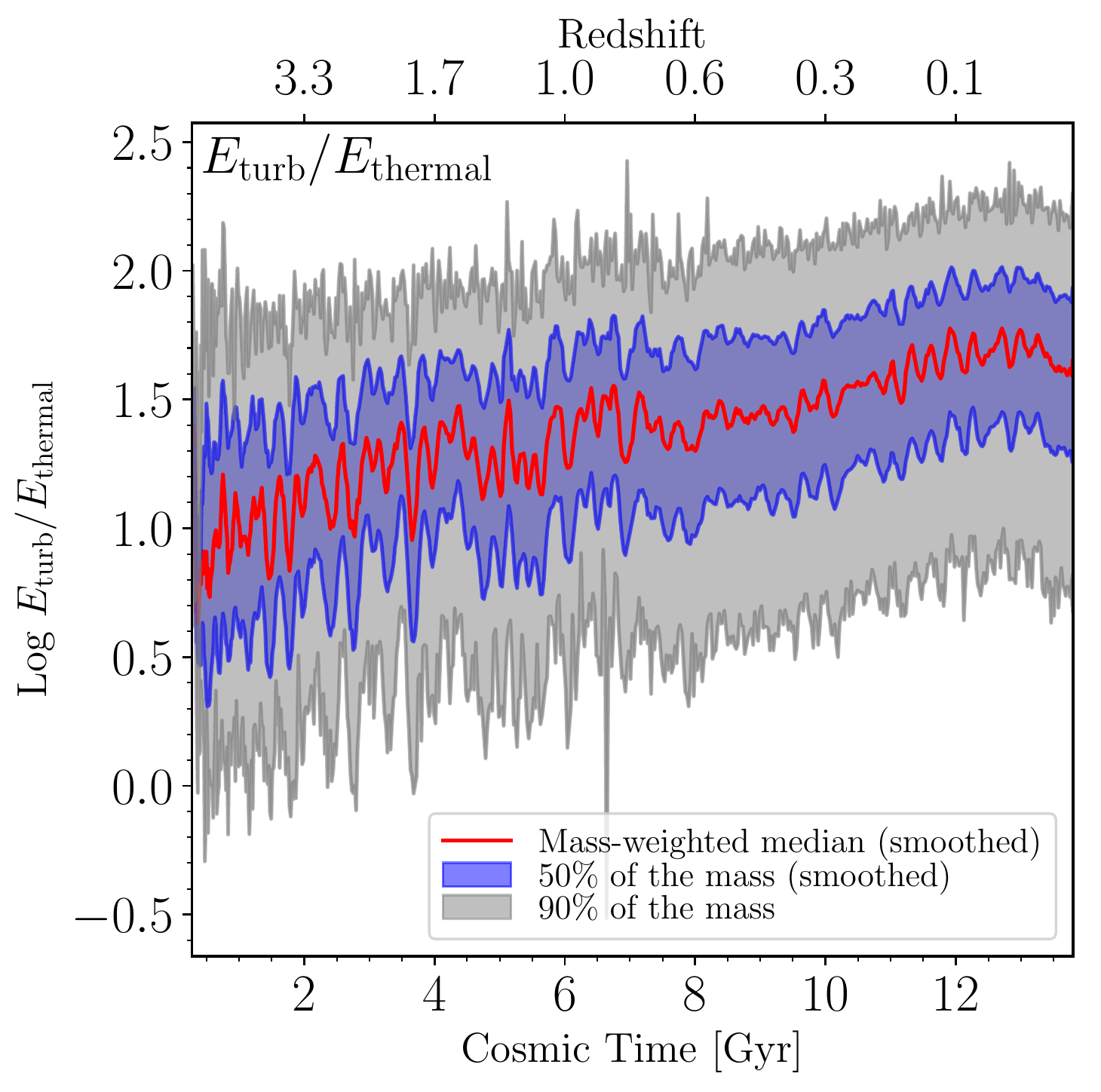}\\
\includegraphics[width=0.33\linewidth]{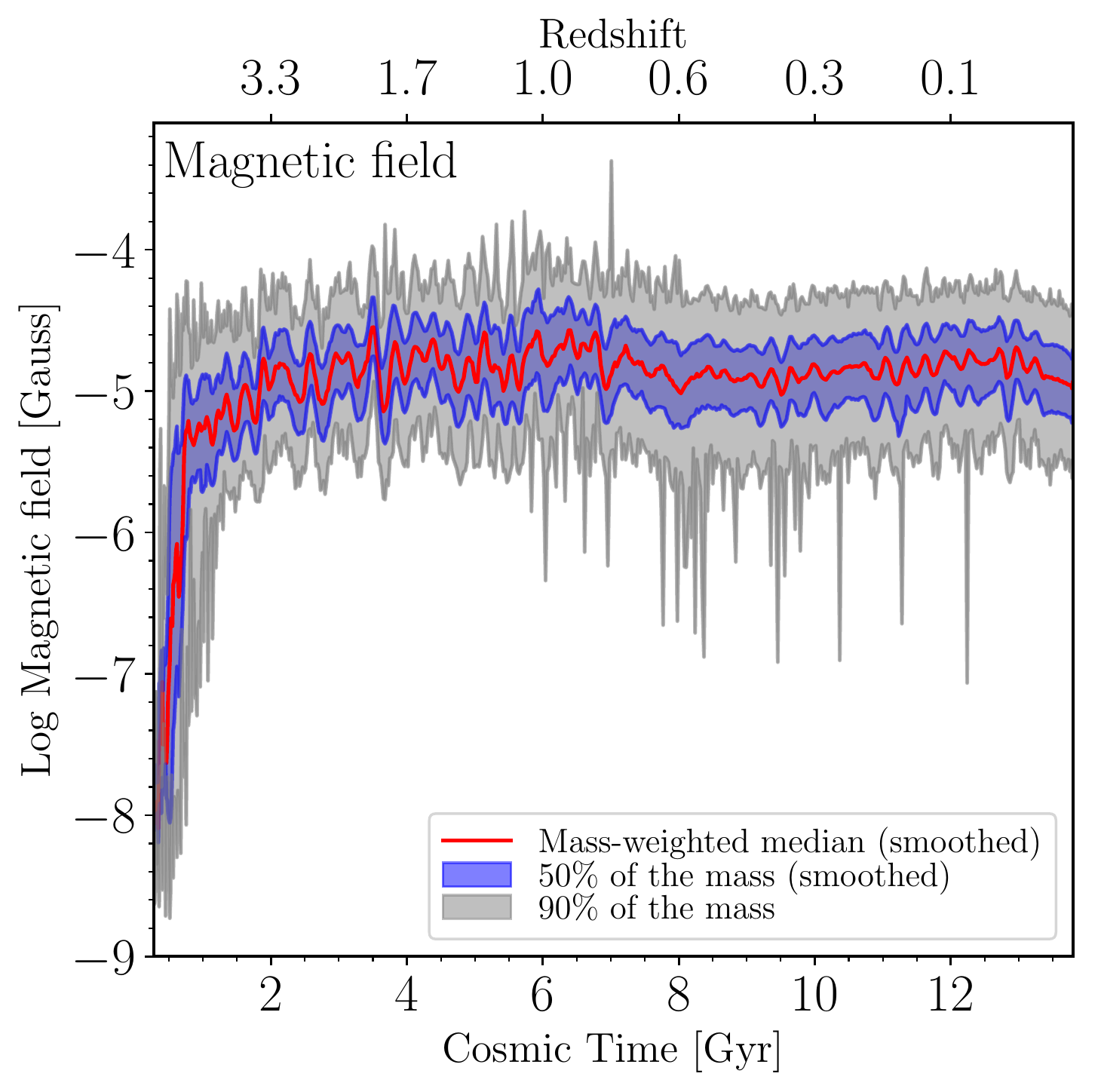}
\includegraphics[width=0.33\linewidth]{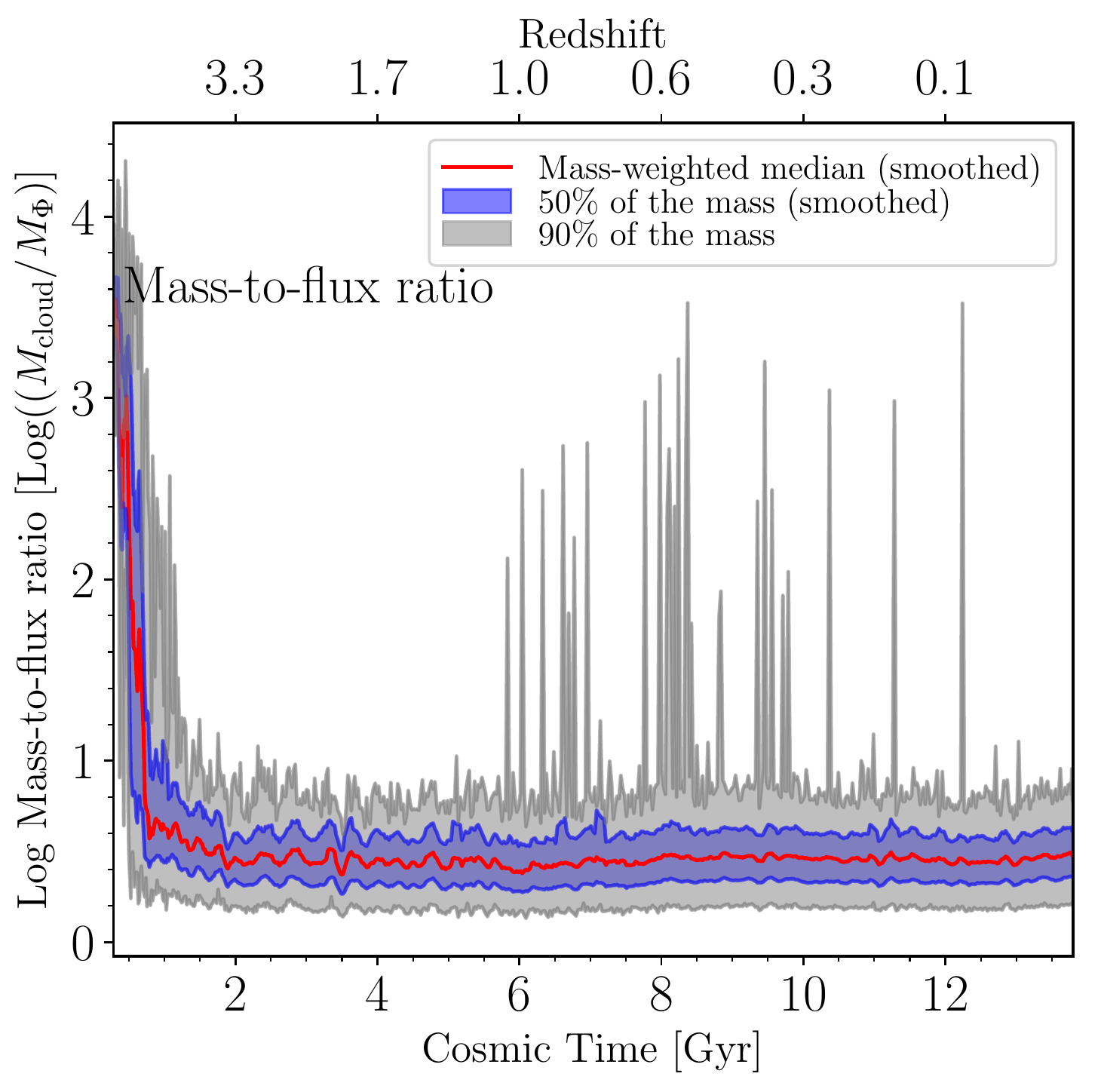}
\includegraphics[width=0.33\linewidth]{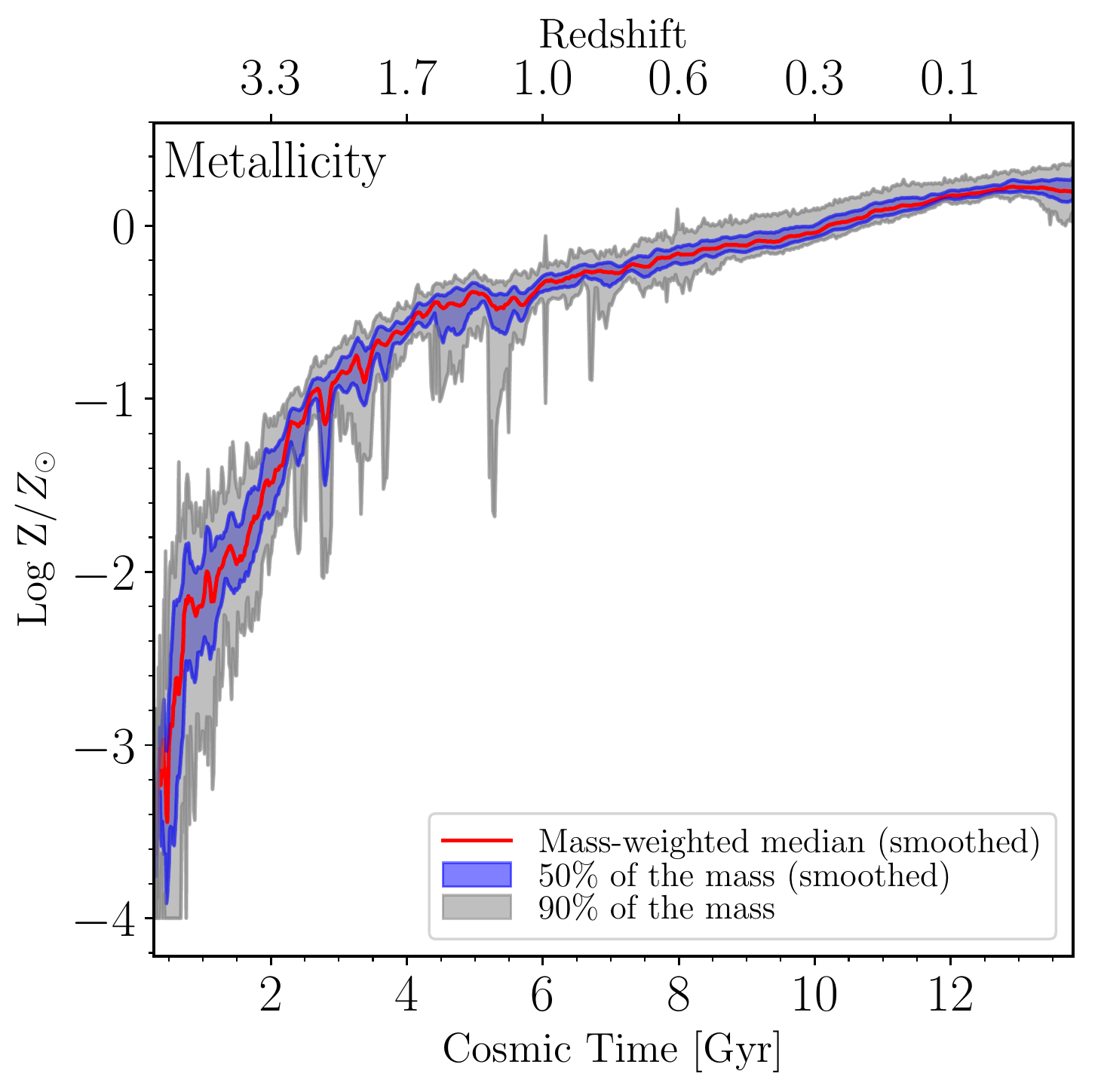}\\
\vspace{-0.5cm}
\caption{Evolution of GMC properties in \textbf{m12i} over cosmic time, including mass, size, surface density, temperature, turbulent velocity dispersion, turbulent to thermal energy ratio, magnetic field, mass-to-flux ratio and metallicity. We find that almost all of these properties remain constant after the galaxy forms. The most obvious exception is metallicity, which rises steadily with time. In fact, this is responsible for the the roughly 0.25 dex decline in temperature as cooling becomes more efficient, which, in turn, leads to the slight increase in the relative importance of turbulent support (as the velocity dispersion stays constant).}
\label{fig:m12i_evolplots}
\vspace{-0.5cm}
\end {center}
\end{figure*}

In Figure \ref{fig:m12i_evolplots}, we show the evolution of bulk cloud properties for the MW-like (\textbf{m12i}) simulated galaxy (see Appendix \ref{sec:m11q} for \textbf{m11q}). We find that the statistics of the bulk GMC properties show little evolution over cosmic time.

We find no global trends with time in the statistics of the GMC bulk mass, size, surface density, velocity dispersion and star formation rate. It should be noted that there are short-lived extreme changes at the tails of the distributions (see spikes in the 90th percentile values in Fig. \ref{fig:m12i_evolplots}), but these events have no long-term effects on the distribution. We find that these spikes are present in about 10\% of our simulation snapshots, with similar frequency in both the \myquote{bursty} and the  more \myquote{quiescent} phases of galaxy evolution.

There are, however, a factor of 2 level short-term variations in the median values of cloud size, surface density and turbulent support during the \myquote{bursty} star formation phase of both galaxies (first 7 Gyr for \textbf{m12i}, all of cosmic time for \textbf{m11q}), leading to somewhat broader distributions in GMC properties during the \myquote{bursty} phase. These trends are consistent with the behavior of the overall ISM in Figure \ref{fig:bulk_properties_evol}.

Most of the GMCs identified in the simulations have weak magnetic support, with a typical mass-to-flux ratio of 3, comparable to observed GMCs \citep{crutcher_2009_mc_magnetic_fields}, but some low-mass ($\sim10^{4.5}\,\solarmass$) clouds do approach $M/M_{\Phi}\approx 1$ (see Figure \ref{fig:scaling}). Also, turbulent motions dominate over thermal ones in almost all clouds, indicating that these clouds are supersonic, similar to observed GMCs \citep{Dobbs_2014_GMC_review} and have negligible thermal support.

In both types of galaxies, only the cloud metallicity shows a clear trend. Metallicity rises steadily over cosmic time as previous populations of stars deposit more metals into the ISM (see Section \ref{sec:Z_evol} for comparison with observations). We also find a weakly decreasing trend in bulk GMC temperatures in \textbf{m12i} with constant velocity dispersion, which leads to an increasing ratio of turbulent support as time progresses. This decrease in temperature is due to the increasing metallicity that leads to more efficient cooling. This argument is supported by the absence of this temperature trend in \textbf{m11q}, which has significantly lower metallicity (see Figure \ref{fig:m11q_evolplots}). Meanwhile the magnetic field saturates to about $10\,\mu\mathrm{G}$ in both cases soon after the formation of the galaxy, a value similar to that observed in MW GMCs \citep[see][]{crutcher_2009_mc_magnetic_fields}.

\subsubsection{Metallicity evolution}\label{sec:Z_evol}

\begin{figure}
\begin {center}
\includegraphics[width=0.90\linewidth]{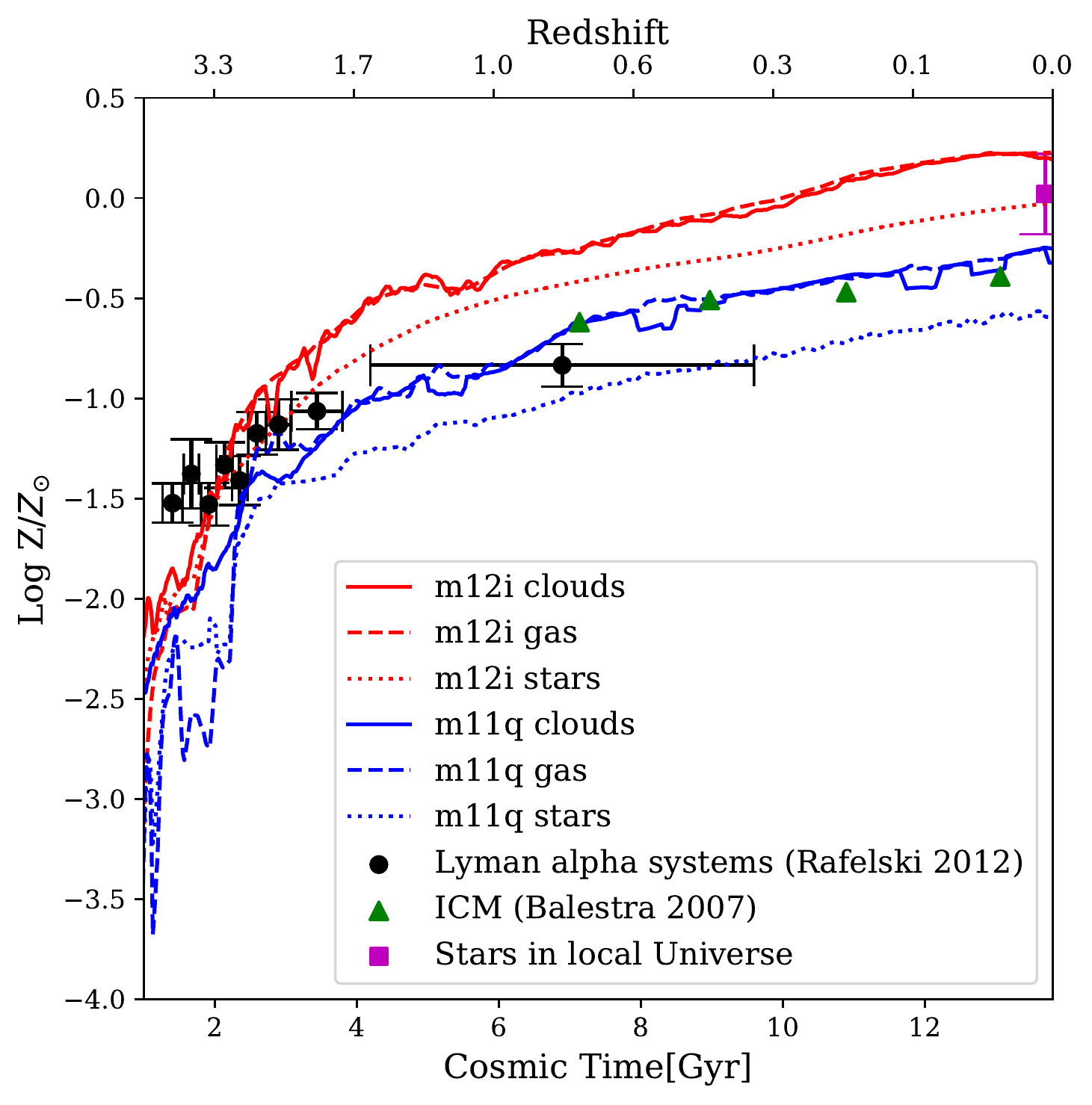}
\vspace{-0.4cm}
\caption{Evolution of the metallicity in \textbf{m12i} and \textbf{m11q} compared with the observed metallicity values in the ICM of massive galaxy clusters \protect\citep{Balestra_2007_ICM_metallicity}, in dampened Ly\protect$\alpha$ systems \protect\citep{Rafelski_2012_Lyalpha_Z} as well as in stars in the local universe \protect\citep{Gallazzi_2008_metal_baryons}. Solid lines show the mass-weighted median metallicity of GMCs, the dashed line shows the average over the galactic ISM, while the dotted line shows the mean stellar metallicity of the galaxy. A moving average over snapshots was applied to these curves to make the figure easier to interpret. We find that the clouds in the simulation follow the same qualitative trend that is observed in these different systems.} 
\label{fig:Z_obs_compare}
\vspace{-0.5cm}
\end {center}
\end{figure}

Figure \ref{fig:m12i_evolplots} show a remarkably tight relation between the metallicities of the identified GMCs and cosmic time\footnote{Note that Figure \protect\ref{fig:m12i_evolplots} only shows the metallicity for GMCs within the primary galaxy of the simulation, but in a few snapshots clouds from satellite galaxies are included, leading to a visible dip in the lower limit of the metallicity.}. Although there is no direct observation of this evolution, it is instructive to compare the metallicity evolution of simulated GMCs with the observed metallicity evolution of bright Ly$\alpha$ absorbing systems and the intracluster medium (ICM) of massive galaxy clusters. Figure \ref{fig:Z_obs_compare} shows that the GMCs in the simulated galaxies follow a similar cosmic evolution as the observed high redshift objects and reach values comparable to the values observed in present day stars in the local Universe. It should be noted that these trends almost perfectly match the mass-metallicity relation of \cite{Ma_2016_mass_metallicity} that was derived using both observations and earlier versions of the FIRE simulations. We find that the clouds essentially follow the evolution of the galactic ISM, and there is no offset between the cold medium and the rest of the ISM, while stellar metallicity lags behind and is consistently lower by roughly a factor of 2. Although the cloud GMC metallicity evolution is not expected to perfectly match that of the cluster ICMs and Ly\protect$\alpha$ systems, we find that the clouds in the simulated galaxies follow the same qualitative trend and our results match the present-day observations in the MW.

\subsection{Trends and scaling relations between GMC properties}

\begin{figure*}
\begin {center}
\includegraphics[width=0.33\linewidth]{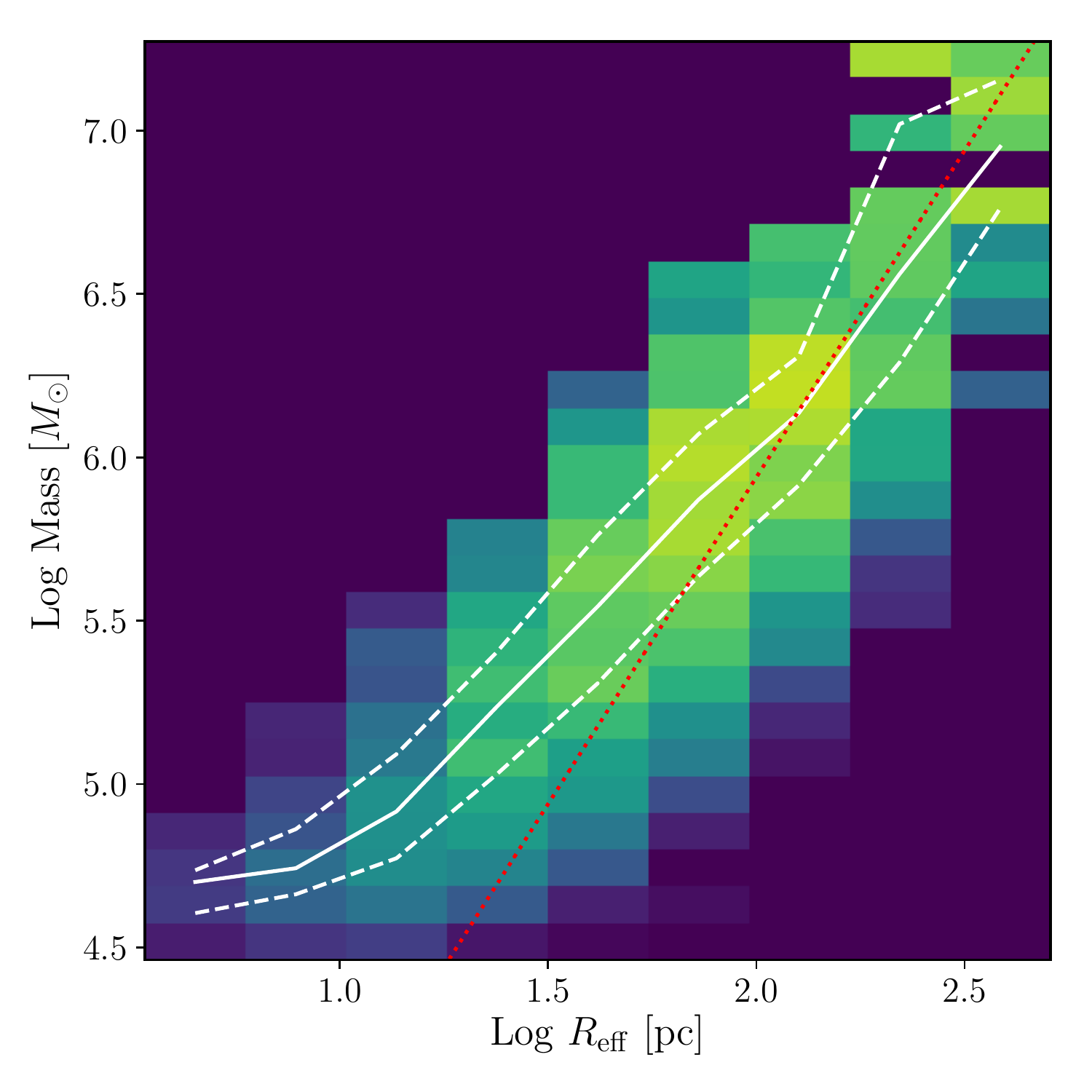}
\includegraphics[width=0.33\linewidth]{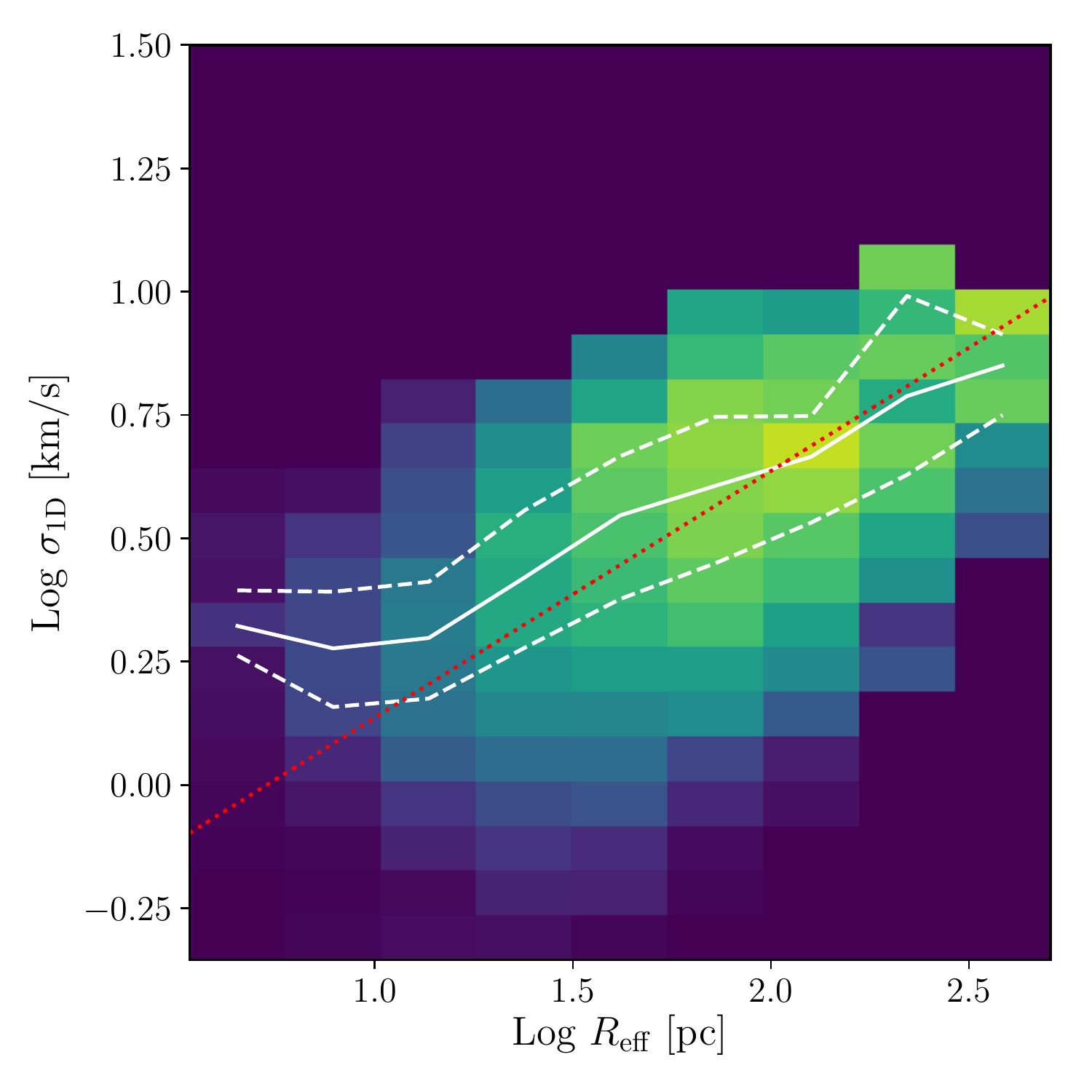}
\includegraphics[width=0.33\linewidth]{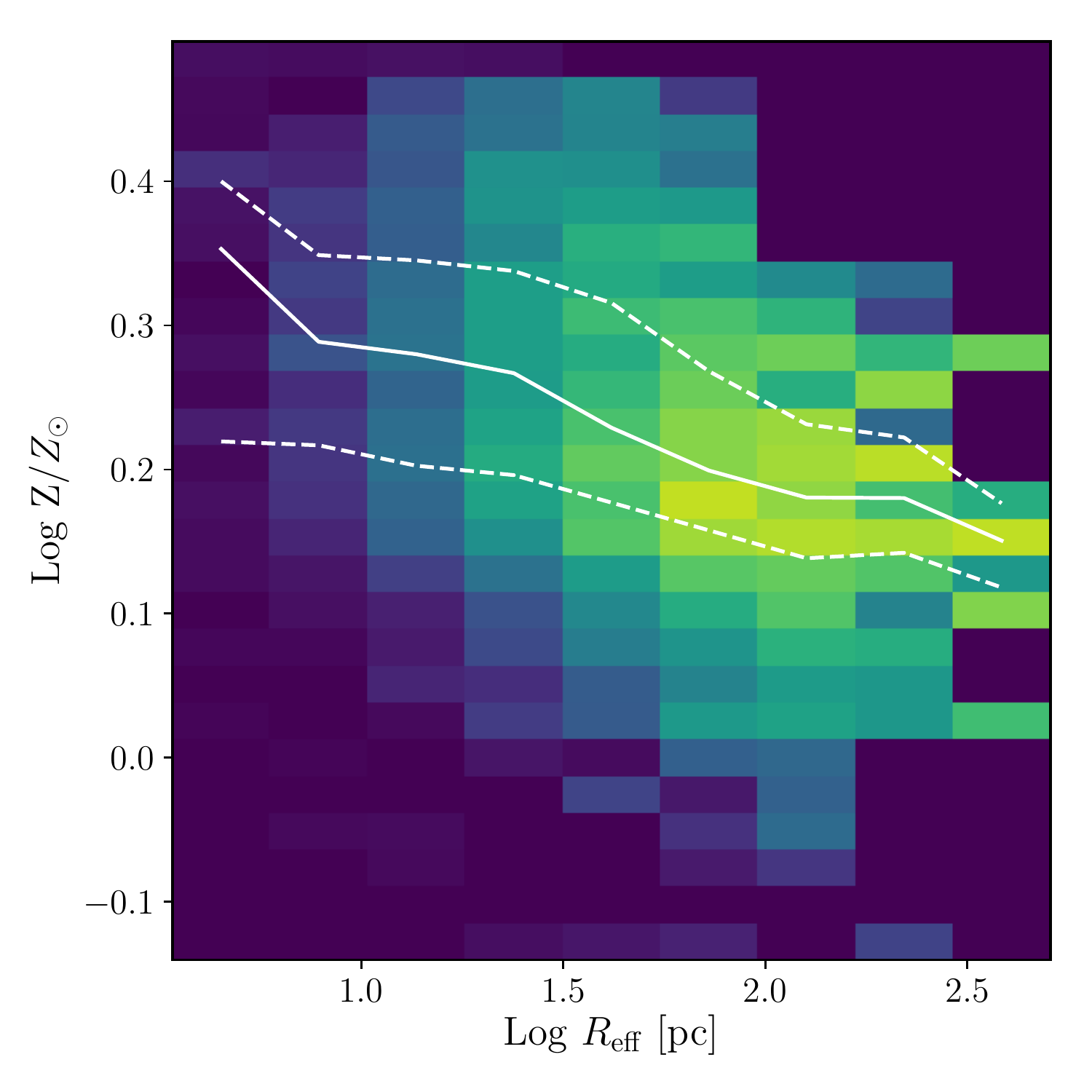}
\\
\includegraphics[width=0.33\linewidth]{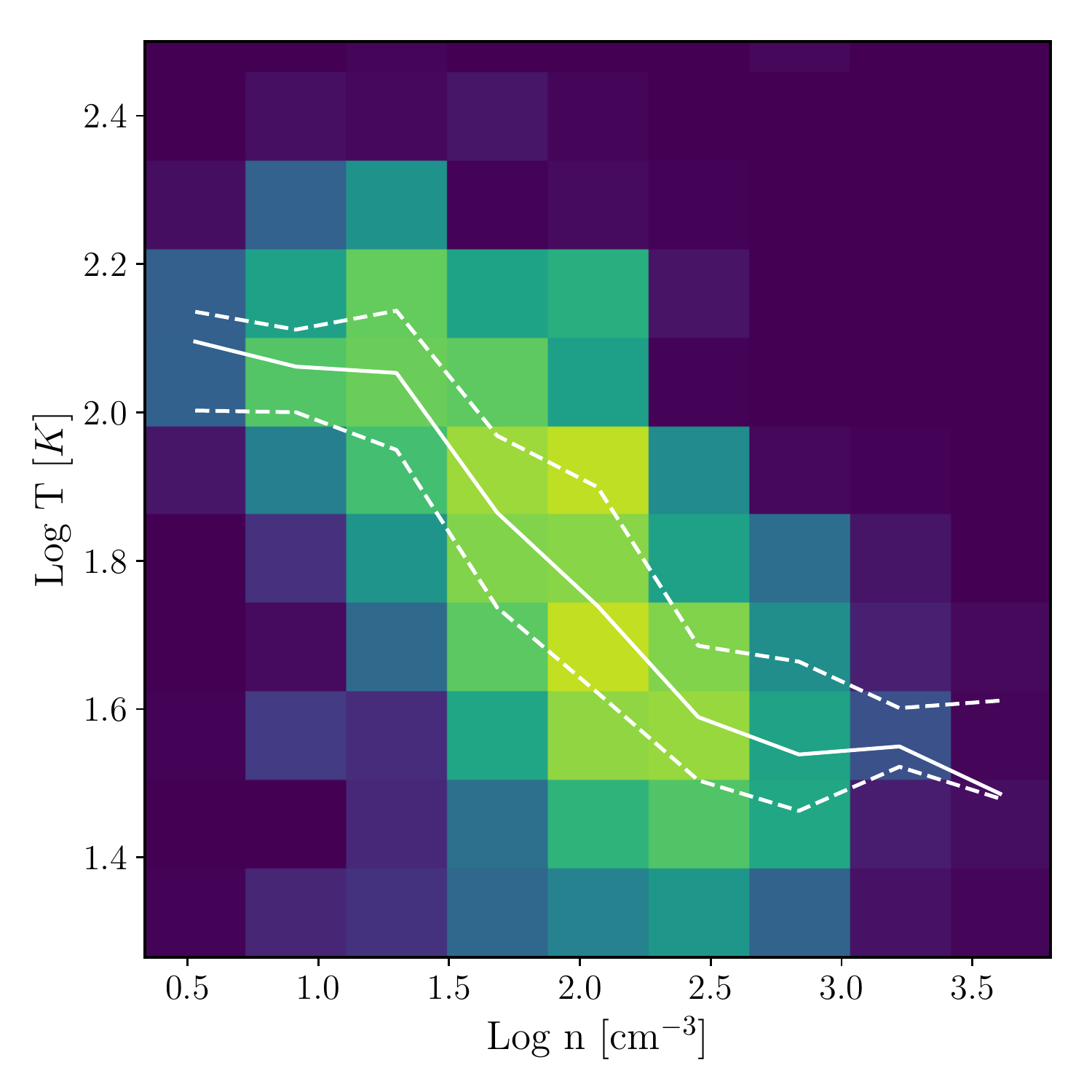}
\includegraphics[width=0.33\linewidth]{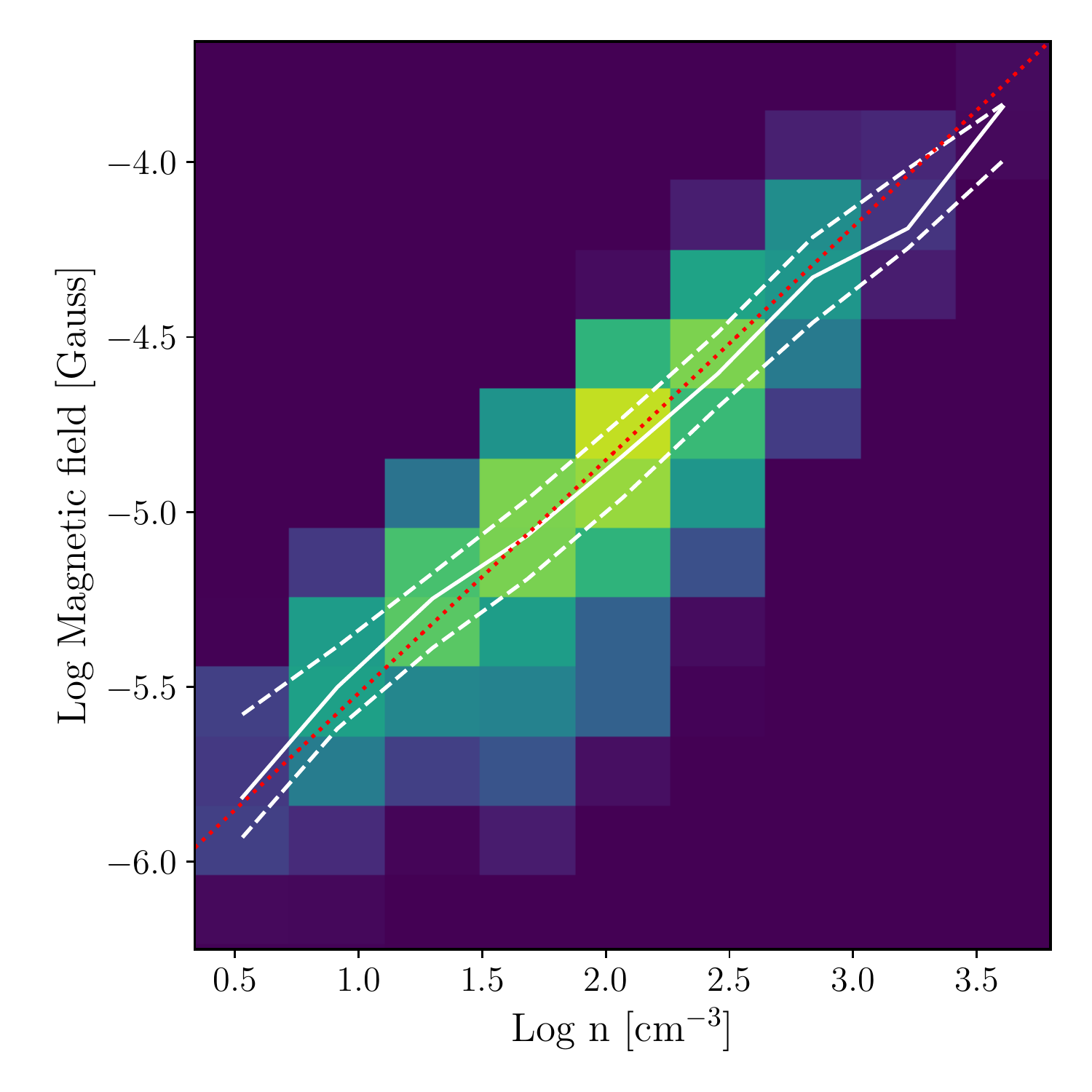}
\includegraphics[width=0.33\linewidth]{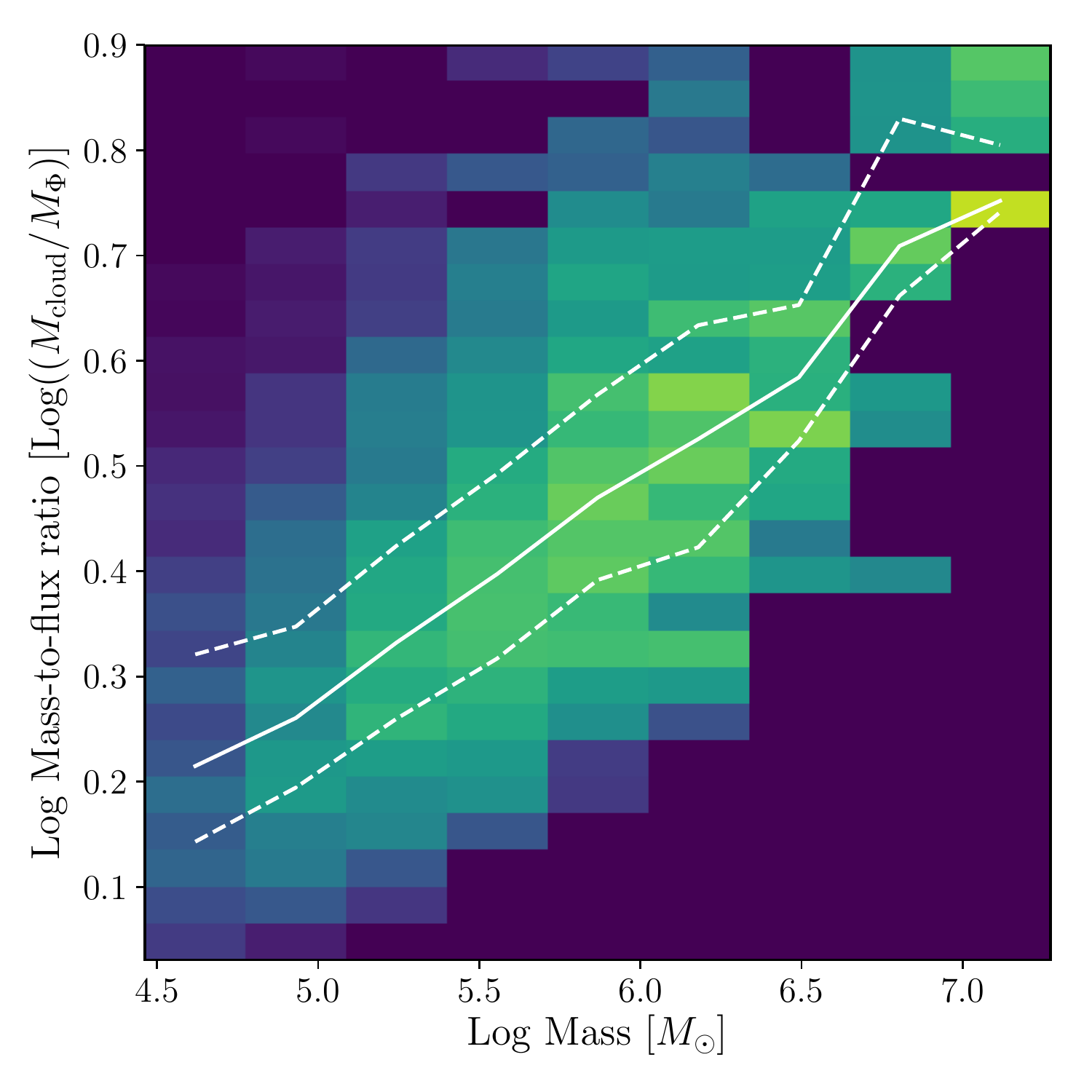}\\
\vspace{-0.4cm}
\caption{Relation between GMCs properties in \textbf{m12i} at z=0. The plotted PDF is color coded logarithmically with a 2 dex stretch (bright yellow/green colors denoting high values, while blue low ones), while the solid white line shows the median value at every size bin with dashed lines showing the inter-quartiles. \emph{Top, left:} Mass-size relation. Massive GMCs have a fixed surface density of $40\,\msun/\pc^2$, shown by a red dotted line. \emph{Top, middle:}  Linewidth-size relation. Massive GMCs roughly follow the Larson-like relation of $\sigma\propto R^{1/2}$ (red dotted line, \protect\citealt{larson_law}), with a flattening at the lowest masses. \emph{Top, right:}  Metallicity-size relation. Metallicity is weakly correlated with the effective radius of the clouds, with larger clouds having lower values. \emph{Bottom, left:} Temperature-density relation. We find that larger (and more massive) clouds tend to be a factor of 3 warmer than smaller ones, mainly due their higher density, which leads to more efficient cooling. \emph{Bottom, middle:} Magnetic field-density relation. There is a clear trend between the average density of the clouds and their mean magnetic field, consistent with a $B\propto \rho^{2/3}$ power-law (red dotted line), similar to the scaling that arises from \myquote{flux-freezing} in isotropic ideal MHD. \emph{Bottom, right:} Mass-to-flux-ratio - mass relation. We find that magnetic fields provide negligible support to massive clouds, but their importance increases for low mass clouds where $M/M_{\Phi}\approx 2$.} 
\label{fig:scaling}
\vspace{-0.5cm}
\end {center}
\end{figure*}

\begin{figure}
\begin {center}
\includegraphics[width=0.90\linewidth]{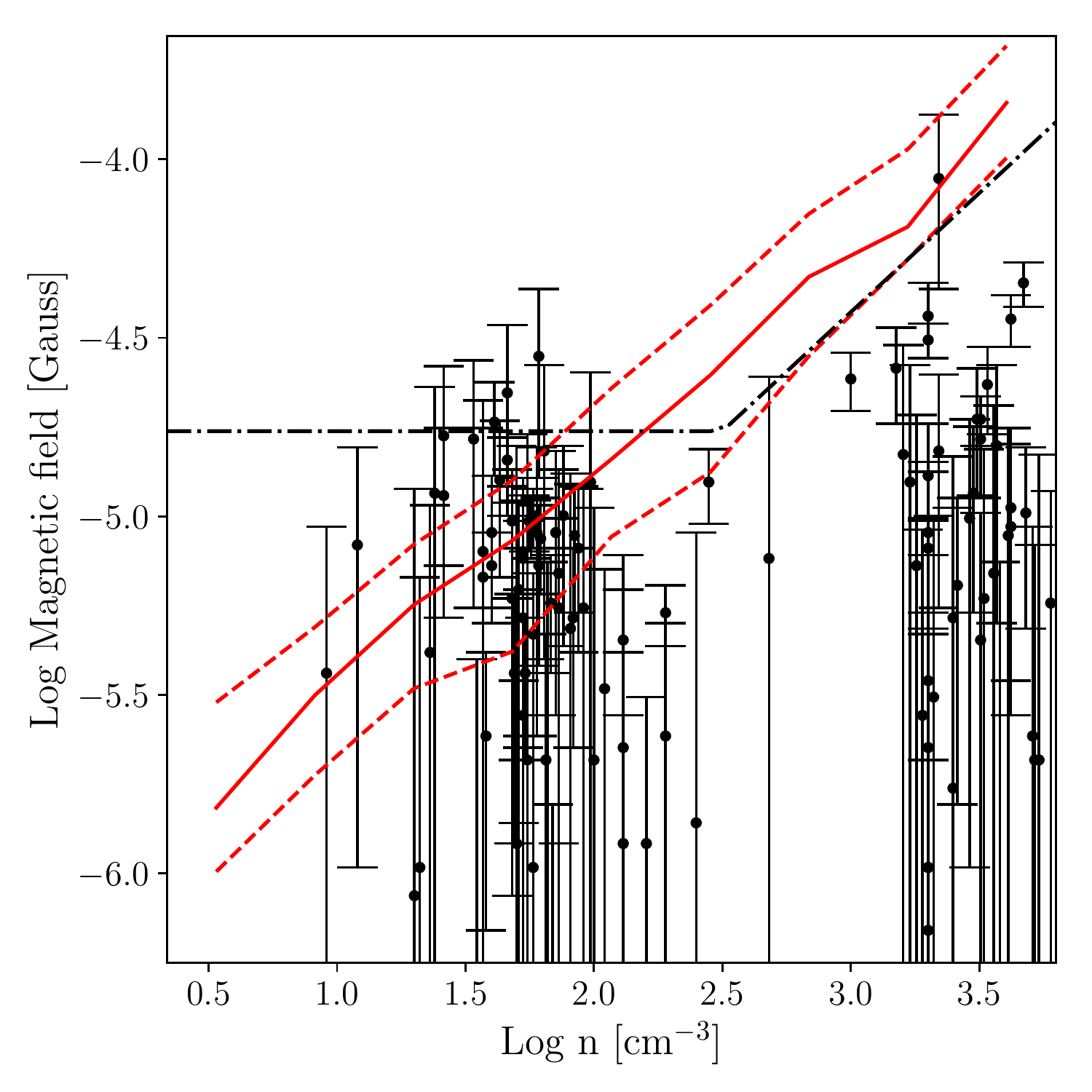}
\vspace{-0.4cm}
\caption{
Magnetic field strength versus density for the clouds of the \textbf{m12i} simulated galaxy compared with observations of Milky Way clouds \protect\citep{Crutcher_2010}. The solid red line shows the median value at every density bin and the dashed lines show the region where 90\% of the mass in the bin is located. Meanwhile, the black dash-dotted line shows the fitting function obtained by \protect\citealt{Crutcher_2010}. We find that the magnetic field in the clouds identified in the simulation follow a similar scaling to observed GMCs, but exhibits no turnover at low densities.} 
\label{fig:magnetic_compare}
\vspace{-0.5cm}
\end {center}
\end{figure}

Figure \ref{fig:scaling} shows several important scaling relations for the GMCs identified in the \textbf{m12i} simulated galaxy:
\begin{itemize}
    \item We find that the present-day GMCs (identified in both galaxies) have a mass-size relation roughly consistent with a constant surface density (albeit a different value, see Table \ref{tab:galaxies_result}), with a deviation to higher surface densities in lower mass clouds, likely due to increased thermal and magnetic support.
    \item GMCs follow a Larson-like linewidth-size scaling relation \citep{larson_law} with a flattening at the lowest masses, similar to what is found by \citealt{Lakhlani_2019_GMC_FIRE_present}.
    \item Metallicity is weakly correlated with the effective radius of the clouds, with larger clouds having lower values. This is likely due to the fact that larger clouds tend not to form in the inner regions of the galactic disk in our simulations, thus they are (on average) less enriched. This is contrary to observed trends in spiral galaxies \citep[e.g.,][]{rice2016_mw_gmc_catalogue, Miville_2017_MWG_GMC, Freeman_2017_M83_GMC}. We will discuss this in more detail in a follow-up paper.
    \item We find that larger (and more massive) clouds tend to be a factor of 3 warmer than smaller ones, mainly due to their lower density, which leads to less efficient cooling. Note that here we take the mass-weighted median gas temperature of the cloud to avoid confusion from the inclusion of the hot ISM envelope.
    \item There is a tight relationship between the magnetic field and the density of the GMC, consistent with the $B\propto \rho^{2/3}$ scaling that arises from flux conservation in the cases of isotropic collapse or energy equipartition, similar to that found in the ISM of simulated galaxies \citep{KungYi_2018_magnetic_fields}. In Figure \ref{fig:magnetic_compare} we compare our clouds with the observations in \citealt{Crutcher_2010}. We find that our clouds follow the same power-law scaling at high densities and are broadly consistent with the low density end.\footnote{The data of \protect\citealt{Crutcher_2010} is for a single direction of the magnetic field, so we shifted those results by a factor of $\sqrt{3}$ to compare with the magnetic field defined by Eq. \ref{eq:B_field}.} Unlike the fitting function of \citealt{Crutcher_2010} the magnetic fields in our simulated clouds do not saturate to $10\,\mu\mathrm{G}$. This could be due to the differences in cloud definitions (see Section \ref{sec:cloudphinder}) as we looked at the magnetic field in bound clouds while low density the Zeeman observations shown in \citealt{Crutcher_2010} have very different selection criteria. Also, re-analysis of these observations have found power-law trends to be more consistent with observations at low densities \cite{Tritsis_2015_magnetic_field_density_relation}. However, some MHD simulations of the ISM do reproduce the observed turnover \protect\citep[e.g.,][]{Padoan_2016_isoT_MHD_SN_driving}. This will be investigated in a follow-up paper. 
    \item Most GMCs have a typical mass-to-flux ratio $>3$, meaning magnetic fields provide little support for these clouds; in particular, the level of magnetic support is not enough to impose a preferred direction of compression (hence the isotropic $2/3$ exponent for the density scaling). Lower-mass clouds ($\sim10^{4.5}\,\solarmass$), however, can attain mass-to-flux ratios comparable to 1. 
\end{itemize}

\section{Implications and Caveats}\label{sec:discussion}

Since GMCs are the primary engines of star formation, the evolution of their properties dramatically influences the star formation histories of their host galaxies and determines age and radial gradients in stellar properties (e.g., the IMF). We find that (with the exception of metallicity and a related weak change in temperature) there is no overall trend in GMC properties as a function of time. This means that \emph{the initial conditions of star formation over cosmic time are essentially constant in a present day MW-like galaxy, with the exception of metallicity and the resulting (less than factor of 2) change in bulk temperature}. Owing to the tight relation between metallicity and cosmic time the average stellar population history can be essentially expressed as a function of metallicity. It should be noted that observations in the local Universe only allow a very weak metallicity dependence for the stellar IMF \citep{guszejnov_extragal_imf_var}.

Although higher-redshift galaxies \emph{at a fixed stellar mass} are observed to have higher velocity dispersions and gas surface densities compared to their $z=0$ counterparts (implying denser, more massive GMCs, see \citealt{Tacconi_2013_massive_galaxy_gas}), when we follow the {\emph progenitor} of a present-day MW-mass galaxy, these trends are offset by the fact that the main progenitor galaxy is also becoming less massive (which, {\emph at fixed redshift} gives a lower velocity dispersion, surface density, and gas mass/Toomre mass). What is surprising is that these two trends quite nearly cancel, giving rise to very weak evolution in the disperion, Toomre mass, and typical cloud properties within that main progenitor.

\subsection{Caveats}

The largest caveat to the interpretation of our results is that our CloudPhinder algorithm identifies the largest bound gas structures in the simulations. Throughout the paper we refer to these objects as analogues of real GMCs. Although it is {\it not}  accurate to claim that all observed GMCs are gravitationally bound (according to any workable observational definition of a cloud), the observed properties of GMCs appear to be broadly consistent with the existence of an underlying population of self-gravitating clouds. Moreover, these self-gravitating structures are responsible for essentially all star formation and hence can be readily thought of as the more direct progenitors of stellar clusters and associations than their observational counterparts. Future work will explore the relationship between the statistics of bound clouds and observed clouds using mock observations on simulated CO emission maps.

The FIRE cosmological galaxy simulations that we are using employ many of approximations to make the problem computationally tractable. While these have been thoroughly checked (see \citealt{hopkins2014_fire, hopkins_fire2, Hopkins_FIRE2MHD_2019} and references therein), there are a several caveats that apply to our results:
\begin{itemize}
\item The simulations presented here have a mass resolution of $7100\,\msun$, which prevents them from resolving low mass ($\sim10^4\,\msun$) GMCs.
\item In  the  simulations,  once gas  elements  satisfy the star formation criteria they are replaced by star particles after a freefall time. This leads to star  formation  happening  in discrete  steps.  This causes no problems in  massive ($\sim 10^6-10^7\,\msun$) GMCs as the first generation of stars formed can continue to alter the GMC properties during subsequent star formation. However, in low-mass clouds star formation  will be artificially abrupt such that the feedback effects from the stars that already formed will not be reflected in the cloud properties. 
\item While the FIRE simulations include a vast number of physical processes, feedback from forming stars, i.e., protostellar outflows, is not explicitly included in the simulations. On GMC scales, feedback from massive stars dominates the energetics \citep[e.g.,][]{matzner02}. On sub-parsec scales, outflows act to reduce the star-formation efficiency of dense gas and determine the masses of individual stars \citep{Offner_Chaban_2017_jets}.  However, here the stellar IMF is an input, since the simulations do not follow the small-scale physics of star formation that produce the IMF. Therefore, neglecting protostellar outflow feedback in our runs should have negligible effect
\item The FIRE simulations assume a fixed IMF identical to the one in the local Universe \citep{kroupa_imf}. Thus the simulations disregard all effects on the GMC properties that might arise from IMF variations. This is also related to the previous point as radiative and outflow feeedback from lower mass stars help to set the IMF, stellar multiplicity and star formation efficiency of dense gas \citep[e.g.,][]{offner2010, guszejnov_feedback_necessity, guszejnov_correlation,Offner_Chaban_2017_jets}, which can significantly alter the long term evolution of the galaxy. 
\item The simulations do not explicitly follow non-equilibrium chemistry (e.g., molecular hydrogen formation/destruction), instead  relying  on  pre-tabulated  equilibrium  cooling  rates as  a  function  of  density,  temperature,  metallicity,  and  the strength of the local radiation field in several bands. These approximations have little to no effect on galactic star formation properties but they could conceivably alter small-scale cloud properties \citep{Hopkins_2012_galaxy_structure}.
\item The simulations include feedback, which models cloud dispersal, but may not resolve the interaction of feedback within clouds, which may impact the details of cloud turbulence and lifetimes.
\end{itemize}

\section{Conclusions}\label{sec:consclusions}
In this work we study the population of the largest gravitationally-bound gas structures that form and disperse dynamically throughout the history of simulated galaxies, analogous to the GMCs observed in galaxies. For our analysis we use two simulated galaxies from the FIRE collaboration \citep{Hopkins_FIRE2MHD_2019}, one (present day) MW-like spiral galaxy (\textbf{m12i}) and an LMC-like dwarf galaxy (\textbf{m11q}). We find that: 
\begin{itemize}
    \item The properties of self-gravitating gas clouds in the simulations are largely consistent with the observed properties of GMCs in the local Universe. Specifically, in a given galaxy at a given time, these clouds have a typical surface density $\Sigma_{\rm GMC} \sim \unit[40]{M_\odot\,pc^{-2}}$, and a typical median mass of $\sim 10^6 \unit{M_\odot}$ and a maximum mass $\sim 10^7 \unit{M_\odot}$. 
    \item The mass function of simulated GMCs is nearly constant throughout cosmic time and is qualitatively similar to the observed present day MW GMC mass function \citep{rice2016_mw_gmc_catalogue}, in agreement with what has been found in other FIRE simulations but with different cloud identification methods (e.g., Fig. 13 in \citealt{hopkins_fire2}). We also find short-lived fluctuations to the high mass tail of the distribution due to the formation of extremely massive clouds. 
    \item We find that the bulk properties of these bound GMCs show little-to-no evolution after the galaxy forms; this is true for both the MW-like \textbf{m12i} galaxy and the \textbf{m11q} dwarf galaxy. This includes the median cloud mass, surface density, size, velocity dispersion and mass-to-flux ratio.
    \item Over cosmic time the only GMC bulk property that shows a systematic change larger than its variance at fixed time is metallicity. The metal content of clouds steadily increases to roughly solar levels, with remarkably little scatter, consistent with theoretical expectations.
    \item In the MW-like (\textbf{m12i}) galaxy we find that over cosmic time the median cloud temperature decreases by a factor of 2, which leads to an increase in the relative importance of turbulence. This is likely due to the more efficient cooling at higher metallicities, which is consistent with the absence of this trend in the simulated dwarf galaxy (\textbf{m11q}) that has 0.5 dex lower metallicity.
    \item We find that the simulated GMCs have a median mass-to-flux ratio of 3 (comparable to observed GMCs), while their median ratio of turbulent to thermal energy is between 10-20. This means that these clouds are turbulence dominated and supersonic. We find a strong correlation between the strength of the magnetic field and the density of the clouds, consistent with the $B\propto \rho^{2/3}$ relation of isotropic collapse in ideal MHD. 
\end{itemize}

\acknowledgments
The authors would like to thank Philip F. Hopkins for his helpful comments.

This work used computational resources of the University of Texas at Austin and the Texas Advanced Computing Center (TACC; http://www.tacc.utexas.edu).
DG is supported by the Harlan J. Smith McDonald Observatory Postdoctoral Fellowship. MYG is supported by a CIERA Postoctoral Fellowship. SSRO is supported by NSF Career Award AST-1650486 and by a Cottrell Scholar Award from the Research Corporation for Science Advancement. MBK acknowledges support from NSF grant AST-1517226 and CAREER grant AST-1752913 and from NASA grants NNX17AG29G and HST-AR-14282, HST-AR-14554, HST-AR-15006, and HST-GO-14191 from the Space Telescope Science Institute, which is operated by AURA, Inc., under NASA contract NAS5-26555. CAFG was supported by NSF through grants AST-1517491, AST-1715216, and CAREER award AST-1652522, by NASA through grants NNX15AB22G and 17-ATP17-0067, and by a Cottrell Scholar Award from the Research Corporation for Science Advancement. AW received support from NASA, through ATP grant 80NSSC18K1097 and HST grants GO-14734 and AR-15057 from STScI, a Hellman Fellowship from UC Davis, and the Heising-Simons Foundation. Support for SRL was provided by NASA through Hubble Fellowship grant \#HST-JF2-$51395.001$-A awarded by the Space Telescope Science Institute, which is operated by the Association of Universities for Research in Astronomy, Inc., for NASA, under contract NAS5-26555.

\vspace{0.25cm}
\bibliographystyle{mnras}
\bibliography{bibliography}

\begin{thebibliography}{}
\makeatletter
\relax
\def\mn@urlcharsother{\let\do\@makeother \do\$\do\&\do\#\do\^\do\_\do\%\do\~}
\def\mn@doi{\begingroup\mn@urlcharsother \@ifnextchar [ {\mn@doi@}
  {\mn@doi@[]}}
\def\mn@doi@[#1]#2{\def\@tempa{#1}\ifx\@tempa\@empty \href
  {http://dx.doi.org/#2} {doi:#2}\else \href {http://dx.doi.org/#2} {#1}\fi
  \endgroup}
\def\mn@eprint#1#2{\mn@eprint@#1:#2::\@nil}
\def\mn@eprint@arXiv#1{\href {http://arxiv.org/abs/#1} {{\tt arXiv:#1}}}
\def\mn@eprint@dblp#1{\href {http://dblp.uni-trier.de/rec/bibtex/#1.xml}
  {dblp:#1}}
\def\mn@eprint@#1:#2:#3:#4\@nil{\def\@tempa {#1}\def\@tempb {#2}\def\@tempc
  {#3}\ifx \@tempc \@empty \let \@tempc \@tempb \let \@tempb \@tempa \fi \ifx
  \@tempb \@empty \def\@tempb {arXiv}\fi \@ifundefined
  {mn@eprint@\@tempb}{\@tempb:\@tempc}{\expandafter \expandafter \csname
  mn@eprint@\@tempb\endcsname \expandafter{\@tempc}}}

\bibitem[\protect\citeauthoryear{{Balestra}, {Tozzi}, {Ettori}, {Rosati},
  {Borgani}, {Mainieri}, {Norman}  \& {Viola}}{{Balestra}
  et~al.}{2007}]{Balestra_2007_ICM_metallicity}
{Balestra} I.,  {Tozzi} P.,  {Ettori} S.,  {Rosati} P.,  {Borgani} S.,
  {Mainieri} V.,  {Norman} C.,   {Viola} M.,  2007, \mn@doi [\aap]
  {10.1051/0004-6361:20065568}, \href
  {https://ui.adsabs.harvard.edu/abs/2007A&A...462..429B} {462, 429}

\bibitem[\protect\citeauthoryear{{Bate} \& {Bonnell}}{{Bate} \&
  {Bonnell}}{2005}]{batebonell2005}
{Bate} M.~R.,  {Bonnell} I.~A.,  2005, \mn@doi [\mnras]
  {10.1111/j.1365-2966.2004.08593.x}, \href
  {http://adsabs.harvard.edu/abs/2005MNRAS.356.1201B} {356, 1201}

\bibitem[\protect\citeauthoryear{{Bertoldi} \& {McKee}}{{Bertoldi} \&
  {McKee}}{1992}]{bertoldi_mckee_1992}
{Bertoldi} F.,  {McKee} C.~F.,  1992, \mn@doi [\apj] {10.1086/171638}, \href
  {https://ui.adsabs.harvard.edu/abs/1992ApJ...395..140B} {395, 140}

\bibitem[\protect\citeauthoryear{{Bolatto}, {Leroy}, {Rosolowsky}, {Walter}  \&
  {Blitz}}{{Bolatto} et~al.}{2008}]{bolatto_2008}
{Bolatto} A.~D.,  {Leroy} A.~K.,  {Rosolowsky} E.,  {Walter} F.,   {Blitz} L.,
  2008, \mn@doi [\apj] {10.1086/591513}, \href
  {http://adsabs.harvard.edu/abs/2008ApJ...686..948B} {686, 948}

\bibitem[\protect\citeauthoryear{{Cava}, {Schaerer}, {Richard},
  {P{\'e}rez-Gonz{\'a}lez}, {Dessauges-Zavadsky}, {Mayer}  \&
  {Tamburello}}{{Cava} et~al.}{2018}]{Cava_2018_giant_clump_highz}
{Cava} A.,  {Schaerer} D.,  {Richard} J.,  {P{\'e}rez-Gonz{\'a}lez} P.~G.,
  {Dessauges-Zavadsky} M.,  {Mayer} L.,   {Tamburello} V.,  2018, \mn@doi
  [Nature Astronomy] {10.1038/s41550-017-0295-x}, \href
  {https://ui.adsabs.harvard.edu/abs/2018NatAs...2...76C} {2, 76}

\bibitem[\protect\citeauthoryear{{Colombo}, {Rosolowsky}, {Ginsburg},
  {Duarte-Cabral}  \& {Hughes}}{{Colombo}
  et~al.}{2015}]{Colombo_2015_GMC_identification_dendo}
{Colombo} D.,  {Rosolowsky} E.,  {Ginsburg} A.,  {Duarte-Cabral} A.,   {Hughes}
  A.,  2015, \mn@doi [\mnras] {10.1093/mnras/stv2063}, \href
  {https://ui.adsabs.harvard.edu/abs/2015MNRAS.454.2067C} {454, 2067}

\bibitem[\protect\citeauthoryear{{Crutcher}}{{Crutcher}}{2012}]{crutcher_2009_mc_magnetic_fields}
{Crutcher} R.~M.,  2012, \mn@doi [\araa] {10.1146/annurev-astro-081811-125514},
  \href {http://adsabs.harvard.edu/abs/2012ARA%26A..50...29C} {50, 29}

\bibitem[\protect\citeauthoryear{{Crutcher}, {Wandelt}, {Heiles}, {Falgarone}
  \& {Troland}}{{Crutcher} et~al.}{2010}]{Crutcher_2010}
{Crutcher} R.~M.,  {Wandelt} B.,  {Heiles} C.,  {Falgarone} E.,   {Troland}
  T.~H.,  2010, \mn@doi [\apj] {10.1088/0004-637X/725/1/466}, \href
  {https://ui.adsabs.harvard.edu/abs/2010ApJ...725..466C} {725, 466}

\bibitem[\protect\citeauthoryear{{Dessauges-Zavadsky}
  et~al.,}{{Dessauges-Zavadsky}
  et~al.}{2015}]{Dessauges_Zavadsky_2015_highz_clumps}
{Dessauges-Zavadsky} M.,  et~al., 2015, \mn@doi [\aap]
  {10.1051/0004-6361/201424661}, \href
  {https://ui.adsabs.harvard.edu/abs/2015A%26A...577A..50D} {577, A50}

\bibitem[\protect\citeauthoryear{{Dessauges-Zavadsky}
  et~al.,}{{Dessauges-Zavadsky}
  et~al.}{2017}]{Dessauges_Zavadsky_2017_molecular_gas_highz}
{Dessauges-Zavadsky} M.,  et~al., 2017, \mn@doi [\aap]
  {10.1051/0004-6361/201628513}, \href
  {https://ui.adsabs.harvard.edu/abs/2017A%26A...605A..81D} {605, A81}

\bibitem[\protect\citeauthoryear{{Dobbs} \& {Pringle}}{{Dobbs} \&
  {Pringle}}{2013}]{dobbs_2013}
{Dobbs} C.~L.,  {Pringle} J.~E.,  2013, \mn@doi [\mnras]
  {10.1093/mnras/stt508}, \href
  {https://ui.adsabs.harvard.edu/abs/2013MNRAS.432..653D} {432, 653}

\bibitem[\protect\citeauthoryear{{Dobbs}, {Burkert}  \& {Pringle}}{{Dobbs}
  et~al.}{2011}]{dobbs_2011}
{Dobbs} C.~L.,  {Burkert} A.,   {Pringle} J.~E.,  2011, \mn@doi [\mnras]
  {10.1111/j.1365-2966.2011.18371.x}, \href
  {https://ui.adsabs.harvard.edu/abs/2011MNRAS.413.2935D} {413, 2935}

\bibitem[\protect\citeauthoryear{{Dobbs} et~al.,}{{Dobbs}
  et~al.}{2014}]{Dobbs_2014_GMC_review}
{Dobbs} C.~L.,  et~al., 2014, in {Beuther} H.,  {Klessen} R.~S.,  {Dullemond}
  C.~P.,   {Henning} T.,  eds, Protostars and Planets VI. p.~3 (\mn@eprint
  {arXiv} {1312.3223}), \mn@doi{10.2458/azu_uapress_9780816531240-ch001}

\bibitem[\protect\citeauthoryear{{Dobbs}, {Rosolowsky}, {Pettitt}, {Braine},
  {Corbelli}  \& {Sun}}{{Dobbs} et~al.}{2019}]{Dobbs_2019_M33_GMC_FoF}
{Dobbs} C.~L.,  {Rosolowsky} E.,  {Pettitt} A.~R.,  {Braine} J.,  {Corbelli}
  E.,   {Sun} J.,  2019, \mn@doi [\mnras] {10.1093/mnras/stz674}, \href
  {https://ui.adsabs.harvard.edu/abs/2019MNRAS.485.4997D} {485, 4997}

\bibitem[\protect\citeauthoryear{{Duarte-Cabral} \& {Dobbs}}{{Duarte-Cabral} \&
  {Dobbs}}{2016}]{Duarte_Cabral_2016_synth_obs_GMC}
{Duarte-Cabral} A.,  {Dobbs} C.~L.,  2016, \mn@doi [\mnras]
  {10.1093/mnras/stw469}, \href
  {https://ui.adsabs.harvard.edu/abs/2016MNRAS.458.3667D} {458, 3667}

\bibitem[\protect\citeauthoryear{{El-Badry} et~al.,}{{El-Badry}
  et~al.}{2018a}]{El_Badry_2018_FIRE_gas}
{El-Badry} K.,  et~al., 2018a, \mn@doi [\mnras] {10.1093/mnras/stx2482}, \href
  {https://ui.adsabs.harvard.edu/abs/2018MNRAS.473.1930E} {473, 1930}

\bibitem[\protect\citeauthoryear{{El-Badry} et~al.,}{{El-Badry}
  et~al.}{2018b}]{El_Badry_2018_FIRE_gas_HI}
{El-Badry} K.,  et~al., 2018b, \mn@doi [\mnras] {10.1093/mnras/sty730}, \href
  {https://ui.adsabs.harvard.edu/abs/2018MNRAS.477.1536E} {477, 1536}

\bibitem[\protect\citeauthoryear{{Elmegreen} \& {Falgarone}}{{Elmegreen} \&
  {Falgarone}}{1996}]{elmegreen_1996_fractal}
{Elmegreen} B.~G.,  {Falgarone} E.,  1996, \mn@doi [\apj] {10.1086/178009},
  \href {http://adsabs.harvard.edu/abs/1996ApJ...471..816E} {471, 816}

\bibitem[\protect\citeauthoryear{{Faucher-Gigu{\`e}re}}{{Faucher-Gigu{\`e}re}}{2018}]{FG_2018_bursty_SFR}
{Faucher-Gigu{\`e}re} C.-A.,  2018, \mn@doi [\mnras] {10.1093/mnras/stx2595},
  \href {https://ui.adsabs.harvard.edu/abs/2018MNRAS.473.3717F} {473, 3717}

\bibitem[\protect\citeauthoryear{{Ferland} et~al.,}{{Ferland}
  et~al.}{2013}]{CLOUDY}
{Ferland} G.~J.,  et~al., 2013, \rmxaa, \href
  {http://adsabs.harvard.edu/abs/2013RMxAA..49..137F} {49, 137}

\bibitem[\protect\citeauthoryear{{Freeman}, {Rosolowsky}, {Kruijssen},
  {Bastian}  \& {Adamo}}{{Freeman} et~al.}{2017}]{Freeman_2017_M83_GMC}
{Freeman} P.,  {Rosolowsky} E.,  {Kruijssen} J.~M.~D.,  {Bastian} N.,   {Adamo}
  A.,  2017, \mn@doi [\mnras] {10.1093/mnras/stx499}, \href
  {http://adsabs.harvard.edu/abs/2017MNRAS.468.1769F} {468, 1769}

\bibitem[\protect\citeauthoryear{{Fujimoto}, {Chevance}, {Haydon}, {Krumholz}
  \& {Kruijssen}}{{Fujimoto} et~al.}{2019}]{fujimoto_2019}
{Fujimoto} Y.,  {Chevance} M.,  {Haydon} D.~T.,  {Krumholz} M.~R.,
  {Kruijssen} J.~M.~D.,  2019, \mn@doi [\mnras] {10.1093/mnras/stz641}, \href
  {https://ui.adsabs.harvard.edu/abs/2019MNRAS.tmp..625F} {p.~625}

\bibitem[\protect\citeauthoryear{{Gallazzi}, {Brinchmann}, {Charlot}  \&
  {White}}{{Gallazzi} et~al.}{2008}]{Gallazzi_2008_metal_baryons}
{Gallazzi} A.,  {Brinchmann} J.,  {Charlot} S.,   {White} S.~D.~M.,  2008,
  \mn@doi [\mnras] {10.1111/j.1365-2966.2007.12632.x}, \href
  {https://ui.adsabs.harvard.edu/abs/2008MNRAS.383.1439G} {383, 1439}

\bibitem[\protect\citeauthoryear{{Garrison-Kimmel} et~al.,}{{Garrison-Kimmel}
  et~al.}{2018}]{Garrison_Kimmel_2018_galaxy_morphology}
{Garrison-Kimmel} S.,  et~al., 2018, \mn@doi [\mnras] {10.1093/mnras/sty2513},
  \href {https://ui.adsabs.harvard.edu/abs/2018MNRAS.481.4133G} {481, 4133}

\bibitem[\protect\citeauthoryear{{Grisdale}, {Agertz}, {Renaud}  \&
  {Romeo}}{{Grisdale} et~al.}{2018}]{Grisdale_2018_simulated_clouds}
{Grisdale} K.,  {Agertz} O.,  {Renaud} F.,   {Romeo} A.~B.,  2018, \mn@doi
  [\mnras] {10.1093/mnras/sty1595}, \href
  {https://ui.adsabs.harvard.edu/abs/2018MNRAS.479.3167G} {479, 3167}

\bibitem[\protect\citeauthoryear{{Grudi{\'c}}, {Hopkins},
  {Faucher-Gigu{\`e}re}, {Quataert}, {Murray}  \& {Kere{\v s}}}{{Grudi{\'c}}
  et~al.}{2016}]{grudic_2016}
{Grudi{\'c}} M.~Y.,  {Hopkins} P.~F.,  {Faucher-Gigu{\`e}re} C.-A.,  {Quataert}
  E.,  {Murray} N.,   {Kere{\v s}} D.,  2016, preprint, \href
  {http://adsabs.harvard.edu/abs/2016arXiv161205635G} {} (\mn@eprint {arXiv}
  {1612.05635})

\bibitem[\protect\citeauthoryear{{Guszejnov}, {Krumholz}  \&
  {Hopkins}}{{Guszejnov} et~al.}{2016}]{guszejnov_feedback_necessity}
{Guszejnov} D.,  {Krumholz} M.~R.,   {Hopkins} P.~F.,  2016, \mn@doi [\mnras]
  {10.1093/mnras/stw315}, \href
  {http://adsabs.harvard.edu/abs/2016MNRAS.458..673G} {458, 673}

\bibitem[\protect\citeauthoryear{{Guszejnov}, {Hopkins}  \&
  {Krumholz}}{{Guszejnov} et~al.}{2017}]{guszejnov_correlation}
{Guszejnov} D.,  {Hopkins} P.~F.,   {Krumholz} M.~R.,  2017, \mn@doi [\mnras]
  {10.1093/mnras/stx725}, \href
  {http://adsabs.harvard.edu/abs/2017MNRAS.468.4093G} {468, 4093}

\bibitem[\protect\citeauthoryear{{Guszejnov}, {Hopkins}  \&
  {Grudi{\'c}}}{{Guszejnov} et~al.}{2018}]{guszejnov_scaling_laws}
{Guszejnov} D.,  {Hopkins} P.~F.,   {Grudi{\'c}} M.~Y.,  2018, \mn@doi [\mnras]
  {10.1093/mnras/sty920}, \href
  {http://adsabs.harvard.edu/abs/2018MNRAS.477.5139G} {477, 5139}

\bibitem[\protect\citeauthoryear{{Guszejnov}, {Hopkins}  \&
  {Graus}}{{Guszejnov} et~al.}{2019}]{guszejnov_extragal_imf_var}
{Guszejnov} D.,  {Hopkins} P.~F.,   {Graus} A.~S.,  2019, arXiv e-prints, \href
  {https://ui.adsabs.harvard.edu/\#abs/2019arXiv190301533G} {p.
  arXiv:1903.01533}

\bibitem[\protect\citeauthoryear{{Hennebelle} \& {Chabrier}}{{Hennebelle} \&
  {Chabrier}}{2008}]{hc08}
{Hennebelle} P.,  {Chabrier} G.,  2008, \mn@doi [\apj] {10.1086/589916}, \href
  {http://adsabs.harvard.edu/abs/2008ApJ...684..395H} {684, 395}

\bibitem[\protect\citeauthoryear{{Heyer} \& {Dame}}{{Heyer} \&
  {Dame}}{2015}]{heyer_dame_2015}
{Heyer} M.,  {Dame} T.~M.,  2015, \mn@doi [\araa]
  {10.1146/annurev-astro-082214-122324}, \href
  {http://adsabs.harvard.edu/abs/2015ARA%26A..53..583H} {53, 583}

\bibitem[\protect\citeauthoryear{{Hopkins}}{{Hopkins}}{2012}]{hopkins_2012_excursion_set}
{Hopkins} P.~F.,  2012, \mn@doi [\mnras] {10.1111/j.1365-2966.2012.20730.x},
  \href {https://ui.adsabs.harvard.edu/abs/2012MNRAS.423.2016H} {423, 2016}

\bibitem[\protect\citeauthoryear{{Hopkins}}{{Hopkins}}{2015}]{Hopkins2015_GIZMO}
{Hopkins} P.~F.,  2015, \mn@doi [\mnras] {10.1093/mnras/stv195}, \href
  {http://adsabs.harvard.edu/abs/2015MNRAS.450...53H} {450, 53}

\bibitem[\protect\citeauthoryear{{Hopkins} \& {Raives}}{{Hopkins} \&
  {Raives}}{2016}]{hopkins_gizmo_mhd}
{Hopkins} P.~F.,  {Raives} M.~J.,  2016, \mn@doi [\mnras]
  {10.1093/mnras/stv2180}, \href
  {http://adsabs.harvard.edu/abs/2016MNRAS.455...51H} {455, 51}

\bibitem[\protect\citeauthoryear{{Hopkins}, {Quataert}  \& {Murray}}{{Hopkins}
  et~al.}{2012}]{Hopkins_2012_galaxy_structure}
{Hopkins} P.~F.,  {Quataert} E.,   {Murray} N.,  2012, \mn@doi [\mnras]
  {10.1111/j.1365-2966.2012.20578.x}, \href
  {http://adsabs.harvard.edu/abs/2012MNRAS.421.3488H} {421, 3488}

\bibitem[\protect\citeauthoryear{{Hopkins}, {Narayanan}  \& {Murray}}{{Hopkins}
  et~al.}{2013}]{hopkins_2013_sf_criteria}
{Hopkins} P.~F.,  {Narayanan} D.,   {Murray} N.,  2013, \mn@doi [\mnras]
  {10.1093/mnras/stt723}, \href
  {https://ui.adsabs.harvard.edu/abs/2013MNRAS.432.2647H} {432, 2647}

\bibitem[\protect\citeauthoryear{{Hopkins}, {Kere{\v s}}, {O{\~n}orbe},
  {Faucher-Gigu{\`e}re}, {Quataert}, {Murray}  \& {Bullock}}{{Hopkins}
  et~al.}{2014}]{hopkins2014_fire}
{Hopkins} P.~F.,  {Kere{\v s}} D.,  {O{\~n}orbe} J.,  {Faucher-Gigu{\`e}re}
  C.-A.,  {Quataert} E.,  {Murray} N.,   {Bullock} J.~S.,  2014, \mn@doi
  [\mnras] {10.1093/mnras/stu1738}, \href
  {http://adsabs.harvard.edu/abs/2014MNRAS.445..581H} {445, 581}

\bibitem[\protect\citeauthoryear{{Hopkins} et~al.,}{{Hopkins}
  et~al.}{2018}]{hopkins_fire2}
{Hopkins} P.~F.,  et~al., 2018, \mn@doi [\mnras] {10.1093/mnras/sty1690}, \href
  {https://ui.adsabs.harvard.edu/abs/2018MNRAS.480..800H} {480, 800}

\bibitem[\protect\citeauthoryear{{Hopkins} et~al.,}{{Hopkins}
  et~al.}{2019}]{Hopkins_FIRE2MHD_2019}
{Hopkins} P.~F.,  et~al., 2019, arXiv e-prints, \href
  {https://ui.adsabs.harvard.edu/abs/2019arXiv190504321H} {p. arXiv:1905.04321}

\bibitem[\protect\citeauthoryear{{Hung} et~al.,}{{Hung}
  et~al.}{2019}]{Hung_2019_gas_evolution}
{Hung} C.-L.,  et~al., 2019, \mn@doi [\mnras] {10.1093/mnras/sty2970}, \href
  {https://ui.adsabs.harvard.edu/abs/2019MNRAS.482.5125H} {482, 5125}

\bibitem[\protect\citeauthoryear{{Ib{\'a}{\~n}ez-Mej{\'{\i}}a}, {Mac Low},
  {Klessen}  \& {Baczynski}}{{Ib{\'a}{\~n}ez-Mej{\'{\i}}a}
  et~al.}{2017}]{ibanez_2017}
{Ib{\'a}{\~n}ez-Mej{\'{\i}}a} J.~C.,  {Mac Low} M.-M.,  {Klessen} R.~S.,
  {Baczynski} C.,  2017, \mn@doi [\apj] {10.3847/1538-4357/aa93fe}, \href
  {https://ui.adsabs.harvard.edu/abs/2017ApJ...850...62I} {850, 62}

\bibitem[\protect\citeauthoryear{Jones, Oliphant, Peterson  et~al.}{Jones
  et~al.}{2001}]{scipy}
Jones E.,  Oliphant T.,  Peterson P.,   et~al., 2001, {SciPy}: Open source
  scientific tools for {Python}, \url {http://www.scipy.org/}

\bibitem[\protect\citeauthoryear{{Katz}, {Weinberg}  \& {Hernquist}}{{Katz}
  et~al.}{1996}]{katz96}
{Katz} N.,  {Weinberg} D.~H.,   {Hernquist} L.,  1996, \mn@doi [\apjs]
  {10.1086/192305}, \href
  {https://ui.adsabs.harvard.edu/abs/1996ApJS..105...19K} {105, 19}

\bibitem[\protect\citeauthoryear{{Kauffmann}, {Pillai}  \&
  {Goldsmith}}{{Kauffmann} et~al.}{2013}]{kauffmann_pillai_2013}
{Kauffmann} J.,  {Pillai} T.,   {Goldsmith} P.~F.,  2013, \mn@doi [\apj]
  {10.1088/0004-637X/779/2/185}, \href
  {http://adsabs.harvard.edu/abs/2013ApJ...779..185K} {779, 185}

\bibitem[\protect\citeauthoryear{{Kroupa}}{{Kroupa}}{2002}]{kroupa_imf}
{Kroupa} P.,  2002, \mn@doi [Science] {10.1126/science.1067524}, \href
  {http://adsabs.harvard.edu/abs/2002Sci...295...82K} {295, 82}

\bibitem[\protect\citeauthoryear{{Krumholz}}{{Krumholz}}{2011}]{krumholz_stellar_mass_origin}
{Krumholz} M.~R.,  2011, \mn@doi [\apj] {10.1088/0004-637X/743/2/110}, \href
  {http://adsabs.harvard.edu/abs/2011ApJ...743..110K} {743, 110}

\bibitem[\protect\citeauthoryear{{Krumholz} \& {Gnedin}}{{Krumholz} \&
  {Gnedin}}{2011}]{krumholz_2011_self_shield}
{Krumholz} M.~R.,  {Gnedin} N.~Y.,  2011, \mn@doi [\apj]
  {10.1088/0004-637X/729/1/36}, \href
  {http://adsabs.harvard.edu/abs/2011ApJ...729...36K} {729, 36}

\bibitem[\protect\citeauthoryear{{Krumholz}, {Dekel}  \& {McKee}}{{Krumholz}
  et~al.}{2012}]{Krumholz_2012_SF_law}
{Krumholz} M.~R.,  {Dekel} A.,   {McKee} C.~F.,  2012, \mn@doi [\apj]
  {10.1088/0004-637X/745/1/69}, \href
  {https://ui.adsabs.harvard.edu/abs/2012ApJ...745...69K} {745, 69}

\bibitem[\protect\citeauthoryear{{Lakhlani} et~al.,}{{Lakhlani}
  et~al.}{2019}]{Lakhlani_2019_GMC_FIRE_present}
{Lakhlani} G.,  et~al., 2019, {The Structure and Properties of GMCs in the FIRE
  Simulations}

\bibitem[\protect\citeauthoryear{{Larson}}{{Larson}}{1981}]{larson_law}
{Larson} R.~B.,  1981, \mnras, \href
  {http://adsabs.harvard.edu/abs/1981MNRAS.194..809L} {194, 809}

\bibitem[\protect\citeauthoryear{{Leitherer} et~al.,}{{Leitherer}
  et~al.}{1999}]{1Leitherer_1999_Starburst99}
{Leitherer} C.,  et~al., 1999, \mn@doi [\apjs] {10.1086/313233}, \href
  {http://adsabs.harvard.edu/abs/1999ApJS..123....3L} {123, 3}

\bibitem[\protect\citeauthoryear{{Ma}, {Hopkins}, {Faucher-Gigu{\`e}re},
  {Zolman}, {Muratov}, {Kere{\v{s}}}  \& {Quataert}}{{Ma}
  et~al.}{2016}]{Ma_2016_mass_metallicity}
{Ma} X.,  {Hopkins} P.~F.,  {Faucher-Gigu{\`e}re} C.-A.,  {Zolman} N.,
  {Muratov} A.~L.,  {Kere{\v{s}}} D.,   {Quataert} E.,  2016, \mn@doi [\mnras]
  {10.1093/mnras/stv2659}, \href
  {https://ui.adsabs.harvard.edu/abs/2016MNRAS.456.2140M} {456, 2140}

\bibitem[\protect\citeauthoryear{{Ma}, {Hopkins}, {Wetzel}, {Kirby},
  {Angl{\'e}s-Alc{\'a}zar}, {Faucher-Gigu{\`e}re}, {Kere{\v s}}  \&
  {Quataert}}{{Ma} et~al.}{2017}]{Ma_2017_fire_morphology}
{Ma} X.,  {Hopkins} P.~F.,  {Wetzel} A.~R.,  {Kirby} E.~N.,
  {Angl{\'e}s-Alc{\'a}zar} D.,  {Faucher-Gigu{\`e}re} C.-A.,  {Kere{\v s}} D.,
   {Quataert} E.,  2017, \mn@doi [\mnras] {10.1093/mnras/stx273}, \href
  {https://ui.adsabs.harvard.edu/abs/2017MNRAS.467.2430M} {467, 2430}

\bibitem[\protect\citeauthoryear{{Matzner}}{{Matzner}}{2002}]{matzner02}
{Matzner} C.~D.,  2002, \mn@doi [\apj] {10.1086/338030}, \href
  {https://ui.adsabs.harvard.edu/abs/2002ApJ...566..302M} {566, 302}

\bibitem[\protect\citeauthoryear{{McKee} \& {Ostriker}}{{McKee} \&
  {Ostriker}}{2007}]{mckee_star_formation}
{McKee} C.~F.,  {Ostriker} E.~C.,  2007, \mn@doi [\araa]
  {10.1146/annurev.astro.45.051806.110602}, \href
  {http://adsabs.harvard.edu/abs/2007ARA%26A..45..565M} {45, 565}

\bibitem[\protect\citeauthoryear{{Miville-Desch{\^e}nes}, {Murray}  \&
  {Lee}}{{Miville-Desch{\^e}nes} et~al.}{2017}]{Miville_2017_MWG_GMC}
{Miville-Desch{\^e}nes} M.-A.,  {Murray} N.,   {Lee} E.~J.,  2017, \mn@doi
  [\apj] {10.3847/1538-4357/834/1/57}, \href
  {https://ui.adsabs.harvard.edu/abs/2017ApJ...834...57M} {834, 57}

\bibitem[\protect\citeauthoryear{{Offner} \& {Chaban}}{{Offner} \&
  {Chaban}}{2017}]{Offner_Chaban_2017_jets}
{Offner} S.~S.~R.,  {Chaban} J.,  2017, \mn@doi [\apj]
  {10.3847/1538-4357/aa8996}, \href
  {https://ui.adsabs.harvard.edu/abs/2017ApJ...847..104O} {847, 104}

\bibitem[\protect\citeauthoryear{{Offner}, {Kratter}, {Matzner}, {Krumholz}  \&
  {Klein}}{{Offner} et~al.}{2010}]{offner2010}
{Offner} S.~S.~R.,  {Kratter} K.~M.,  {Matzner} C.~D.,  {Krumholz} M.~R.,
  {Klein} R.~I.,  2010, \mn@doi [\apj] {10.1088/0004-637X/725/2/1485}, \href
  {http://adsabs.harvard.edu/abs/2010ApJ...725.1485O} {725, 1485}

\bibitem[\protect\citeauthoryear{{Oklop{\v{c}}i{\'c}}, {Hopkins}, {Feldmann},
  {Kere{\v{s}}}, {Faucher-Gigu{\`e}re}  \& {Murray}}{{Oklop{\v{c}}i{\'c}}
  et~al.}{2017}]{Oklopvic_2017_FIRE_highz_clumps}
{Oklop{\v{c}}i{\'c}} A.,  {Hopkins} P.~F.,  {Feldmann} R.,  {Kere{\v{s}}} D.,
  {Faucher-Gigu{\`e}re} C.-A.,   {Murray} N.,  2017, \mn@doi [\mnras]
  {10.1093/mnras/stw2754}, \href
  {https://ui.adsabs.harvard.edu/abs/2017MNRAS.465..952O} {465, 952}

\bibitem[\protect\citeauthoryear{{Orr} et~al.,}{{Orr}
  et~al.}{2017}]{orr_fire_ks}
{Orr} M.,  et~al., 2017, preprint, \href
  {http://adsabs.harvard.edu/abs/2017arXiv170101788O} {} (\mn@eprint {arXiv}
  {1701.01788})

\bibitem[\protect\citeauthoryear{{Padoan} \& {Nordlund}}{{Padoan} \&
  {Nordlund}}{2011}]{padoan_nordlund_2011_imf}
{Padoan} P.,  {Nordlund} {\AA}.,  2011, \mn@doi [\apjl]
  {10.1088/2041-8205/741/1/L22}, \href
  {http://adsabs.harvard.edu/abs/2011ApJ...741L..22P} {741, L22}

\bibitem[\protect\citeauthoryear{{Padoan}, {Pan}, {Haugb{\o}lle}  \&
  {Nordlund}}{{Padoan} et~al.}{2016}]{Padoan_2016_isoT_MHD_SN_driving}
{Padoan} P.,  {Pan} L.,  {Haugb{\o}lle} T.,   {Nordlund} {\r{A}}.,  2016,
  \mn@doi [\apj] {10.3847/0004-637X/822/1/11}, \href
  {https://ui.adsabs.harvard.edu/abs/2016ApJ...822...11P} {822, 11}

\bibitem[\protect\citeauthoryear{{Pan}, {Fujimoto}, {Tasker}, {Rosolowsky},
  {Colombo}, {Benincasa}  \& {Wadsley}}{{Pan}
  et~al.}{2015}]{Pan2015_synth_obs_GMC}
{Pan} H.-A.,  {Fujimoto} Y.,  {Tasker} E.~J.,  {Rosolowsky} E.,  {Colombo} D.,
  {Benincasa} S.~M.,   {Wadsley} J.,  2015, \mn@doi [\mnras]
  {10.1093/mnras/stv1843}, \href
  {https://ui.adsabs.harvard.edu/abs/2015MNRAS.453.3082P} {453, 3082}

\bibitem[\protect\citeauthoryear{{Pavesi}, {Riechers}, {Faisst}, {Stacey}  \&
  {Capak}}{{Pavesi} et~al.}{2018}]{Pavesi_2018_low_sfe_high_z_galaxies}
{Pavesi} R.,  {Riechers} D.~A.,  {Faisst} A.~L.,  {Stacey} G.~J.,   {Capak}
  P.~L.,  2018, arXiv e-prints, \href
  {https://ui.adsabs.harvard.edu/abs/2018arXiv181200006P} {p. arXiv:1812.00006}

\bibitem[\protect\citeauthoryear{{Pettitt}, {Egusa}, {Dobbs}, {Tasker},
  {Fujimoto}  \& {Habe}}{{Pettitt} et~al.}{2018}]{Pettitt_2018_GMC_galaxy_sim}
{Pettitt} A.~R.,  {Egusa} F.,  {Dobbs} C.~L.,  {Tasker} E.~J.,  {Fujimoto} Y.,
   {Habe} A.,  2018, \mn@doi [\mnras] {10.1093/mnras/sty2040}, \href
  {https://ui.adsabs.harvard.edu/abs/2018MNRAS.480.3356P} {480, 3356}

\bibitem[\protect\citeauthoryear{{Rafelski}, {Wolfe}, {Prochaska}, {Neeleman}
  \& {Mendez}}{{Rafelski} et~al.}{2012}]{Rafelski_2012_Lyalpha_Z}
{Rafelski} M.,  {Wolfe} A.~M.,  {Prochaska} J.~X.,  {Neeleman} M.,   {Mendez}
  A.~J.,  2012, \mn@doi [\apj] {10.1088/0004-637X/755/2/89}, \href
  {https://ui.adsabs.harvard.edu/abs/2012ApJ...755...89R} {755, 89}

\bibitem[\protect\citeauthoryear{{Rice}, {Goodman}, {Bergin}, {Beaumont}  \&
  {Dame}}{{Rice} et~al.}{2016}]{rice2016_mw_gmc_catalogue}
{Rice} T.~S.,  {Goodman} A.~A.,  {Bergin} E.~A.,  {Beaumont} C.,   {Dame}
  T.~M.,  2016, \mn@doi [\apj] {10.3847/0004-637X/822/1/52}, \href
  {http://adsabs.harvard.edu/abs/2016ApJ...822...52R} {822, 52}

\bibitem[\protect\citeauthoryear{{Richings} \& {Schaye}}{{Richings} \&
  {Schaye}}{2016}]{Riching_2016_GMC_chemical_evol_in_sims}
{Richings} A.~J.,  {Schaye} J.,  2016, \mn@doi [\mnras]
  {10.1093/mnras/stw1135}, \href
  {https://ui.adsabs.harvard.edu/abs/2016MNRAS.460.2297R} {460, 2297}

\bibitem[\protect\citeauthoryear{{Rosolowsky} \& {Leroy}}{{Rosolowsky} \&
  {Leroy}}{2006}]{Rosolowsky_2006_GMC_CPROPS}
{Rosolowsky} E.,  {Leroy} A.,  2006, \mn@doi [\pasp] {10.1086/502982}, \href
  {https://ui.adsabs.harvard.edu/abs/2006PASP..118..590R} {118, 590}

\bibitem[\protect\citeauthoryear{{Rosolowsky}, {Pineda}, {Kauffmann}  \&
  {Goodman}}{{Rosolowsky} et~al.}{2008}]{rosolowsky2008_dendogram}
{Rosolowsky} E.~W.,  {Pineda} J.~E.,  {Kauffmann} J.,   {Goodman} A.~A.,  2008,
  \mn@doi [\apj] {10.1086/587685}, \href
  {http://adsabs.harvard.edu/abs/2008ApJ...679.1338R} {679, 1338}

\bibitem[\protect\citeauthoryear{{Sharma}, {Richard}, {Yuan}, {Gupta},
  {Kewley}, {Patr{\'\i}cio}, {Leethochawalit}  \& {Jones}}{{Sharma}
  et~al.}{2018}]{Sharma_2018_hires_lensed_highz_galaxy}
{Sharma} S.,  {Richard} J.,  {Yuan} T.,  {Gupta} A.,  {Kewley} L.,
  {Patr{\'\i}cio} V.,  {Leethochawalit} N.,   {Jones} T.~A.,  2018, \mn@doi
  [\mnras] {10.1093/mnras/sty2352}, \href
  {https://ui.adsabs.harvard.edu/abs/2018MNRAS.481.1427S} {481, 1427}

\bibitem[\protect\citeauthoryear{{Sparre}, {Hayward}, {Feldmann},
  {Faucher-Gigu{\`e}re}, {Muratov}, {Kere{\v s}}  \& {Hopkins}}{{Sparre}
  et~al.}{2017}]{Sparre_2017_fire_starbursts}
{Sparre} M.,  {Hayward} C.~C.,  {Feldmann} R.,  {Faucher-Gigu{\`e}re} C.-A.,
  {Muratov} A.~L.,  {Kere{\v s}} D.,   {Hopkins} P.~F.,  2017, \mn@doi [\mnras]
  {10.1093/mnras/stw3011}, \href
  {https://ui.adsabs.harvard.edu/abs/2017MNRAS.466...88S} {466, 88}

\bibitem[\protect\citeauthoryear{{Springel}, {White}, {Tormen}  \&
  {Kauffmann}}{{Springel} et~al.}{2001}]{Springel_2001_SUBFIND}
{Springel} V.,  {White} S.~D.~M.,  {Tormen} G.,   {Kauffmann} G.,  2001,
  \mn@doi [\mnras] {10.1046/j.1365-8711.2001.04912.x}, \href
  {http://adsabs.harvard.edu/abs/2001MNRAS.328..726S} {328, 726}

\bibitem[\protect\citeauthoryear{{Su}, {Hopkins}, {Hayward}, {Faucher-Giguere},
  {Keres}, {Ma}  \& {Robles}}{{Su} et~al.}{2016}]{kungyi_weak_mhd_2016}
{Su} K.-Y.,  {Hopkins} P.~F.,  {Hayward} C.~C.,  {Faucher-Giguere} C.-A.,
  {Keres} D.,  {Ma} X.,   {Robles} V.~H.,  2016, preprint, \href
  {http://adsabs.harvard.edu/abs/2016arXiv160705274S} {} (\mn@eprint {arXiv}
  {1607.05274})

\bibitem[\protect\citeauthoryear{{Su}, {Hayward}, {Hopkins}, {Quataert},
  {Faucher-Gigu{\`e}re}  \& {Kere{\v{s}}}}{{Su}
  et~al.}{2018}]{KungYi_2018_magnetic_fields}
{Su} K.-Y.,  {Hayward} C.~C.,  {Hopkins} P.~F.,  {Quataert} E.,
  {Faucher-Gigu{\`e}re} C.-A.,   {Kere{\v{s}}} D.,  2018, \mn@doi [\mnras]
  {10.1093/mnrasl/slx172}, \href
  {https://ui.adsabs.harvard.edu/abs/2018MNRAS.473L.111S} {473, L111}

\bibitem[\protect\citeauthoryear{{Tacconi} et~al.,}{{Tacconi}
  et~al.}{2013}]{Tacconi_2013_massive_galaxy_gas}
{Tacconi} L.~J.,  et~al., 2013, \mn@doi [\apj] {10.1088/0004-637X/768/1/74},
  \href {https://ui.adsabs.harvard.edu/abs/2013ApJ...768...74T} {768, 74}

\bibitem[\protect\citeauthoryear{{Tritsis}, {Panopoulou}, {Mouschovias},
  {Tassis}  \& {Pavlidou}}{{Tritsis}
  et~al.}{2015}]{Tritsis_2015_magnetic_field_density_relation}
{Tritsis} A.,  {Panopoulou} G.~V.,  {Mouschovias} T.~C.,  {Tassis} K.,
  {Pavlidou} V.,  2015, \mn@doi [\mnras] {10.1093/mnras/stv1133}, \href
  {https://ui.adsabs.harvard.edu/abs/2015MNRAS.451.4384T} {451, 4384}

\bibitem[\protect\citeauthoryear{{Ward}, {Benincasa}, {Wadsley}, {Sills}  \&
  {Couchman}}{{Ward} et~al.}{2016}]{Ward_2016_simulated_clouds}
{Ward} R.~L.,  {Benincasa} S.~M.,  {Wadsley} J.,  {Sills} A.,   {Couchman}
  H.~M.~P.,  2016, \mn@doi [\mnras] {10.1093/mnras/stv2360}, \href
  {https://ui.adsabs.harvard.edu/abs/2016MNRAS.455..920W} {455, 920}

\bibitem[\protect\citeauthoryear{{Wiersma}, {Schaye}  \& {Smith}}{{Wiersma}
  et~al.}{2009}]{Wiersma2009_cooling}
{Wiersma} R.~P.~C.,  {Schaye} J.,   {Smith} B.~D.,  2009, \mn@doi [\mnras]
  {10.1111/j.1365-2966.2008.14191.x}, \href
  {http://adsabs.harvard.edu/abs/2009MNRAS.393...99W} {393, 99}

\bibitem[\protect\citeauthoryear{{Wisnioski} et~al.,}{{Wisnioski}
  et~al.}{2015}]{Wisnioski_2015_KMOS3D}
{Wisnioski} E.,  et~al., 2015, \mn@doi [\apj] {10.1088/0004-637X/799/2/209},
  \href {https://ui.adsabs.harvard.edu/abs/2015ApJ...799..209W} {799, 209}

\makeatother
\end{thebibliography}

\appendix

\section{CloudPhinder algorithm}
\label{appendix:cloudphinder}
The CloudPhinder algorithm identifies gravitationally-bound iso-density contours of a given particle type in output snapshots from {\small GADGET}, {\small GIZMO}, {\small AREPO}, or related codes. Concisely, its approach is to examine successively lower iso-density contours surrounding density peaks, calculating the virial parameter (Equation \ref{eq:alphavir}) at each step, and looks for the lowest density value for which the particles in the contour satisfy $\alpha_{\rm vir} < \alpha_{\rm crit}$ and designates this structure as a bound cloud. This algorithm effectively finds the largest self-gravitating density contour around each density peak.

The initial construction of density iso-contours from unstructured particle data largely follows {\small SUBFIND} \citep{Springel_2001_SUBFIND}. First, the particles are sorted in decreasing order of density. Then, for each particle $i$, starting with the densest particle, the order of operations is:
\begin{enumerate}
    \item Determine the $N_{\rm neighbor}\sim 32$ nearest neighbours of particle $i$.
    \item Of those neighbours, determine the subset that are denser than that particle's density $\rho_i$. 
    \item Consider three possibilities for assigning particle $i$ to a group: 
    \begin{enumerate}
        \item If there are no denser neighbours, particle $i$ is located at a density peak, so create a new group that contains only that particle.
        \item If there is exactly one denser neighbour $j$, or the two closest denser neighbours $j$ and $k$ belong to the same group, assign particle $i$ to the group to which particle $j$ belongs.
        \item If the two closest denser neighbours $j$ and $k$ belong to different groups, particle $i$ is located at a saddle point in the density field. Merge the groups to which $j$ and $k$ belong, and add particle $i$ to that group.
    \end{enumerate}
    \item Evaluate the virial parameter of the group to which particle $i$ was assigned (Equation \ref{eq:alphavir}), measuring the kinetic energy in the centre-of-mass frame of the group. If $\alpha_{\rm vir} \leq \alpha_{\rm crit}$, save the group as a bound cloud, and delete any previously-found bound clouds that are subsets of the present group.
\end{enumerate}

Thus, the algorithm will proceed to successively lower density contours, merging together density peaks at saddle points, until it either reaches a pre-defined density minimum ($n=1\,\mathrm{cm}^{-3}$ in our case) or eventually the entire ISM mass is considered as a potential bound group. The final output is the set of bound clouds that are not substructures of any larger cloud. In practice, this algorithm requires fast tree-based methods for nearest-neighbour searches ({\small cKDTree} from {\small scipy}, \citealt{scipy}) and evaluating the gravitational potential (using the Python package {\small pykdgrav}\footnote{\url{https://www.github.com/omgspace/pykdgrav}}). 

\section{Results for m11q dwarf galaxy}\label{sec:m11q}

This appendix contains the equivalents of Figures \ref{fig:bulk_properties_evol} and \ref{fig:m12i_evolplots} for the \textbf{m11q} simulated dwarf galaxy.

\begin{figure*}
\begin {center}
\includegraphics[width=0.33\linewidth]{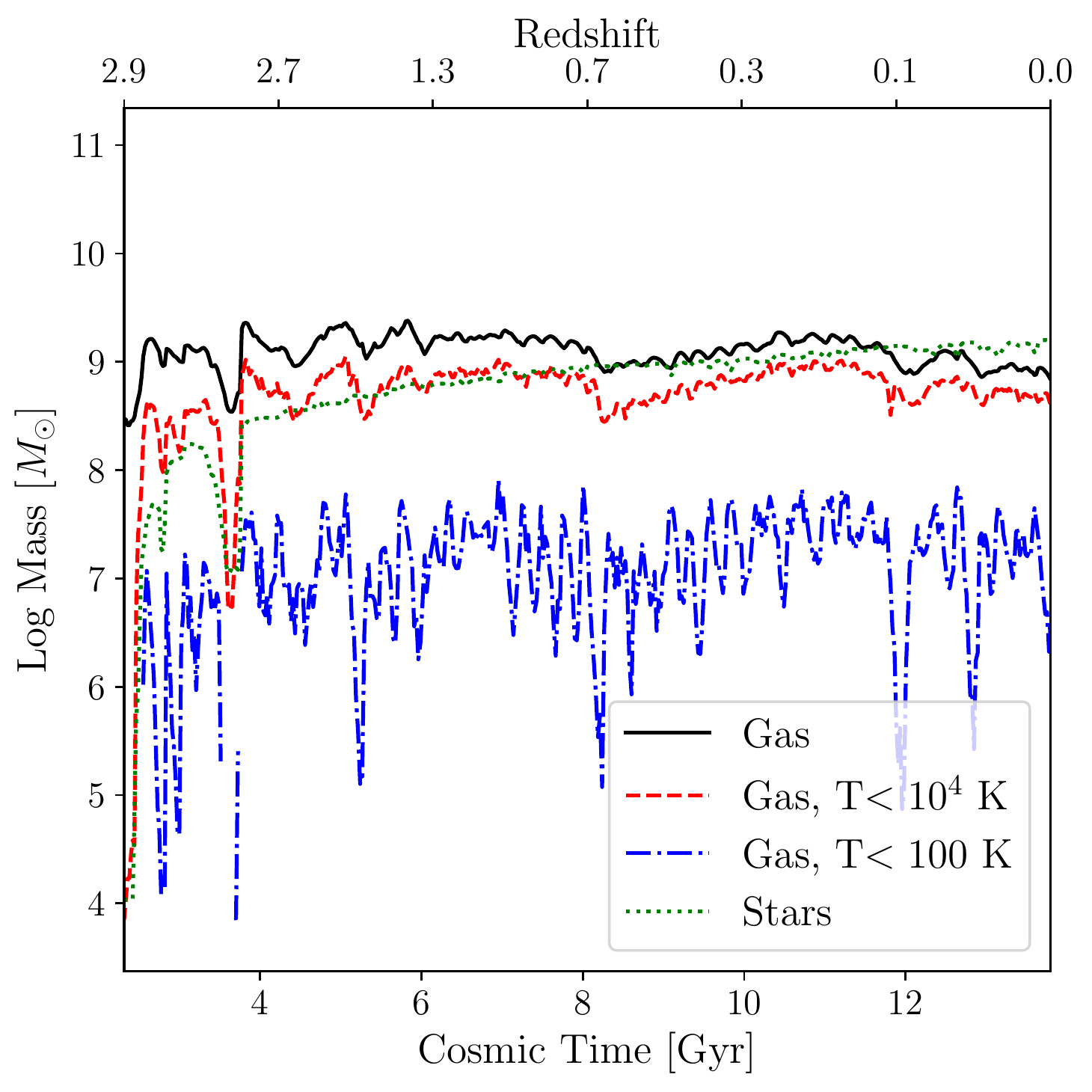}
\includegraphics[width=0.33\linewidth]{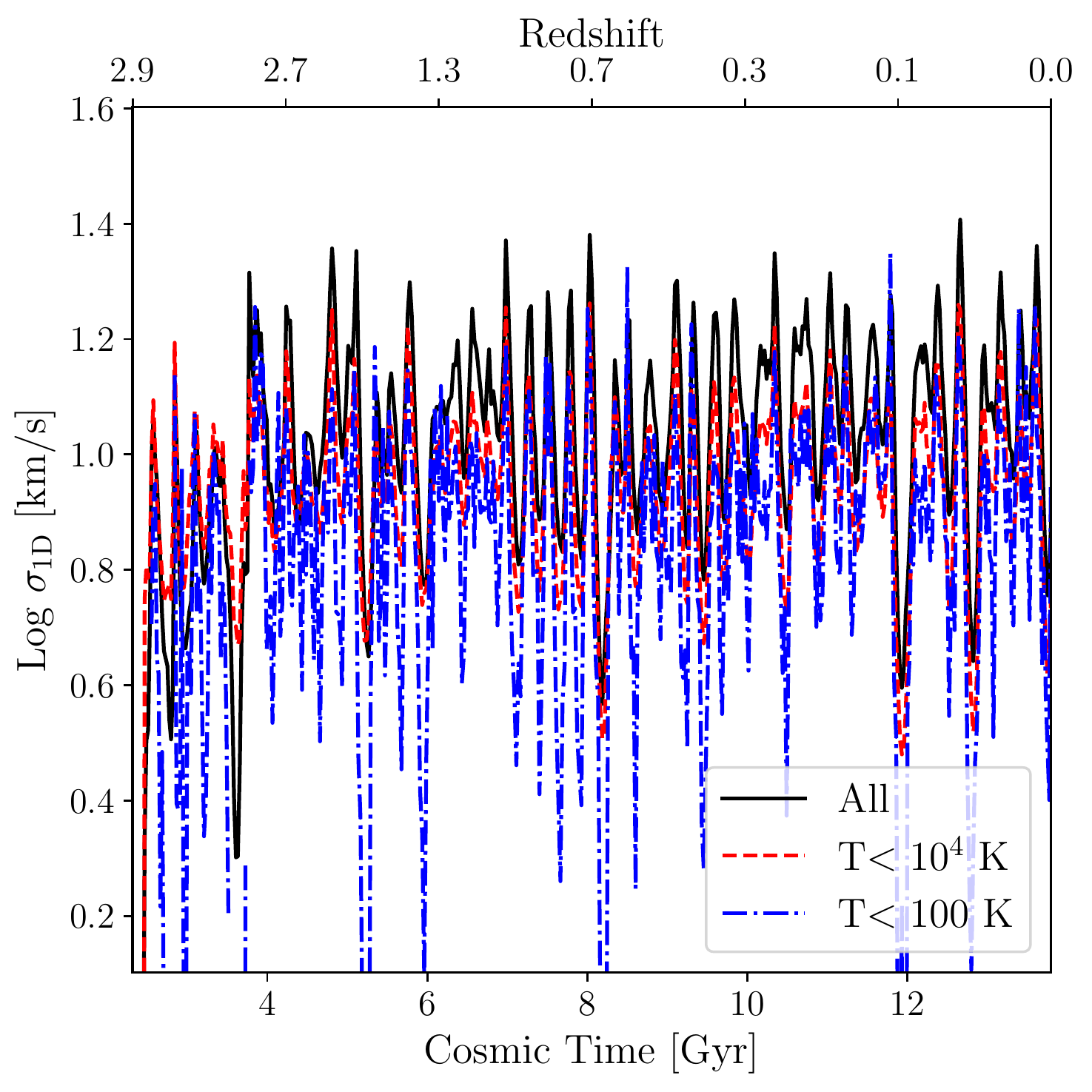}
\includegraphics[width=0.33\linewidth]{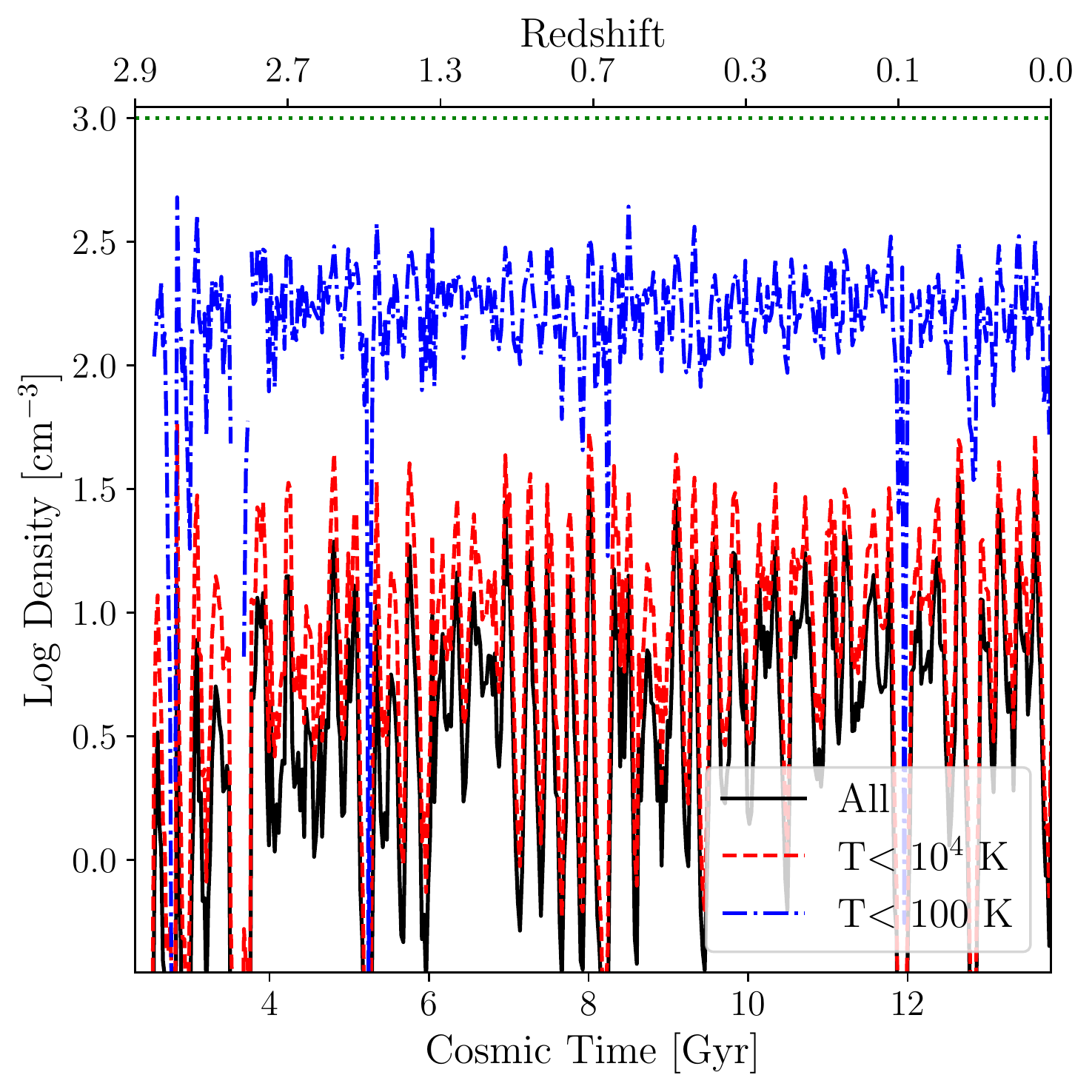}\\
\vspace{-0.4cm}
\caption{Evolution of average galactic properties in \textbf{m11q}, including galactic gas and stellar mass (left), average 1D velocity dispersion on 500 pc scale (middle) and gas density over cosmic time right, $n_{\rm crit}$ for star formation noted with horizontal line). Except for the stellar and cold gas mass these galactic properties appear to have no trend on large timescales, but they exhibit factor of 2 level variations on shorter timescales due to the \myquote{burstiness} of star formation in the galaxy.} 
\label{fig:bulk_properties_evol_m11q}
\vspace{-0.5cm}
\end {center}
\end{figure*}

\begin{figure*}
\begin {center}
\includegraphics[width=0.33\linewidth]{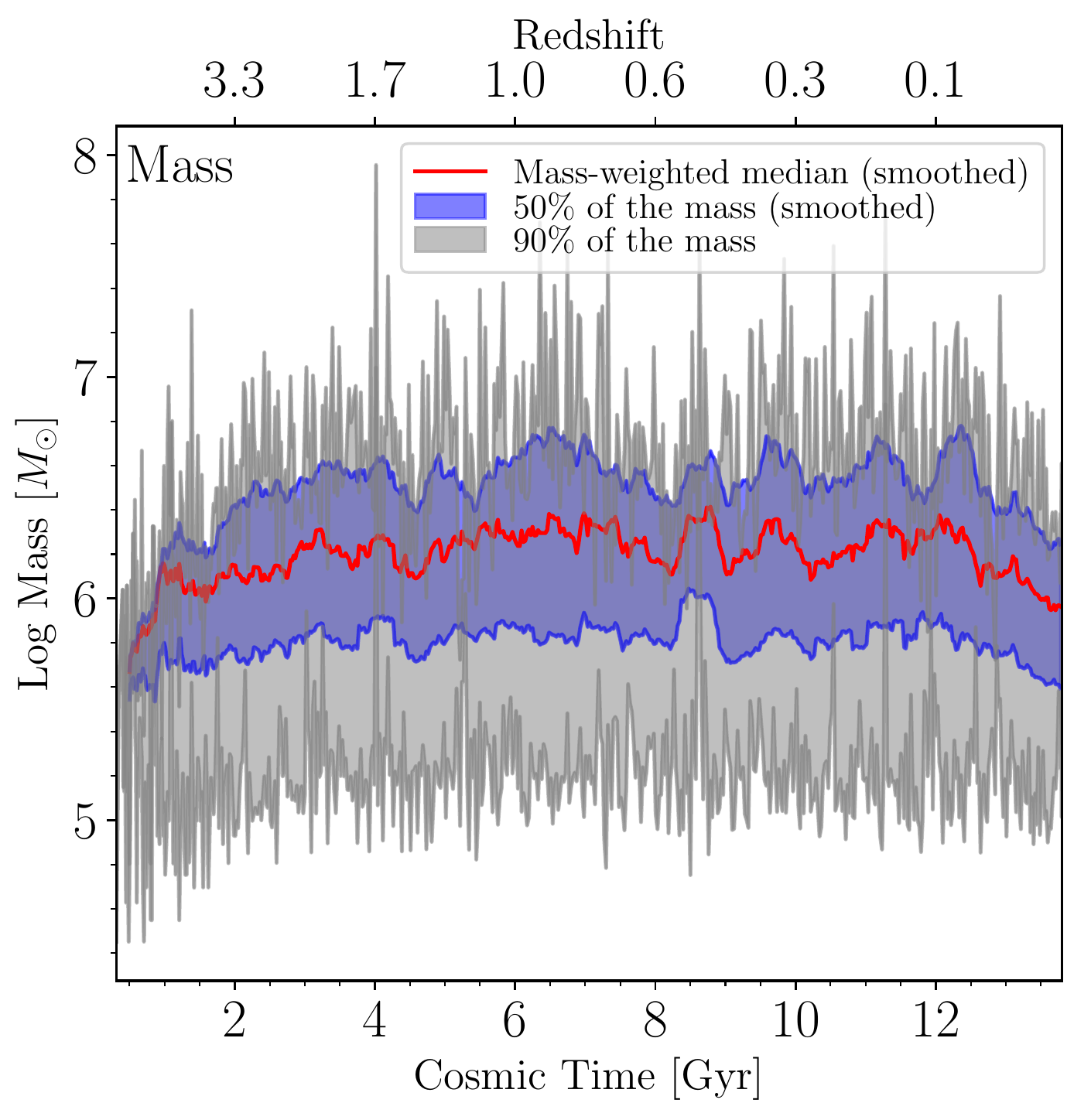}
\includegraphics[width=0.33\linewidth]{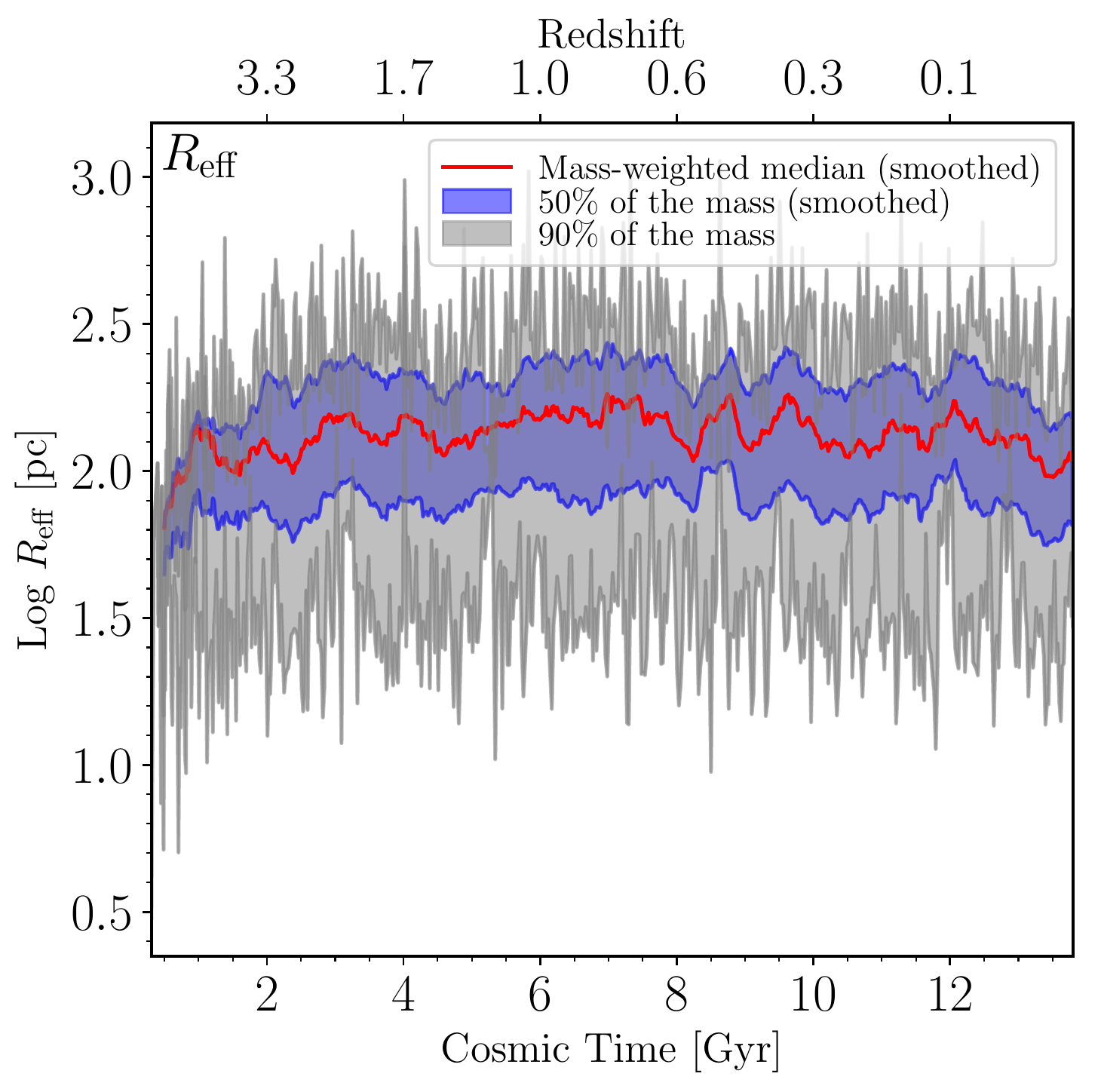}
\includegraphics[width=0.33\linewidth]{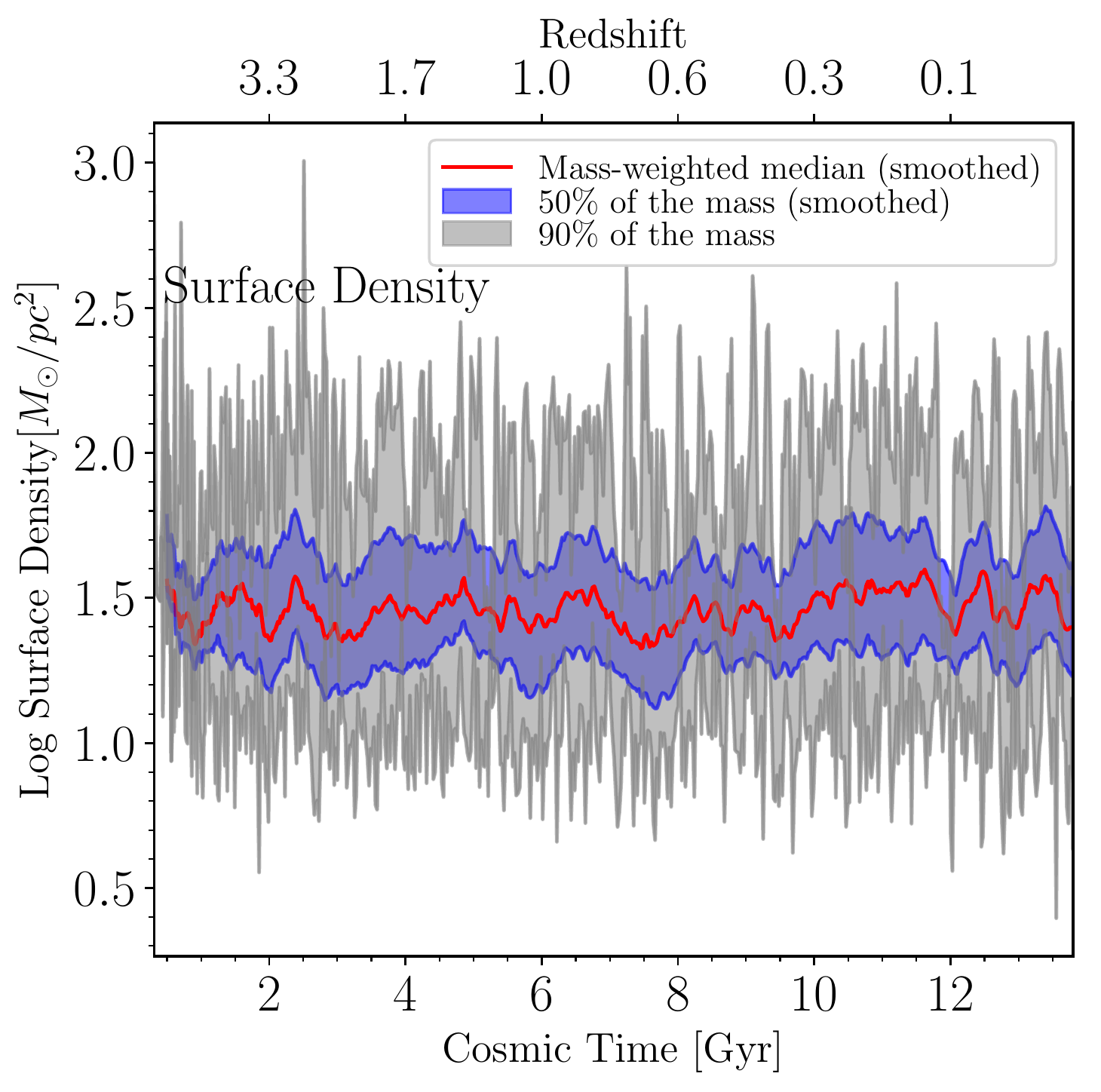} \\
\includegraphics[width=0.33\linewidth]{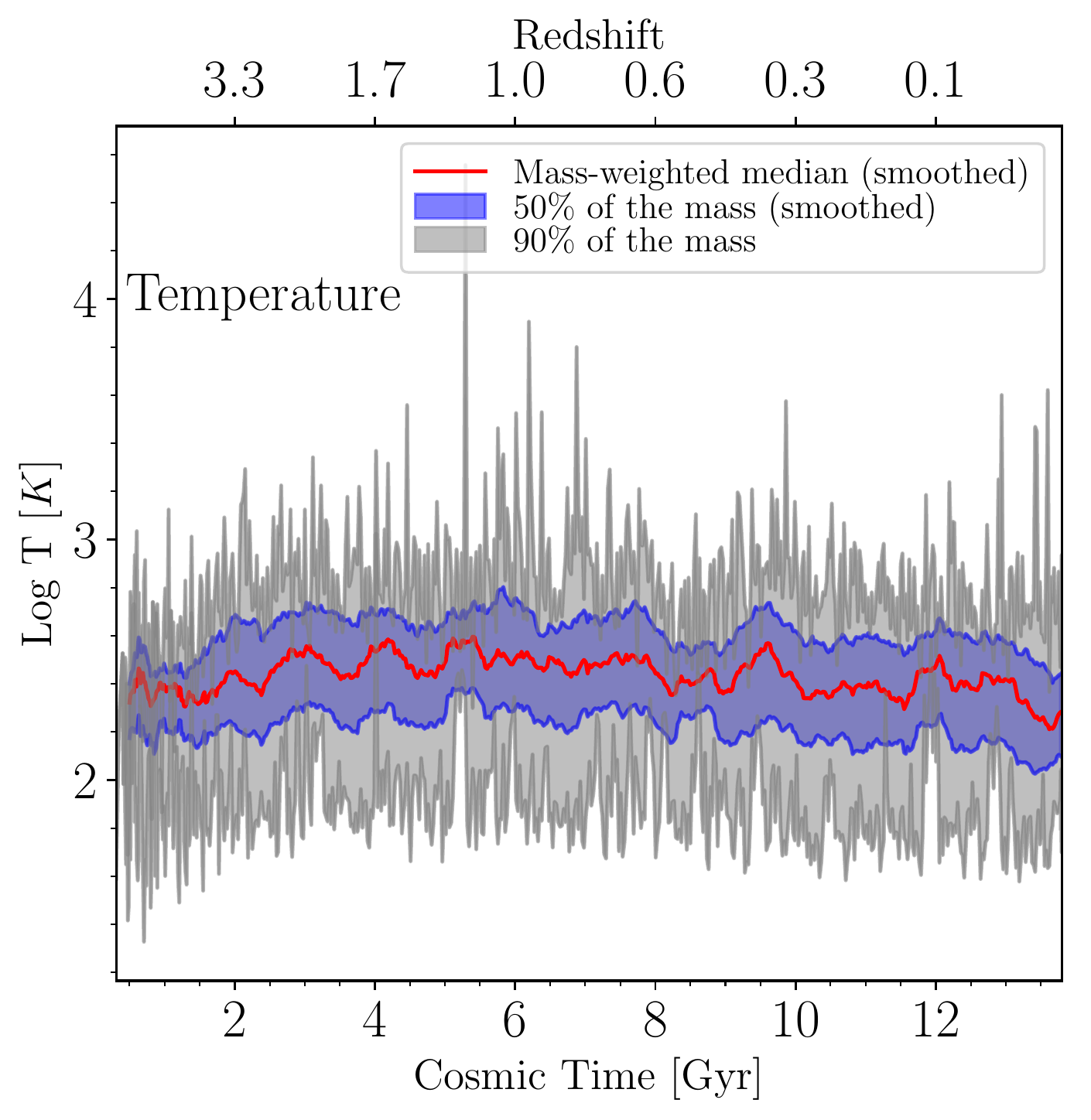}
\includegraphics[width=0.33\linewidth]{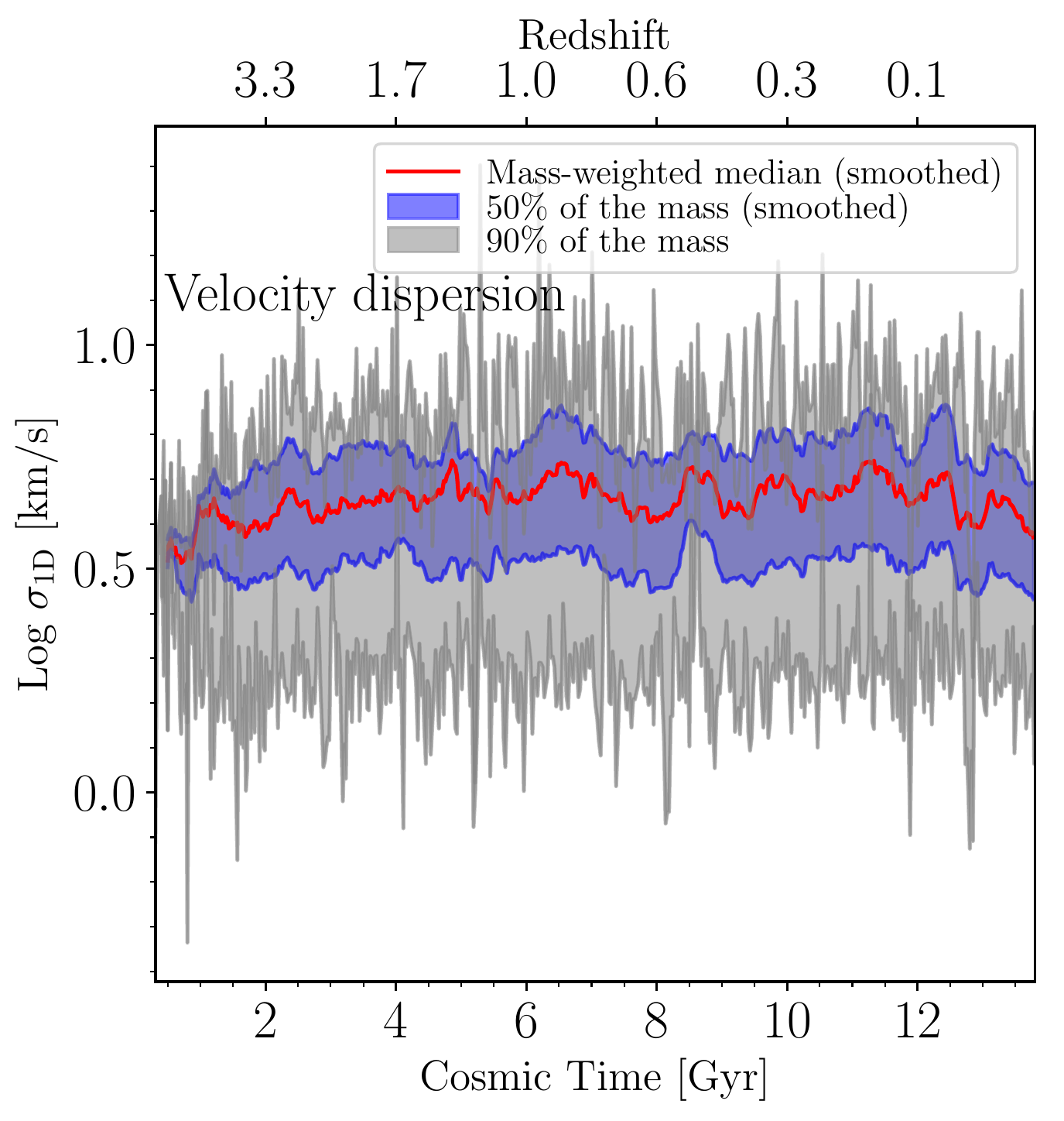}
\includegraphics[width=0.33\linewidth]{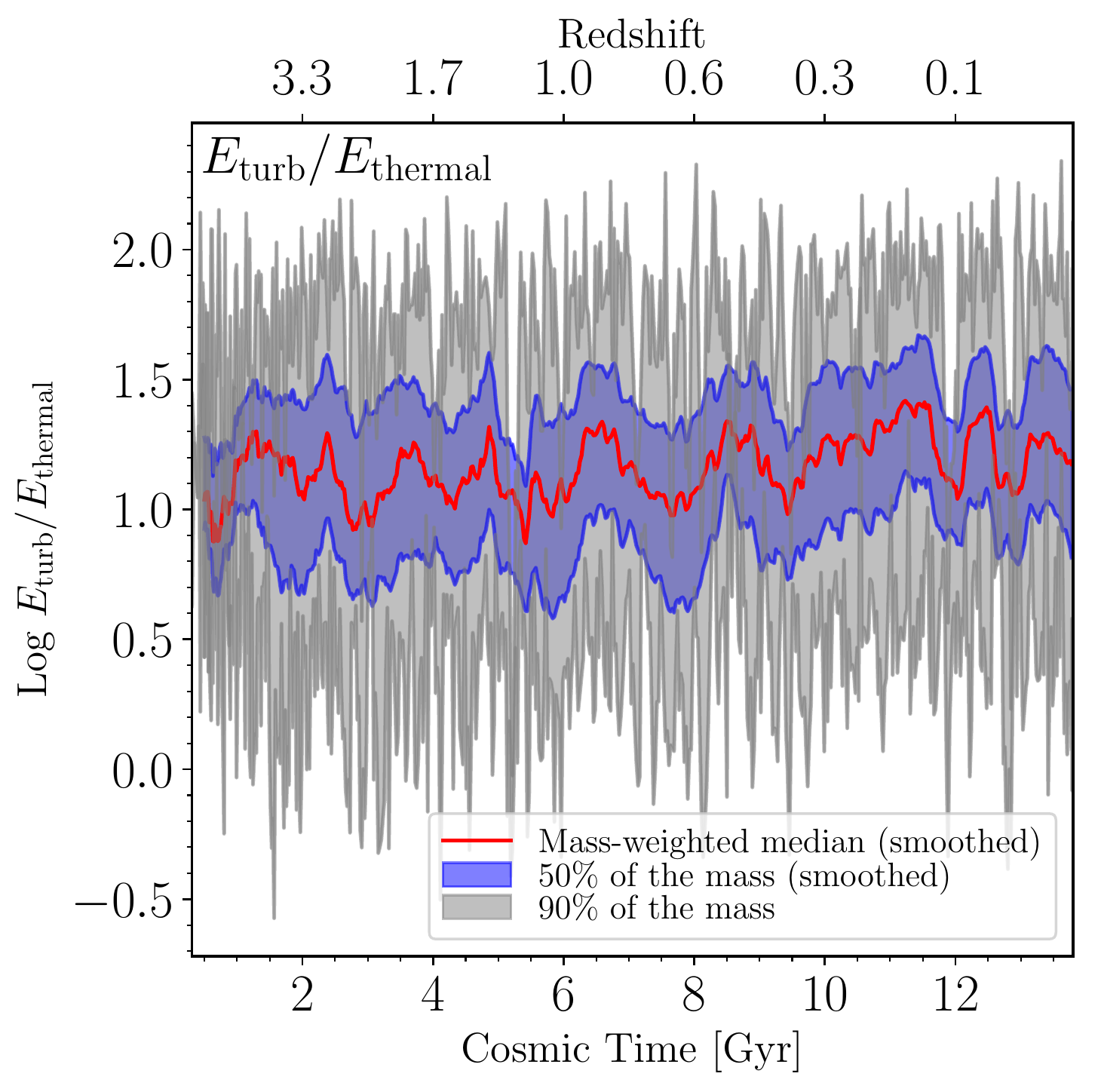}\\
\includegraphics[width=0.33\linewidth]{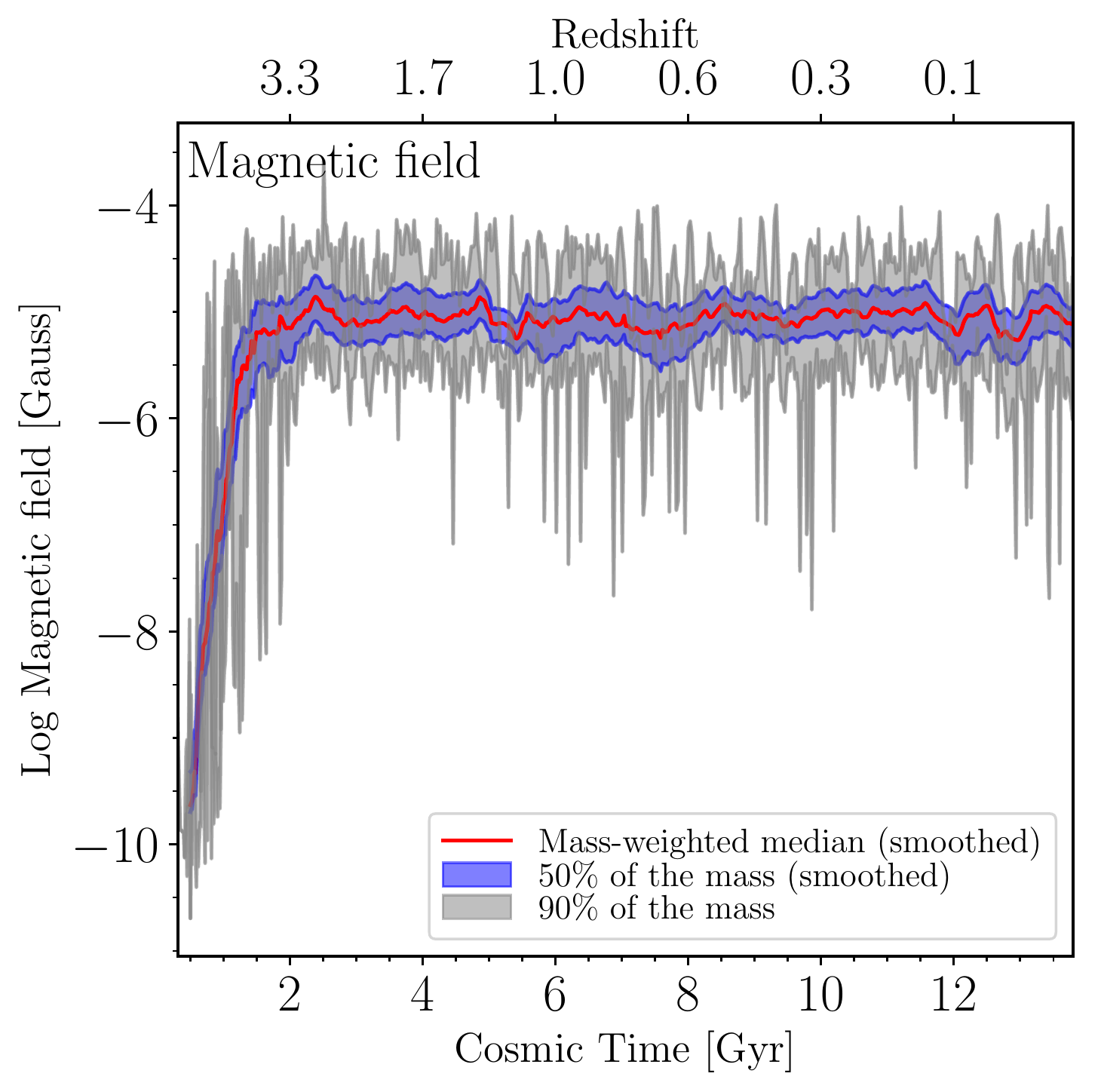}
\includegraphics[width=0.33\linewidth]{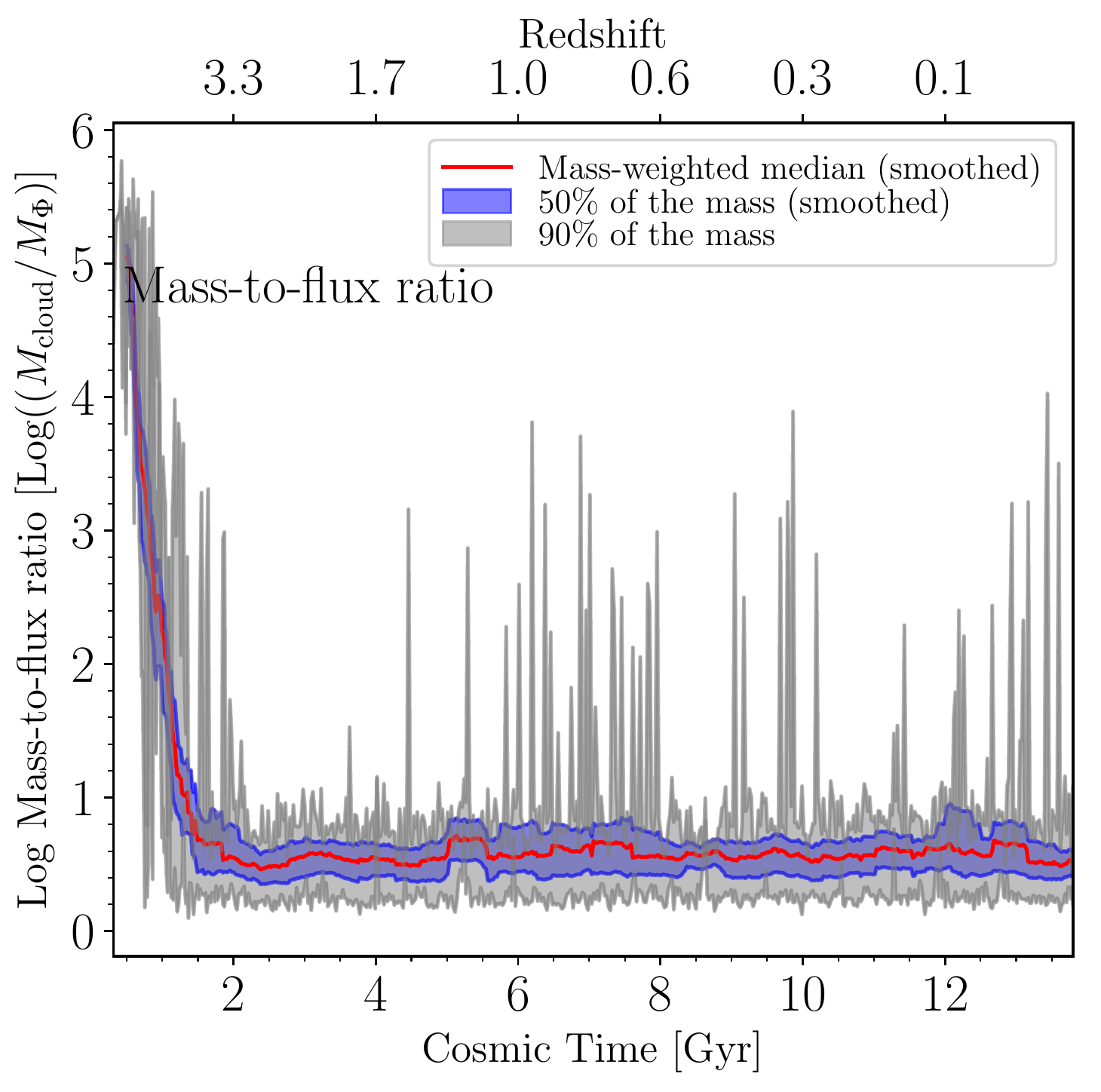}
\includegraphics[width=0.33\linewidth]{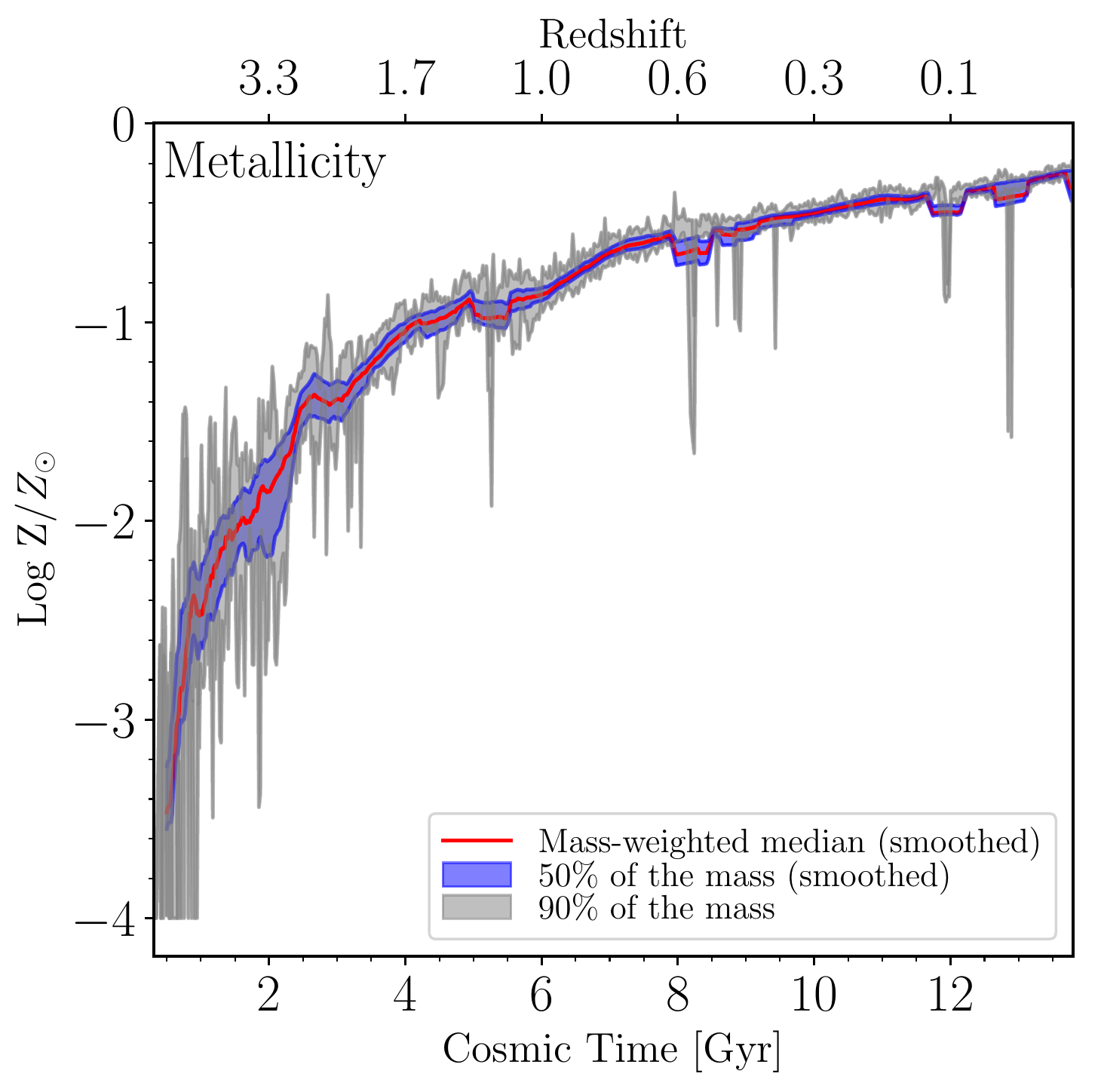}\\
\vspace{-0.5cm}
\caption{Evolution of GMC properties in the \textbf{m11q} simulated dwarf galaxy over cosmic time, including mass, size, surface density, temperature, turbulent velocity dispersion, turbulent to thermal energy ratio, magnetic field, mass-to-flux ratio and metallicity. Similar to the results for \textbf{m12i} (see Figure \ref{fig:m12i_evolplots}), we find that almost all of these properties remain constant after the galaxy forms. Similar to \textbf{m12i} metallicity is an exception as it rises it leads to more efficient cooling, which in turn leads to a slight decrease in temperature. Interestingly the actual values of bulk properties are similar to those in the MW-like \textbf{m12i}, with the exception of metallicity that is a factor of 2 lower and, as a result, the temperatures are about 50\% higher.}
\label{fig:m11q_evolplots}
\vspace{-0.5cm}
\end {center}
\end{figure*}

\end{document}